\documentclass[bibyear]{aa} 

\usepackage{natbib}
\usepackage{color}
\usepackage{graphicx}
\usepackage{ae,aecompl}
\usepackage{txfonts}
\usepackage{subcaption}

\usepackage{amsmath}
\usepackage{graphicx}
\usepackage{txfonts}
\usepackage[normalem]{ulem}
\usepackage{overpic}
\usepackage{dblfloatfix} 
\usepackage{subcaption}
\newcommand{\ramses}{\texttt{RAMSES}}
\newcommand{\agama}{\texttt{AGAMA}}
\newcommand{\epsELN}{\varepsilon_{{}_{\rm gELN}}}

\begin{document} 

   \title{Mapping parameters of idealised hydrodynamic galaxy simulations to bar properties: a case study with the RAMSES code}
   \titlerunning{Mapping parameters of galaxy simulations to bar properties}

   \subtitle{}

    \author{Srikanth T. Nagesh
          \inst{1,2}
          \and
          Benoit Famaey\inst{2}
          \and
          Jonathan Freundlich\inst{2}
          \and
          Yves Revaz\inst{1}
          \and 
          Giacomo Monari\inst{2}
          \and
          Arnaud Siebert\inst{2}
          \and
        Rodrigo Ibata\inst{2}   
          }

   \institute{
        Laboratoire d'Astrophysique, EPFL, Observatoire de Sauverny, 1290 Versoix, Switzerland\\
        \email{srikanth.nagesh@epfl.ch}
         \and
         Université de Strasbourg, CNRS, Observatoire astronomique de Strasbourg, UMR 7550, F-67000 Strasbourg, France\\ 
             }

   \date{} 

 
  \abstract
   {Cosmological simulations with high spatial resolution display a missing galactic bars problem, i.e., not reproducing the fraction of bars as a function of redshift, and/or a too-short bars problem. In order to take a step back on these problems, we run and analyse a grid of idealised hydrodynamic simulations of disc galaxies with stellar mass $\sim 10^{10} \, {\rm M}_\odot$, performed with {\tt RAMSES}. We vary the resolution, gas mass, stellar velocity dispersion, bulge mass, halo mass, and concentration in order to identify which parameters inhibit bar formation in such an idealised framework, from a purely Newtonian dynamical perspective, without cooling, star formation nor feedback. More precisely, we check whether an initially axisymmetric disc forms a bar within the typical time elapsed between $z\sim 1$ and $z \sim 0.2$, and show that most diagnostics proposed in the literature are too simplistic to reliably predict the outcome. Nevertheless, a region of parameter space that seems to efficiently inhibit bar formation is still identified in our simulations: a high Romeo-Falstad stability parameter, with a threshold decreasing quadratically with the bulge mass, and a high generalised Efstathiou-Lake-Negroponte (ELN) parameter, including the bulge mass. Moreover, decreasing the gas fraction and increasing the number of dark matter particles both tend to decrease the growth rate of the bar. We argue that, in cosmological simulations, if the bar is destroyed, e.g. by the formation of a bulge, once the galaxy has already moved to such a bar-inhibiting region of parameter space, it is likely difficult to re-form. It is plausible that this modern version of the angular momentum catastrophe is still happening in large-volume cosmological simulations. Finally, we confirm previous results indicating that only baryon-dominated discs away from the stellar-to-halo-mass relation expected from abundance matching can form large enough bars with respect to their corotation radius.}
   \maketitle
%

\section{Introduction}

Galactic bars are the dominant non-axisymmetric structures within disc galaxies, and are present in about two thirds of them in the local Universe. The existence of bars has been known for over a century, with their first recorded observational identification dating back to the work of \citet{Curtis1918}, as a ``band of matter extending diametrically across the inner parts'' of some spiral ``nebulae'', before \citet{Hubble1926} introduced the nomenclature of ``barred spiral'' galaxies. Observations nowadays show that up to 70\% (depending a bit on the detection criteria) of disc galaxies in the local Universe do host a bar \citep{Erwin_2018}. The fraction of barred galaxies also appears to strongly decrease with redshift \citep[e.g.,][]{Sheth_2008, Simmons_2014, Euclid_bars_2025}, but barred galaxy candidates are detected up to at least $z \sim 3$ \citep[e.g.,][]{Costantin_2023, LeConte_2024, Geron_2025}, a redshift at which their fraction stabilizes at the level of 10-20\%. 

Galactic bars have also naturally been among the first instabilities to be studied within rotationally-supported cold discs \citep{Hohl_1971, Ostriker_Peebles_1973, Lynden_1979, Tremaine_1984, Athanassoula_Sellwood_1986}, but despite more than five decades of research and their quasi-ubiquity in the local Universe, the physics regulating their formation, survival, inhibition, or dissolution within the cosmological context still remains rather poorly understood. Indeed, while the physics of bar formation within pure $N$-body disc-halo systems \citep[e.g.,][]{Petersen_2019,Sellwood_2025} is steadily gaining more understanding, despite the exact mechanisms and resonances at play still not being fully under control, the situation is even worse in state-of-the-art large-volume cosmological simulations based on the $\Lambda$ Cold Dark Matter ($\Lambda$CDM) cosmological model and standard baryonic physics: those simulations actually fail to produce the observed decreasing fraction of barred galaxies across redshift \citep[e.g.,][]{Algorry_2017, Peschken_Lokas_2019, Rosas_2020, Zhao_2020, Roshan2021_2, Reddish_2022}. Furthermore, when they form, these simulated bars are often too short with respect to to their corotation radius \citep[e.g.,][]{Roshan2021_2, Frankel_2022, Ansar_2025}.

From a theoretical point of view, a bar is a global non-axisymmetric instability mode of the inner disc, developing spontaneously through a feedback loop of repeated swing-amplification, which requires the disc to be locally self-gravitating enough and cold enough for the mode not to be damped by random motions. On top of these local conditions, another global one is that the bar mode grows by shedding angular momentum outward, through exchanges with the disc and halo at resonances. This resonant redistribution drives a secular strengthening of the bar \citep[e.g.,][]{Athanassoula_2003}, whilst the presence of an inner Lindblad resonance (ILR) instead absorbs the wave and suppresses it by deflecting the orbits that would otherwise sustain the bar. This resonant coupling makes the role of the surrounding mass distribution twofold. A massive central spheroidal concentration inhibits bar formation by rendering the disc less self-gravitating and by fostering an ILR that deflects the bar-sustaining orbits. Massive bulges are therefore typical culprits for bar inhibition. Massive or highly concentrated dark matter (DM) halos act similarly by lowering the self-gravity, but their role is ambiguous, since a responsive halo can also \emph{amplify} the bar by absorbing the angular momentum it sheds, typically slowing the bar pattern speed at the same time. Once formed, bars play a crucial role in the secular evolution of galaxies through this same angular-momentum redistribution via resonances \citep[e.g.,][]{Athanassoula_2003, Donghia_2020}, by driving the inflow of gas towards the central regions via gravitational torques, and by forming boxy/peanut (pseudo-)bulges through the buckling instability and/or vertical resonances \citep[e.g.,][]{Combes_1981, Combes_1990}. Bars are, however, fragile: once formed they can be destroyed by the build-up of a central mass concentration \citep{Pfenniger_1991, Shen_2004}, typically through the gas inflow they themselves drive or through mergers. In summary, bars are both natural and fragile features of cold, self-gravitating discs. After destruction, the subsequent re-accretion of cold gas onto the disc can nevertheless lead to the re-formation of the bar \citep{Bournaud_2002, Bournaud_2005, Combes_2008, Barway_2020}, if the galaxy sits in a region of parameter space dynamically favorable to bar re-formation.

The combination of all the above mechanisms is highly complex, non-linear, and still not fully under control from a theoretical point of view. The presence of gas, with cooling, star formation, feedback, and the effects of environment, all render the situation even more difficult to track, especially in cosmological simulations where everything is coupled. In order to take a step back regarding the current failure of most large-volume cosmological simulations to produce the correct fraction of bars as a function of redshift, and to make a first step away from pure $N$-body disc-halo systems, we propose here, as a very first stage, to explore a large grid of 62 idealised galaxy disc models with hydrodynamic simulations involving no gas cooling nor star formation (nor associated feedback), in order to systematically re-examine the conditions of bar formation in such idealised cases. Ideally, one should also compare the effect of different integration schemes, numerical codes, etc., 
but we first concentrate here only on simulations performed with the {\tt RAMSES} code, with resolution and particle mass resembling those of the {\tt NewHorizon} simulation \citep{Dubois_2021,Reddish_2022}. Given that the observed peak of the bar fraction as a function of stellar mass at $z=0$ lies at about $10^{10} \, {\rm M}_\odot$ \citep{Erwin_2018}, and that the time for the observed bar fraction to increase sharply between $z\sim 1$ and $z \sim 0.2$ \citep{Sheth_2008} is about 5~Gyr, our grid of simulations concentrates on this stellar mass and this duration.

Our goal is to map the parameters of our idealised hydrodynamic simulations to the bar properties from a purely dynamical perspective, without gas cooling, star formation, nor feedback. The simulations are performed with the \ramses\ code, and we will study in particular the effect of resolution, number of particles, gas mass, disc velocity dispersion, bulge mass, DM halo mass and concentration. Despite not delving into the additional complications of varying other parameters such as halo spin, we will show that, already with such a simplified treatment, most criteria proposed in the literature for bar formation are actually too simplistic to reliably predict the outcome. We will nevertheless attempt to identify the regions of parameter space that inhibit bar formation in our grid our simulations, in order to isolate the zero order effect before considering more complex scenarii in further works. This study can therefore be seen as a step further from the recent study of \citet{Frosst}, by exploring a different code and integration scheme, as well as the effect of the presence of both gas and massive bulges on bar formation, before including further physical processes such as DM halo spin, gas cooling, star formation and feedback in future works. In section \ref{sec:Numerical_setup}, we present our numerical setup and our analysis methods, including the measure of the bar strength and pattern speed. In section \ref{sec:num_convg}, we explore the effect of the number of particles and the resolution on the evolution of the bar amplitude. In section \ref{sec:critfromgrid}, we present our grid of simulations, and show with concrete counter-examples that most of the proposed criteria for bar formation in the literature {\it fail} to capture the conditions prone to bar formation in our own grid of idealised simulations. We nevertheless obtain a diagnostic for the most relevant parameters inhibiting bar formation in our simulations: in particular, we show that the limiting \citet{Romeo_2013}'s stability parameter, related to the (stellar+gaseous) disc random motions, below which a bar can form, decreases quadratically with the bulge mass/bulge fraction. We also note that a galaxy may evolve in the corresponding parameter space as the bar forms, such that this diagnostic tool is relevant for identifying the conditions existing {\it prior} to bar formation, but not when the bar has already formed. We also show that fast bars (i.e., large enough bars w.r.t. the corotation radius) are typically embedded in haloes whose mass makes the galaxy lie {\it above} the stellar-to-halo mass relation from abundance matching. Then, in section \ref{sec:feedback}, we present two simulations where a bar grows at first but is destroyed in the presence of important amounts of material inflow towards the centre, once including gas cooling, star formation, and feedback, whilst the galaxy has already moved to the region of parameter space prone to inhibiting its re-formation. Finally, section \ref{sec:Conclusions} summarizes the article and discusses a possible physical interpretation for the dependence on the bulge mass that we found in the limiting minimum \citet{Romeo_2013}'s stability parameter, above which bars cannot form within 5~Gyr in our simulations.   

\section{Numerical setup and analysis methods}
\label{sec:Numerical_setup}

\subsection{Numerical code}

We run idealised galaxy simulations using \ramses, the adaptive mesh refinement (AMR) grid code developed by \citet{Teyssier_2002}, which has both a Poisson solver for the gravitational field and a second-order Godunov solver for hydrodynamics. 
In this paper, we mostly focus, as a first step in our research programme, on a purely dynamical perspective: this means that the main simulations are run without gas cooling, star formation, nor feedback. 
The simulations are all run for a duration of 5~Gyr. This timescale is motivated by the sharp observational increase of the barred galaxies fraction as a function of redshift between $z \sim 1$ and $z \sim 0.2$ \citep[e.g.,][]{Sheth_2008}, corresponding to a typical 5~Gyr duration.

\subsection{Initial conditions}

The initial conditions (ICs) for our idealised simulations are generated using the Action-based Galaxy Modelling Architecture \citep[\texttt{AGAMA}; ][]{AGAMA} code for the stellar and DM particles. Within a guess-potential, we specify in \texttt{AGAMA} the action-based distribution function of each collisionless component of the model (stellar disc, DM halo, and bulge if choosing to use one), while the gas disc potential is set as an external potential. Then the total potential is recomputed from the densities associated to these distribution functions, and the guess-potential is updated to the new one. The actions are then mapped to positions and velocities in the updated potential, and the code recomputes the new density of each component, a procedure which is repeated until convergence. Positions and velocities of stellar and DM particles are then sampled from the distribution functions\footnote{We use the `\texttt{QuasiIsothermal}' action-based distribution function for the stellar disc and the `\texttt{QuasiSpherical}' action-based distribution function for the bulge (when using one) and for the DM halo.}. The gas disc, on the other hand, is set up directly in \texttt{RAMSES} as an isothermal disc component\footnote{Starting from the circular velocity curve obtained from the \texttt{AGAMA} model, a fixed gas temperature, and a double exponential profile in $R$ and $z$ (or a Gaussian profile in $z$), the rotational velocity profile of the gas is computed following \citet[][Chapter 3]{Chapon_2011}. Hereafter we refer to the 1D gas velocity dispersion as $\sigma_{\rm gas} = \sqrt{kT/\mu m_H}$, where $k_b$ is the Boltzmann constant, $\mu = 7/4$ is the mean molecular weight of a typical neutral gas mixture of hydrogen and helium, $m_H$ is the mass of a hydrogen atom, and $T$ is the temperature.}. 

The stellar disc has an exponential density profile, $\propto~\exp(-R/R_d)$ in galactocentric radius $R$, and an isothermal $ \propto \mathrm {sech}^2  \left(z/2h_z\right)$ profile in height $z$, where $R_d$ and $h_z$ are the scale-length and scale-height, respectively. Its radial velocity dispersion profile also follows an exponential profile, typically with a larger scale-length, $R_{\sigma,d}$, and a central velocity dispersion $\sigma_{r,d,0}$. 

The gas disc also has an exponential profile in $R$, but can have either an exponential profile in $z$, $\propto {{\rm exp}(-|z|/h)}$, or a Gaussian one, $\propto {{\rm exp}(-z^2/(2h^2))}$. In the former case, the exact same gas vertical exponential is used as an external potential in \texttt{AGAMA}, while in the latter case, a $\mathrm {sech}^2$ profile is used, in which case the initial stellar distribution function is slightly away from equilibrium. We note that galaxies in cosmological conditions are seldom at perfect equilibrium, such that this latter setup may actually better reflect such conditions. 

The DM halo, on the other hand, has a truncated NFW \citep{Navarro_1996} profile,
\begin{equation}
    \rho_{\rm DM} ({r}) = \frac{\rho_0}{(r/r_s)\left(1 + r/r_s\right)^2}\exp\left(-\left(\dfrac{{r}}{r_{\rm cut}}\right)^{\xi}\right),
    \label{eq:halo_density}
\end{equation}
where $\rho_0$ and $r_s$ are the characteristic density and scale radius while $\xi$ and $r_{\rm cut}$ are the cutoff strength and radius, respectively. Exactly as for the stellar disc, it is iteratively adjusted from an action-space distribution function.

Finally, when introducing a bulge, a more general truncated $\alpha-\beta-\gamma$ double power-law density profile is used for its density profile \citep[e.g.,][]{Hernquist1990, Zhao1996}. It is also based on an action-space distribution function.

\subsection{Fiducial Models}
\label{sec:Fiducial_model}

\begin{table*}[]

    \caption{Parameters of the Fiducial Models, together with some of the parameters discussed in Sect.~4, $\epsELN$ (Eq.~\ref{eq:ELN}), the minimum $Q_{\rm RF}$ (Eq.~\ref{eq:Q_RF}) within the disc, the bulge-to-total mass ratio $p \equiv B/T$, the parameter $\zeta_{\rm RF}$ (Eq.~\ref{eq:zeta}), and the status of the bar when a bar is formed. The gas fraction $f_g$ is in \%, masses are in units of $10^{10}M_\odot$, lengths in kpc, and velocity dispersion in km/s. For the latter, both the theoretical input and the actual value after iterating to a quasi-equilibrium model and evolving it for 100~Myr (in brackets) are listed here. Common parameters between all the models are discussed in section~\ref{sec:Fiducial_model}.
    }
    \label{tab:Fiducial_paramss}
    \centering
    \begin{tabular}{lcccccccccccccl}
    \hline\hline
\noalign{\smallskip}
\# & $M_{\rm 200}$ & $c_{\rm 200}$ & $M_{\rm *,d}$  & $M_{\rm g}$ & $M_{\rm b}$ & $R_{\rm d}$ & $\sigma_{r,d,0}$ & $R_{\sigma,d}$ & $f_g$ & $\epsELN$ & $Q_{\rm RF}$ & $p \equiv B/T$ & $\zeta_{\rm RF}$ & Bar status\\ 
\hline
\noalign{\smallskip}
    I & 48.6 & 8.94 & 1.03 & 0.25 & 0.00 & 2.00  & 25 (\small{22.07}) & 4.00 & 20 & 0.95 & 1.12 & 0.0 & 1.12 & Fast growth \\ 
    II & 48.6 & 8.94 & 1.03 & 0.25 & 0.00 & 2.00  & 80 (\small{59.01)} & 4.00 & 20 & 0.95 & 1.85 & 0.00 & 1.85 & Slow growth \\
    III & 236. & 7.62 & 1.03 & 0.25 & 0.00 & 2.00 & 25 (\small{24.15}) & 4.00 & 20 & 1.31 & 1.30 & 0.00 & 1.30 & \\
    IV & 48.6 & 8.94 & 1.03 & 0.25 & 0.00 & 2.00 & 125 (\small{89.11})& 10.00 & 20 & 0.95 & 3.48 & 0.00 & 3.48 & \\
    V & 48.6 & 8.94 & 0.65 & 0.25 & 0.38 & 2.00 & 25 (\small{46.94}) & 4.00 & 20 & 0.95 & 1.57 & 0.37 & 2.93 & \\
    
    \hline
    \end{tabular}
    
\end{table*}

Given that the peak of the bar fraction at $z=0$ is at stellar masses of about $\sim 10^{10} \, {\rm M}_\odot$ \citep{Erwin_2018}, we concentrate on this stellar mass. We first set up the initial conditions (ICs) for five models that will serve as the basis from which we will vary the parameters in the grid of simulations. Their parameters are decribed in Tab.~\ref{tab:Fiducial_paramss}.

As Fiducial Model~I, we first consider an extremely cold bulgeless disc galaxy with the following components: 
\begin{itemize}
\item[(i)] a cold thin stellar disc with stellar mass $M_ *= 1.03 \times 10^{10} \, {\rm M}_\odot$, scale-length $R_d=2 \, {\rm kpc}$, sech$^2$-scale-height $h_z = 200$ pc (such that $R_d/h_z = 10$), an input central velocity dispersion $\sigma_{r,d,0}=25 \, \rm km s^{-1}$ and a dispersion scale-length $R_{\sigma,d}=4$~kpc. The central velocity dispersion mentioned here is the theoretical input value used in \texttt{AGAMA} for the distribution function. The actual value after iterating to a quasi-equilibrium model typically differs from this input parameter (it can be slightly larger or smaller). We always let the models settle for 100~Myr before using any galaxy parameter as a diagnostic in the rest of our study. We therefore give the value at 100 Myr next to the input parameter for the central velocity dispersion in Tab.~\ref{tab:Fiducial_paramss}.

\item[(ii)] a gaseous disc making up 20\% of the total baryonic mass, hence with mass $M_{\rm g}= 2.54 \times 10^9 M_\odot$, with a  scale-length of  $4 \, {\rm kpc}$, a Gaussian scale-height\footnote{We have also tested an exponential vertical profile for the gas with the exact same vertical profile for the external potential in \texttt{AGAMA}, and found that the results were unaffected.} of $400$~pc, and a gas velocity dispersion $\sigma_{\rm gas} = 6.8$ km/s; 
\item[(iii)] a truncated NFW DM halo with a virial mass chosen so that the galaxy lies on the \citet{Behroozi_2013} stellar-to-halo-mass relation (SHMR), $M_{\rm 200}= 4.86 \times 10^{11} M_\odot$ (i.e., a virial radius $r_{200} = 160.75 ~\rm kpc$). Then the concentration was chosen so as to obey the mass-concentration relation of \citet{Dutton_Maccio_2014}, $c_{\rm 200} = 8.94$. The truncation strength and radius were chosen as $\xi=2$ and $r_{\rm cut} = 2 \times r_{\rm 200}$, to ensure a smooth truncation of the halo. 
\end{itemize}
Then, as Fiducial Model~II, we consider the exact same galaxy, except that the theoretical input for the central stellar velocity dispersion is now raised to $\sigma_{r,d,0}=80 \, \rm km s^{-1}$. This is motivated by the fact that, when the Fiducial Model~I is run, the actual central velocity dispersion increases to this value at the end of the simulation, once the bar is formed. Hence, it is interesting to consider what happens if it starts as axisymmetric with such a central velocity dispersion. For Fiducial Model~III, we explore the effect of decreasing the self-gravity of the ultra-cold disc, by keeping a central velocity dispersion of $\sim 25 \, \rm km s^{-1}$ but increasing the virial mass of the DM halo to $M_{\rm 200}= 2.36 \times 10^{12} M_\odot$, with concentration  \citep{Dutton_Maccio_2014}, $c_{\rm 200} = 7.62$. For Fiducial Model~IV, we go back to the virial mass $M_{\rm 200}= 4.86 \times 10^{11} M_\odot$, but we make the stellar disc even hotter than Fiducial Model~II, and with a larger velocity dispersion scale-length of $R_{\sigma,d}=10$~kpc, to explore whether a hot enough stellar disc can damp the instability modes. Finally, for Fiducial Model~V, we swap 37\% of the original stellar disc for a stellar bulge, in order to check the capability of a massive bulge ($B/T \equiv p = 0.37$) to inhibit bar formation. 

All five model ICs are then generated within a 500~kpc box, with the galaxies placed at the centre. Within the \ramses\ code, we use refinement levels from 7 to 14, which corresponds to a maximum spatial resolution of 30~pc, while each run has $10^6$ stellar particles and $10^6$ DM particles. This means that the stellar and DM particles have a mass of $\sim 10^4 ~ \rm M_\odot$ and between $\sim 5 \times 10^5 ~ \rm M_\odot$ and $\sim 2 \times 10^6 ~ \rm M_\odot$, respectively. These choices (resolution and particle mass) are motivated by a close resemblance to the {\tt NewHorizon} cosmological simulation setup. All five models are first relaxed for 100~Myr, which is taken as the initial axisymmetric state of all models in the remainder of our analyses.

\begin{figure*}[htbp]
    \centering

    \begin{subfigure}{0.45\textwidth}
        \centering
        \includegraphics[width=1.0\textwidth,height=0.75\textwidth]{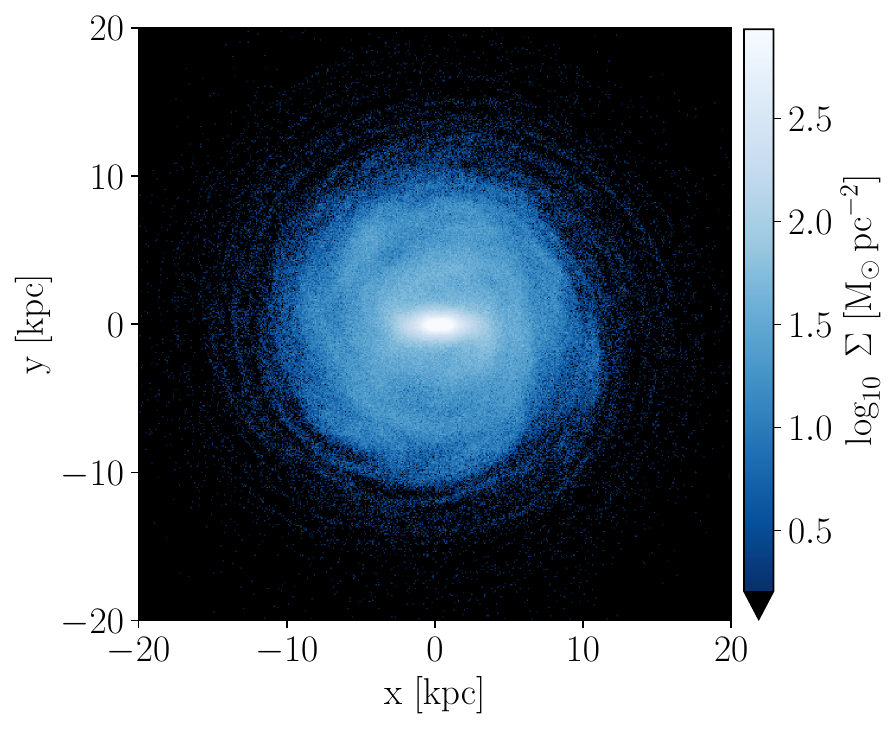}
        \caption{2~Gyr Fiducial~I}
        \label{fig:SD_FidI_2Gyr}
    \end{subfigure}
    \hfill
    \begin{subfigure}{0.45\textwidth}
        \centering
        \includegraphics[width=1.0\textwidth,height=0.75\textwidth]{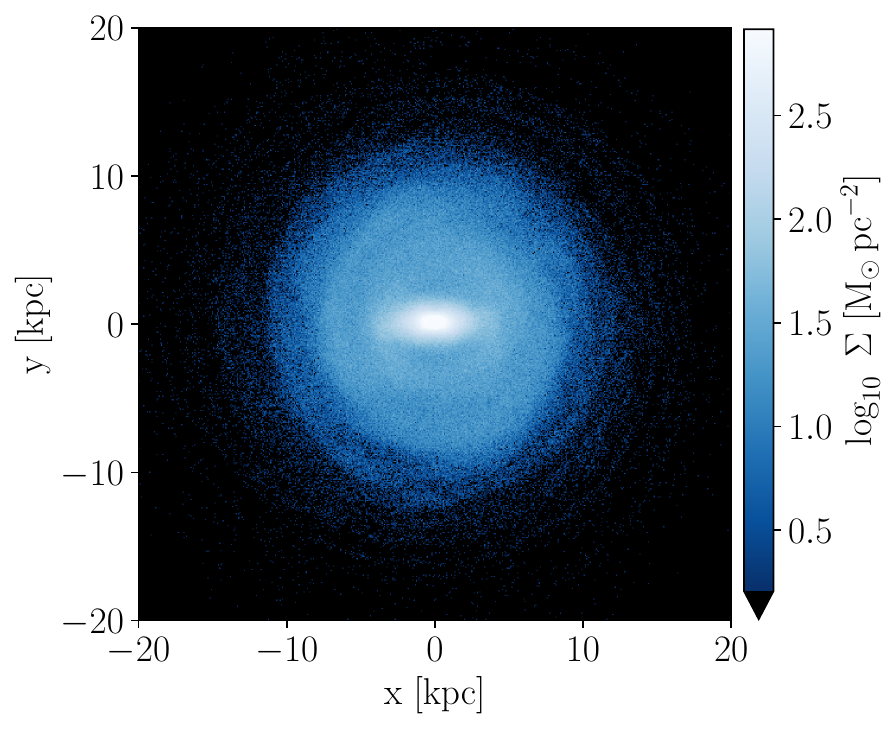}
        \caption{4.5~Gyr Fiducial~I}
        \label{fig:SD_FidI_4_5Gyr}
    \end{subfigure}

    \begin{subfigure}{0.45\textwidth}
        \centering
        \includegraphics[width=1.0\textwidth,height=0.75\textwidth]{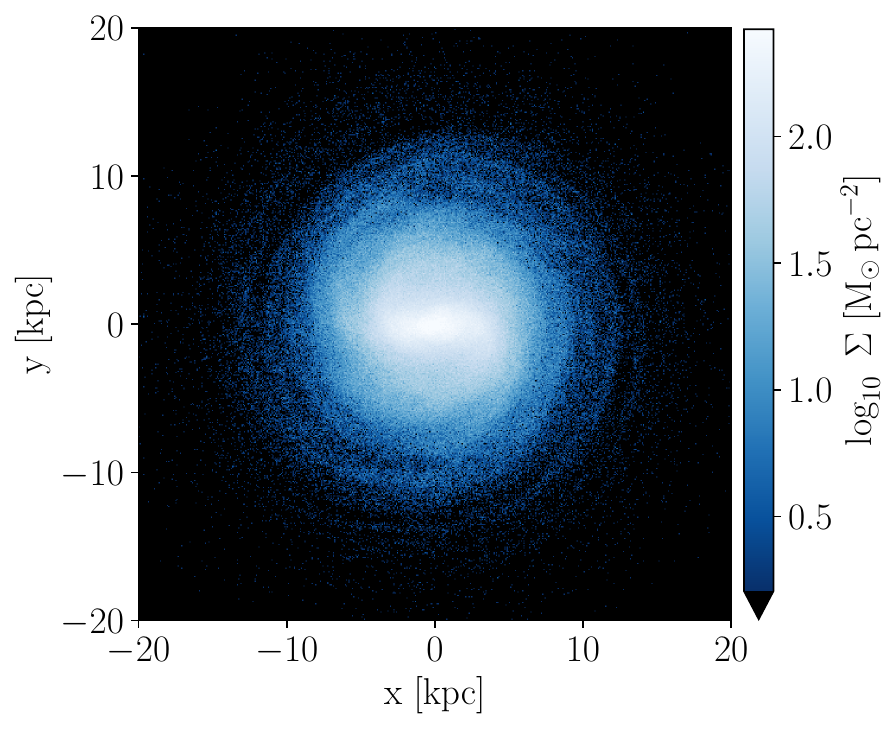}
        \caption{2~Gyr Fiducial~II}
        \label{fig:SD_FidII_2Gyr}
    \end{subfigure}
    \hfill
    \begin{subfigure}{0.45\textwidth}
        \centering
        \includegraphics[width=1.0\textwidth,height=0.75\textwidth]{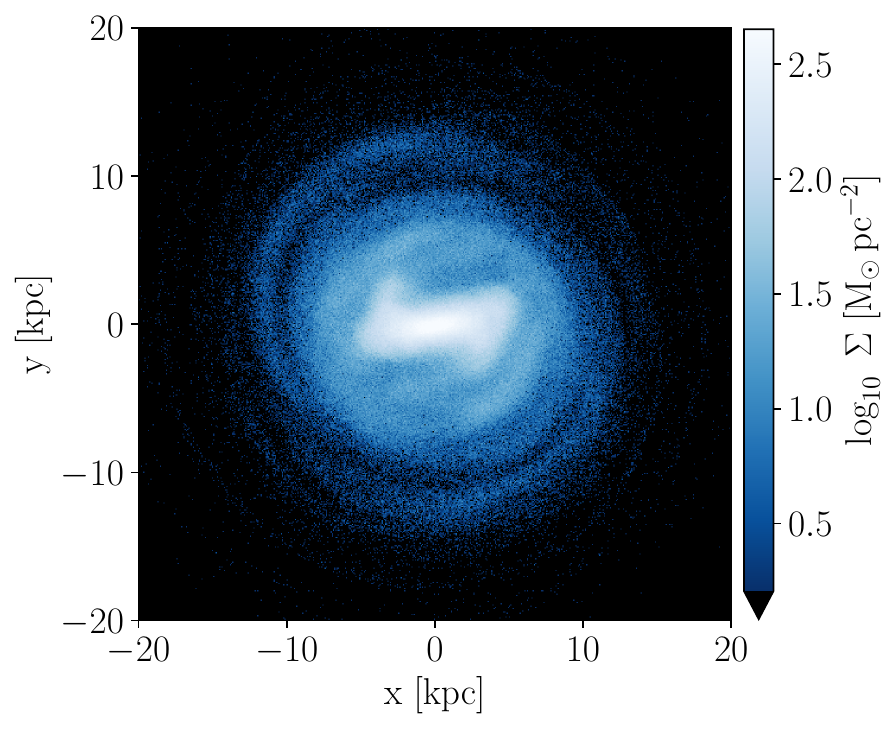}
        \caption{4.5~Gyr Fiducial~II}
        \label{fig:SD_FidII_4_5Gyr}        
    \end{subfigure}

    \caption{Face-on stellar surface density for Fiducial I (top a-b) and II (bottom c-d) Models at $t\sim 2\rm~ Gyr$ and $t\sim 4.5~\rm Gyr$, highlighting the presence of a bar. The amplitude of the bar is stable in the Fiducial Model I (top panels) while it grows in the Fiducial Model II (bottom panels).}
    \label{fig:SD}
\end{figure*}

\begin{figure}[htbp]
    \centering

    \begin{subfigure}{0.45\textwidth}
        \centering
        \includegraphics[width=1.0\textwidth,height=0.75\textwidth]{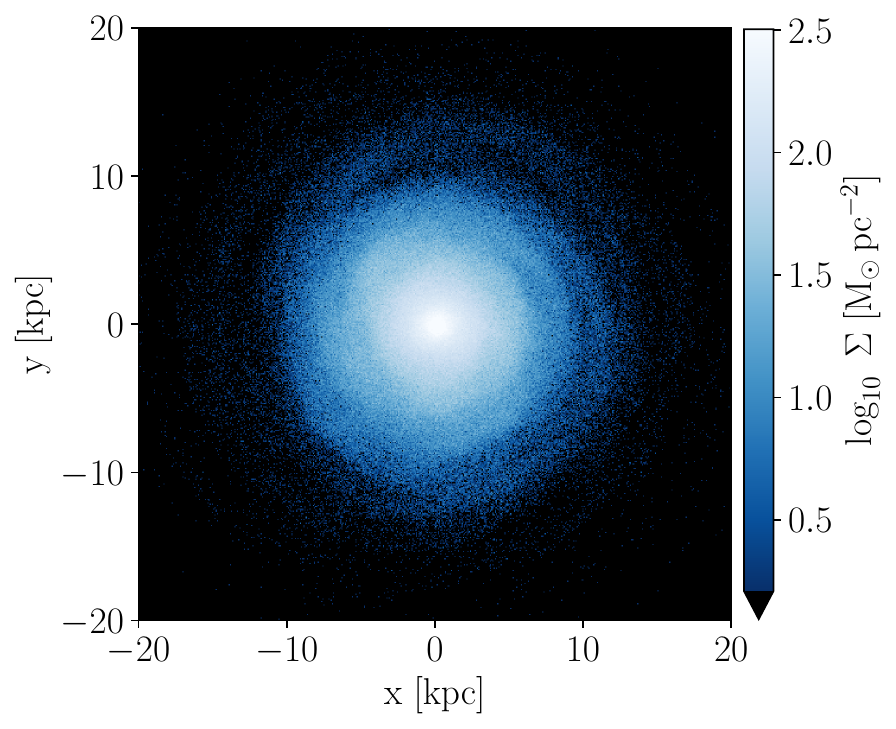}
        \caption{4.5~Gyr Fiducial~III}
        \label{fig:SD_FidIII_4_5Gyr}
    \end{subfigure}
    \vfill
    \begin{subfigure}{0.45\textwidth}
        \centering
        \includegraphics[width=1.0\textwidth,height=0.75\textwidth]{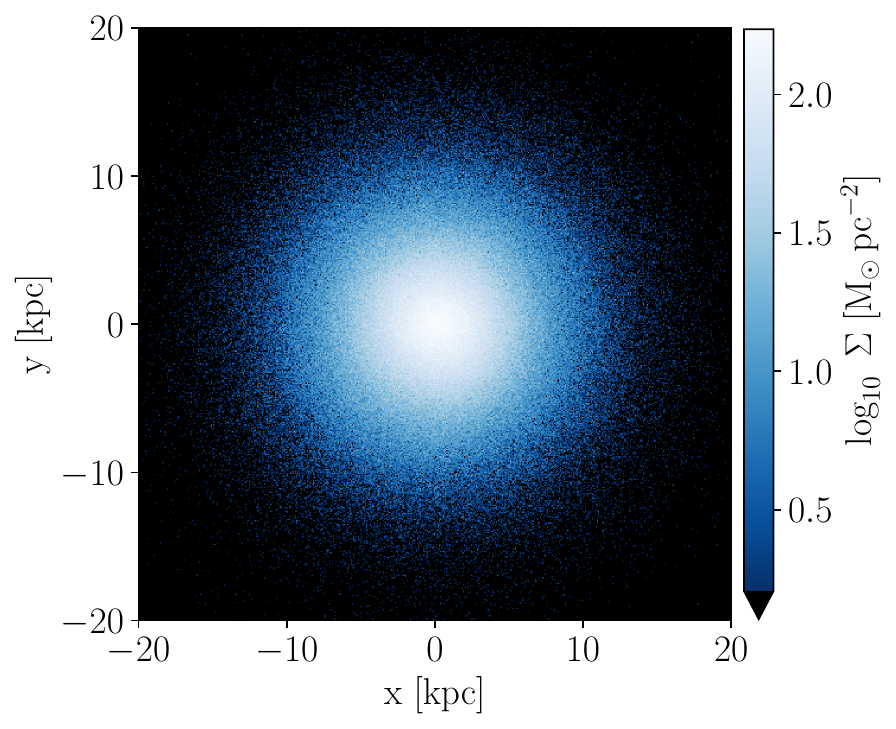}
        \caption{4.5~Gyr Fiducial~IV}
        \label{fig:SD_FidIV_4_5Gyr}
    \end{subfigure}

    \begin{subfigure}{0.45\textwidth}
        \centering
        \includegraphics[width=1.0\textwidth,height=0.75\textwidth]{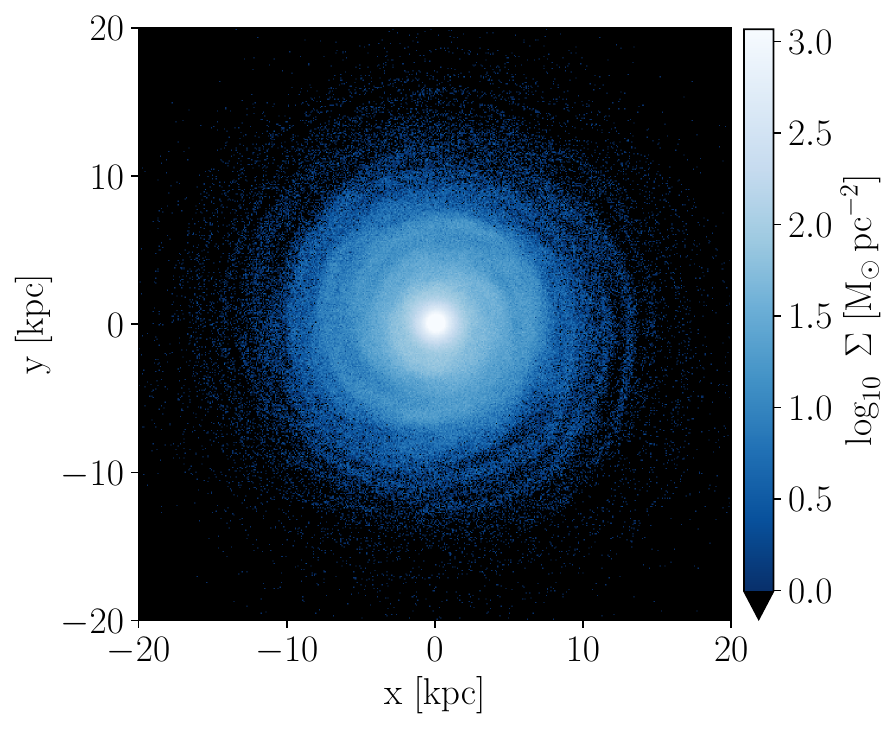}
        \caption{4.5~Gyr Fiducial~V}
        \label{fig:SD_FidV_4_5Gyr}
    \end{subfigure}
    \hfill
    \caption{Face-on stellar surface density for Fiducial III (top: massive DM halo - a), Fiducial IV (middle: hot stellar disc - b) and Fiducial V (bottom: massive central bulge - c) at $t\sim 4.5\rm~ Gyr$, highlighting the absence of a bar in all three cases.}
    \label{fig:SD_part2}
\end{figure}

In Fig.~\ref{fig:SD}, we display two snapshots of the stellar surface density seen face-on at $t \sim 2 \, {\rm Gyr}$ and $t \sim 4.5 \, {\rm Gyr}$ for both Fiducial Model~I and Fiducial Model~II. As can be clearly seen, a bar develops in both cases, but its amplitude is stable in Fiducial Model~I, while it is slowly growing in Fiducial Model~II, reaching a higher amplitude at the end. In Fig.~\ref{fig:SD_part2}, we display the snapshots at $t \sim 4.5 \, {\rm Gyr}$ of the Fiducial Models III, IV, and V. Remarkably, none of them forms a bar, meaning that all three bar-inhibiting mechanisms explored here, a massive DM halo, a massive bulge, or a hot stellar disc, can work independently in order to damp the bar. The cold discs (III and V) still develop linear instabilities in the form of spiral arms, but the central regions remain immune to bar formation. The hot disc (IV), on the other hand, remains perfectly axisymmetric along the whole duration of the simulation.

\begin{figure*}[!ht] 

\centering 

\includegraphics[height=0.18\textheight,width = 0.45\textwidth]{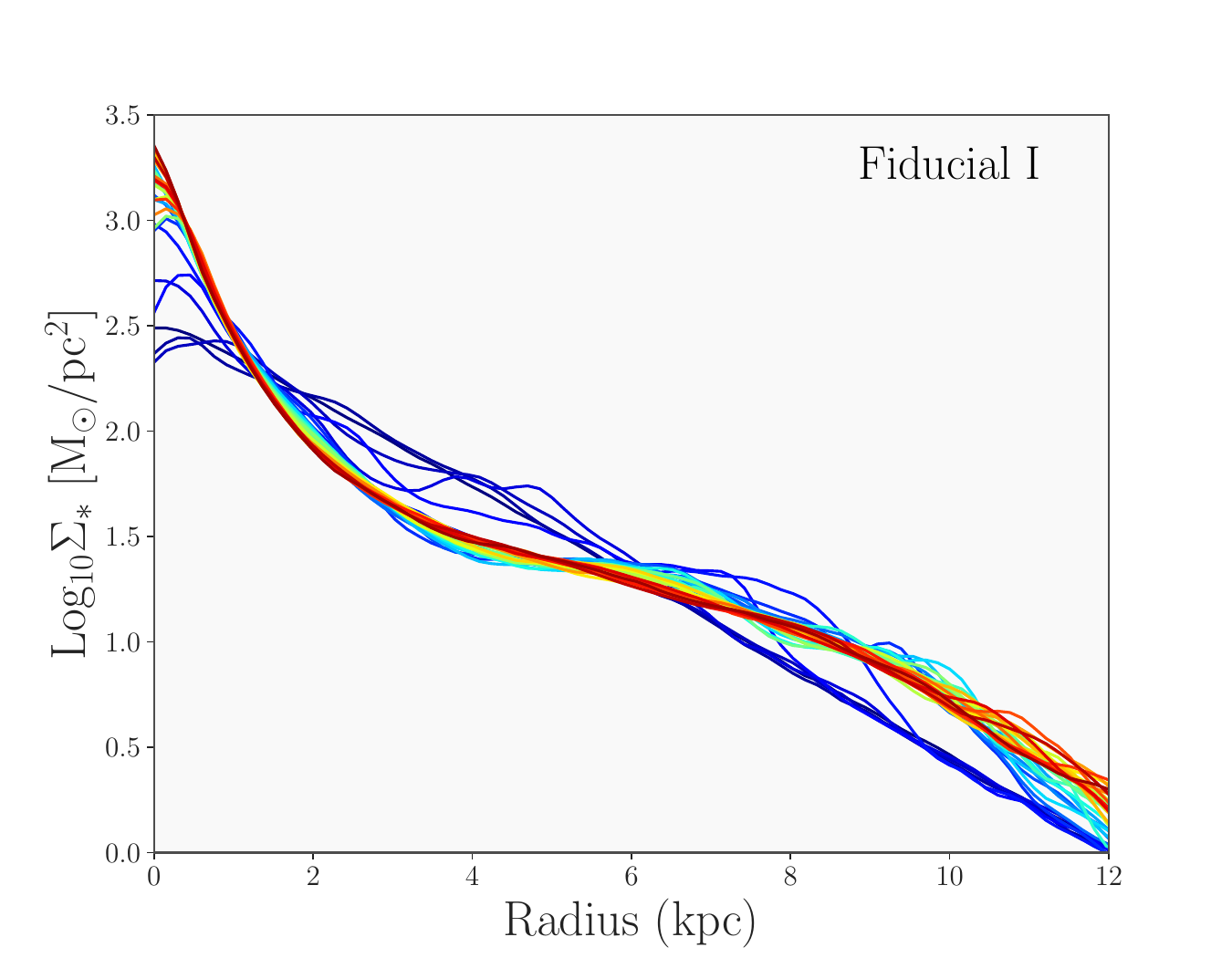}
\hfill 
\includegraphics[height=0.18\textheight,width = 0.45\textwidth]{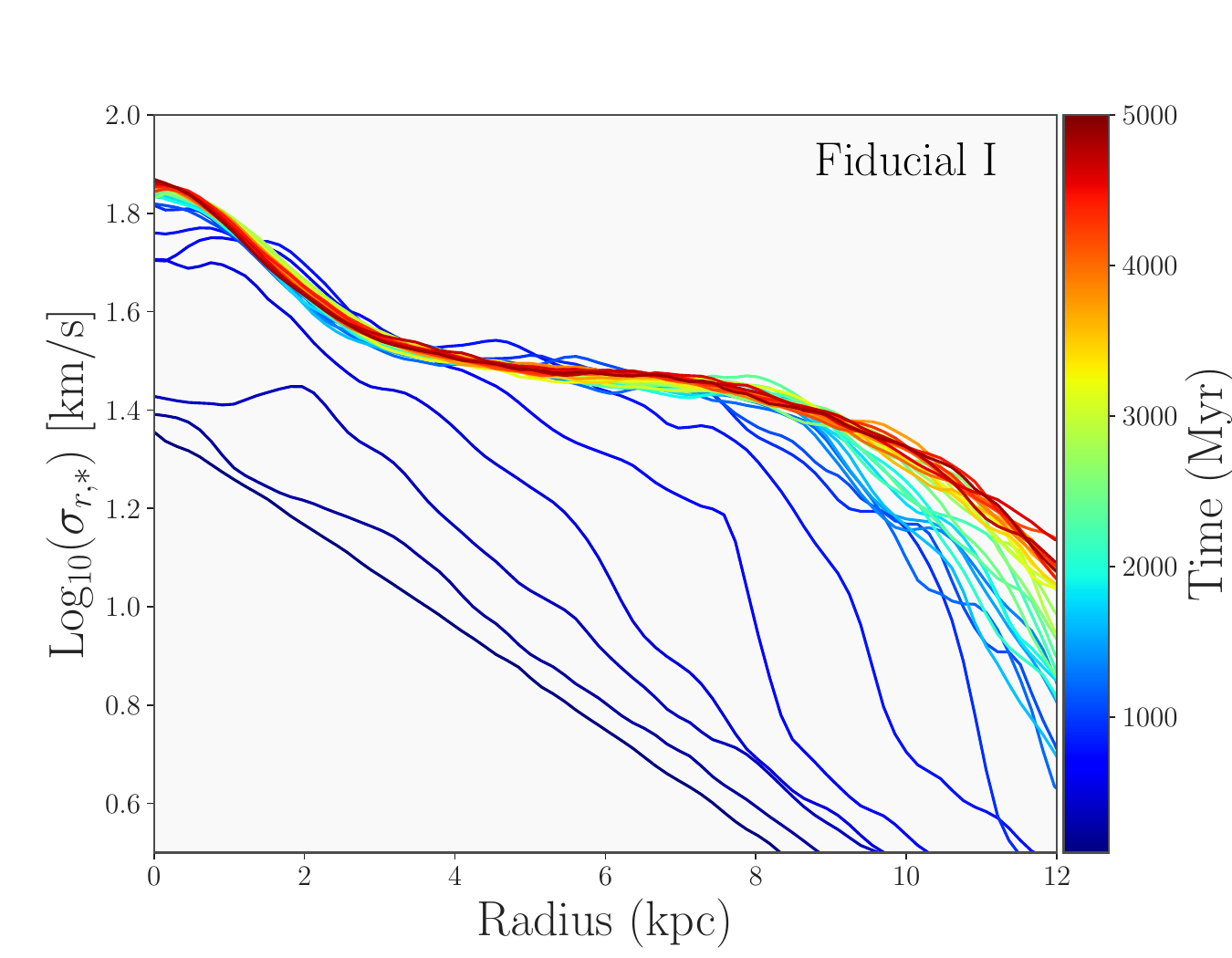}

\includegraphics[height=0.18\textheight,width = 0.45\textwidth]{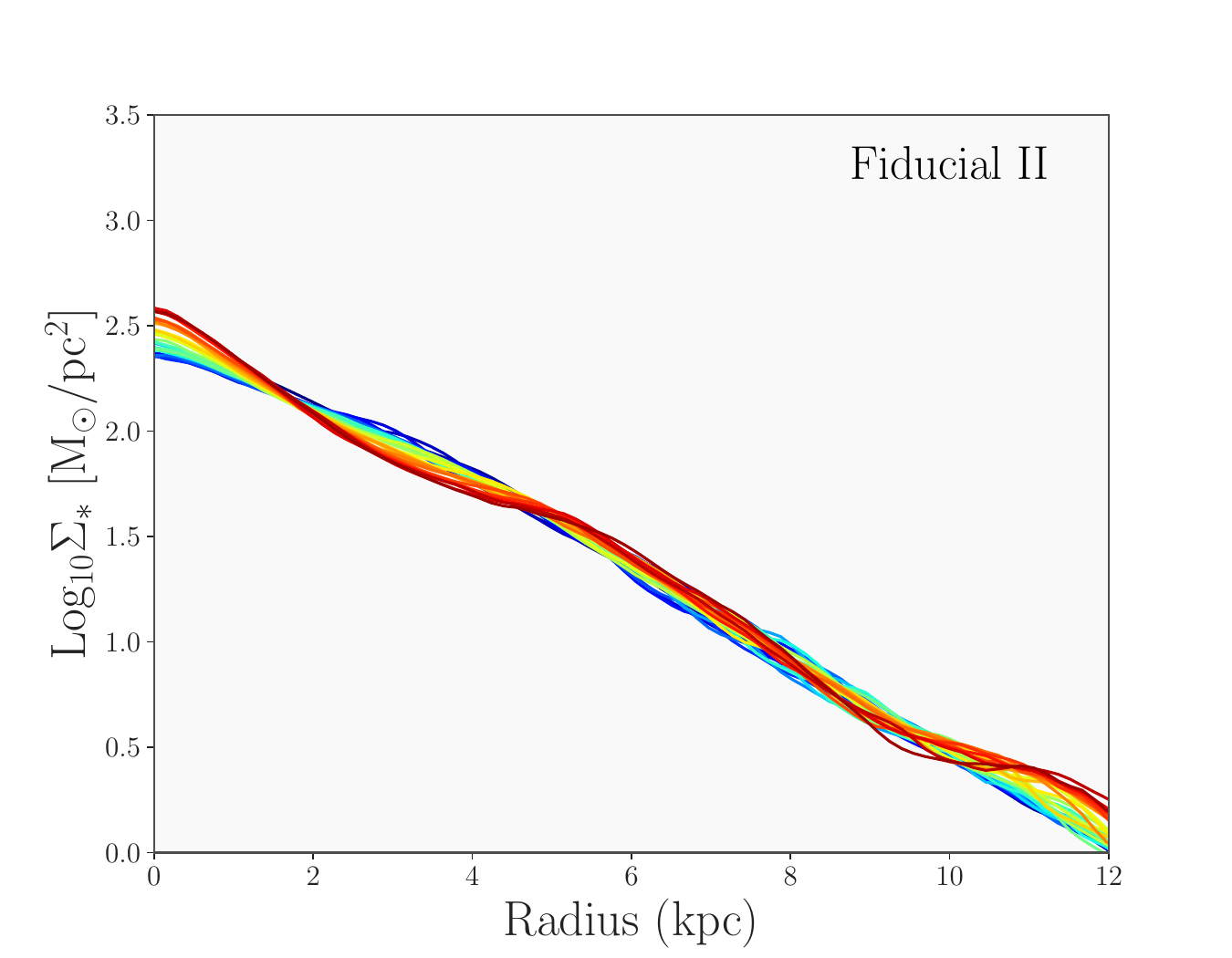}
\hfill 
\includegraphics[height=0.18\textheight,width = 0.45\textwidth]{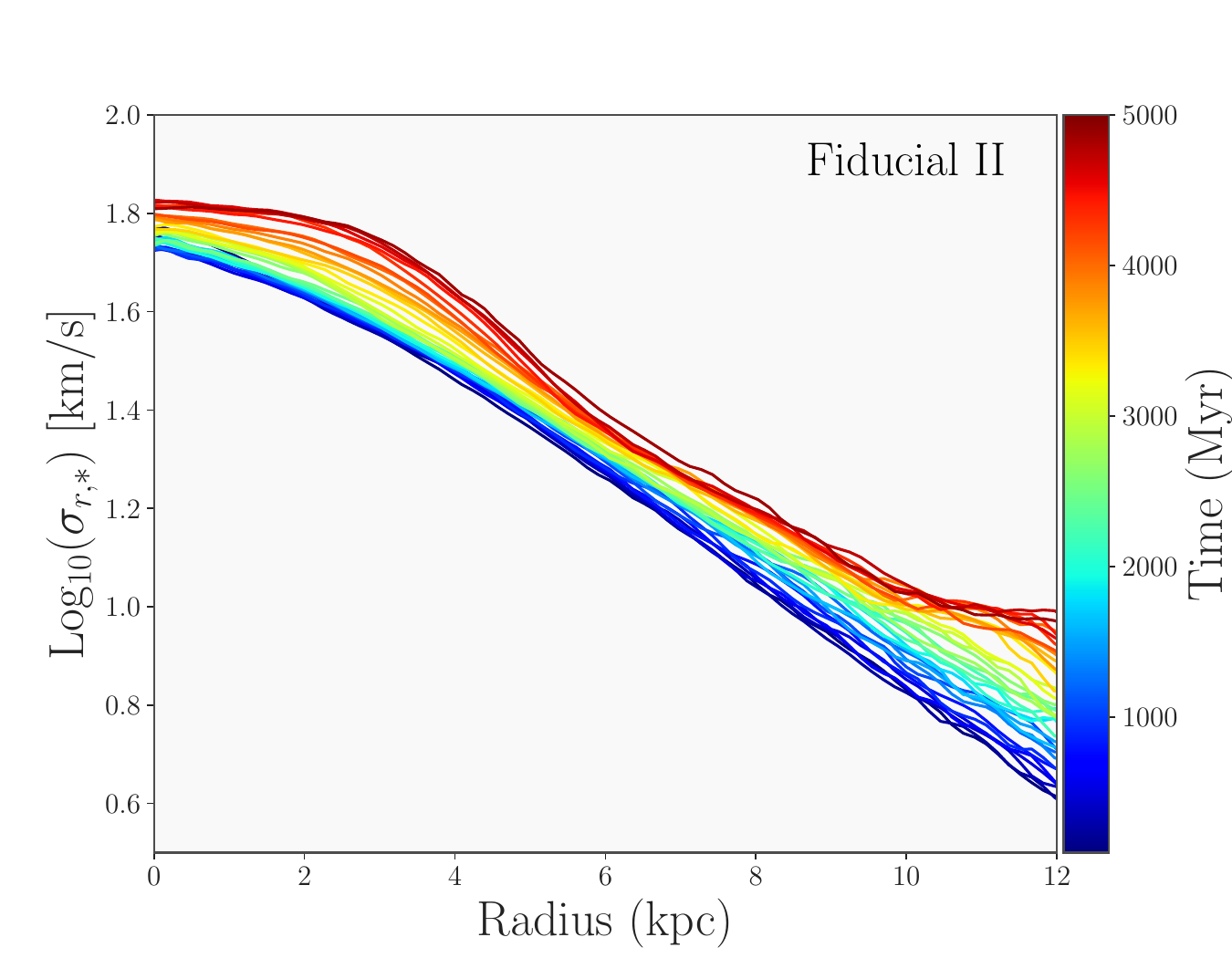}

\includegraphics[height=0.18\textheight,width = 0.45\textwidth]{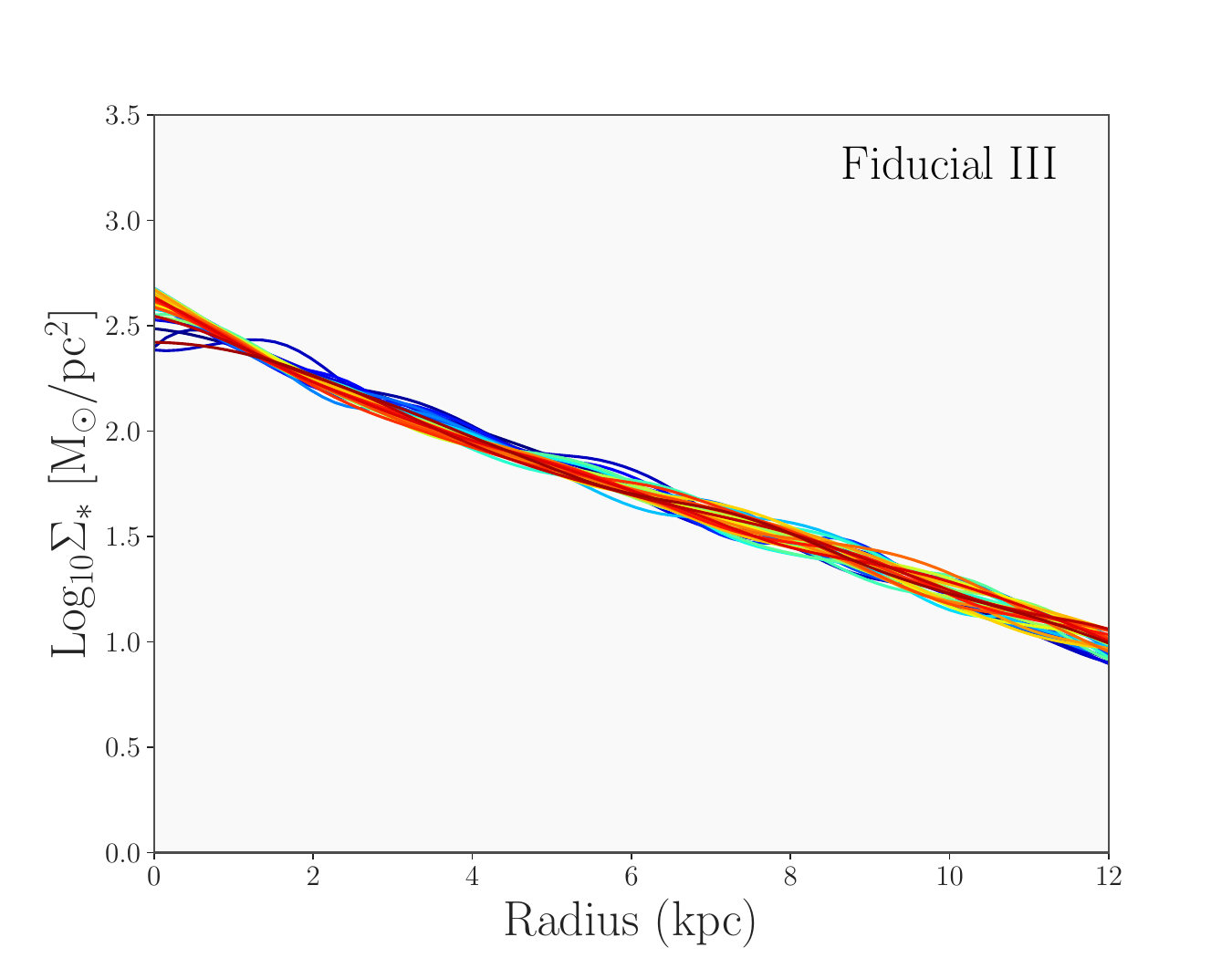} \hfill 
\includegraphics[height=0.18\textheight,width = 0.45\textwidth]{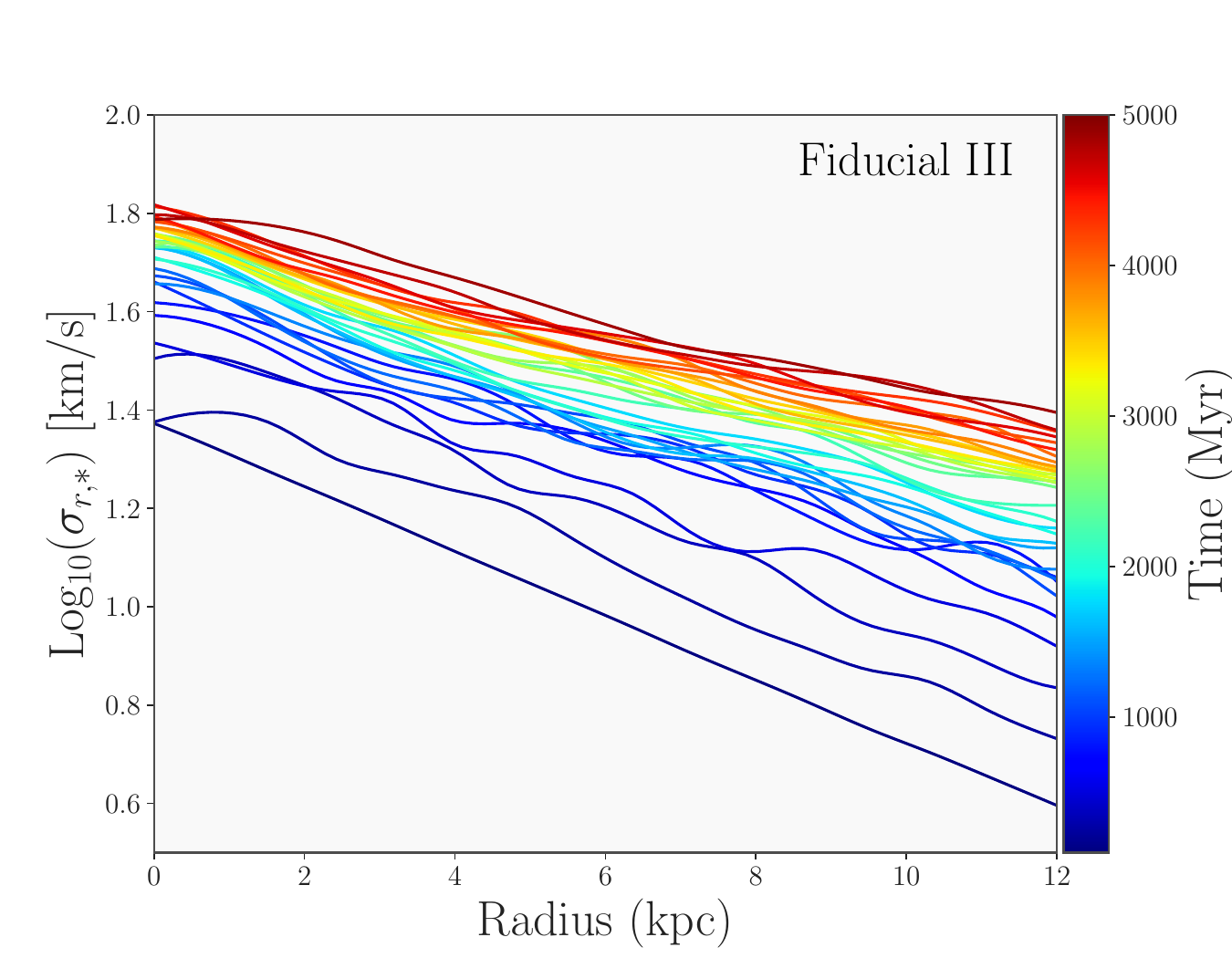} 

\includegraphics[height=0.18\textheight,width = 0.45\textwidth]{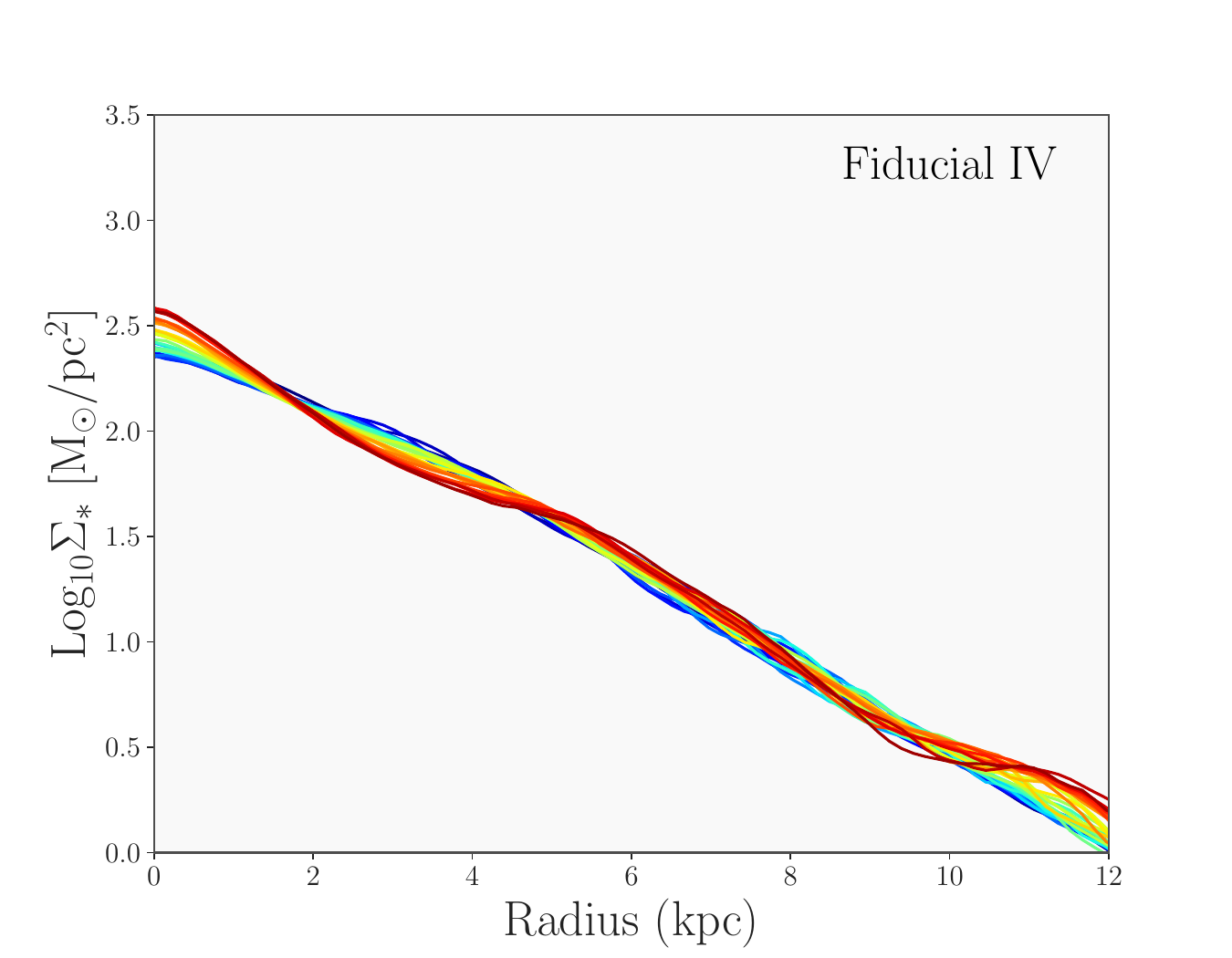} \hfill 
\includegraphics[height=0.18\textheight,width = 0.45\textwidth]{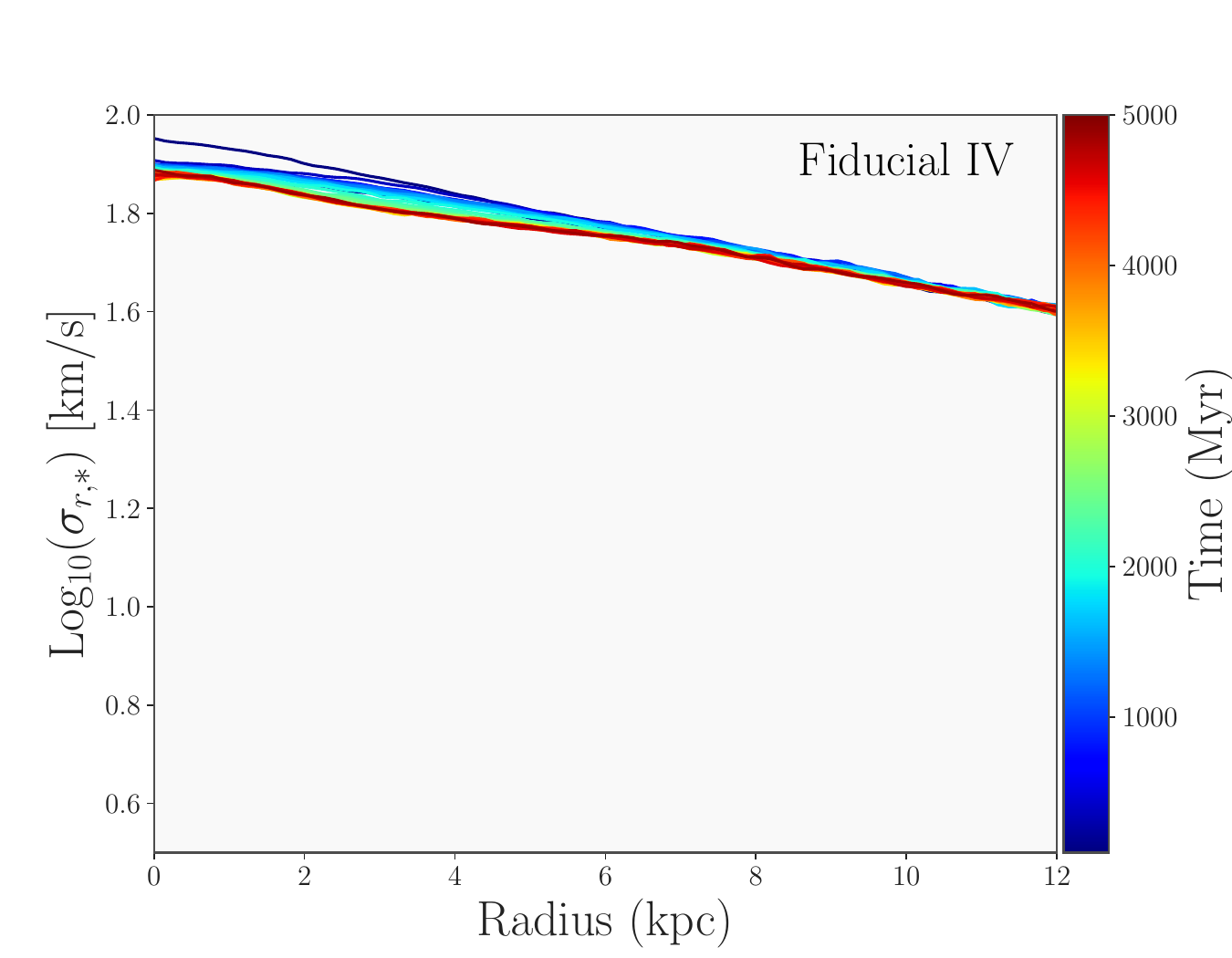} 

\includegraphics[height=0.18\textheight,width = 0.45\textwidth]{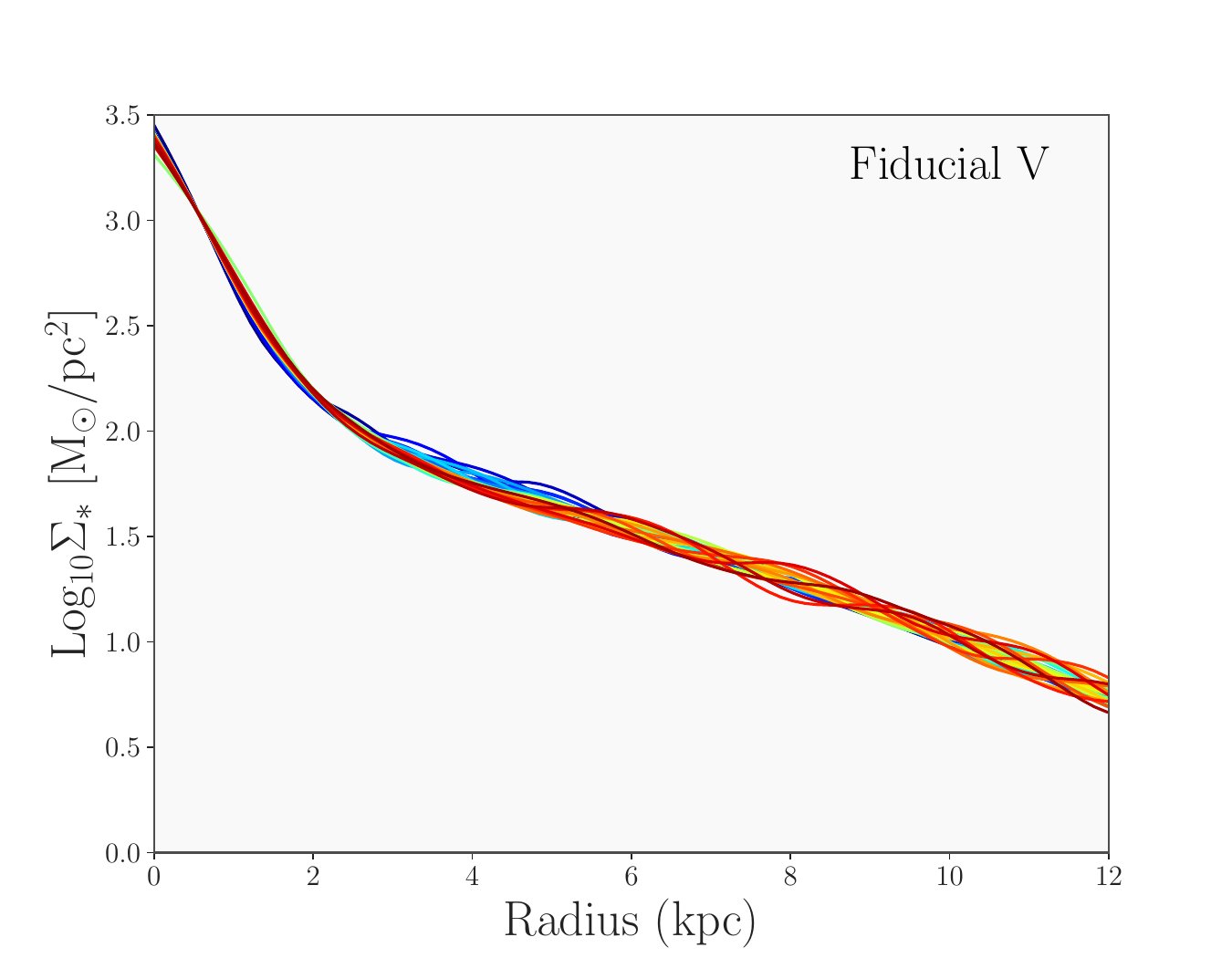} 
\hfill 
\includegraphics[height=0.18\textheight,width = 0.45\textwidth]{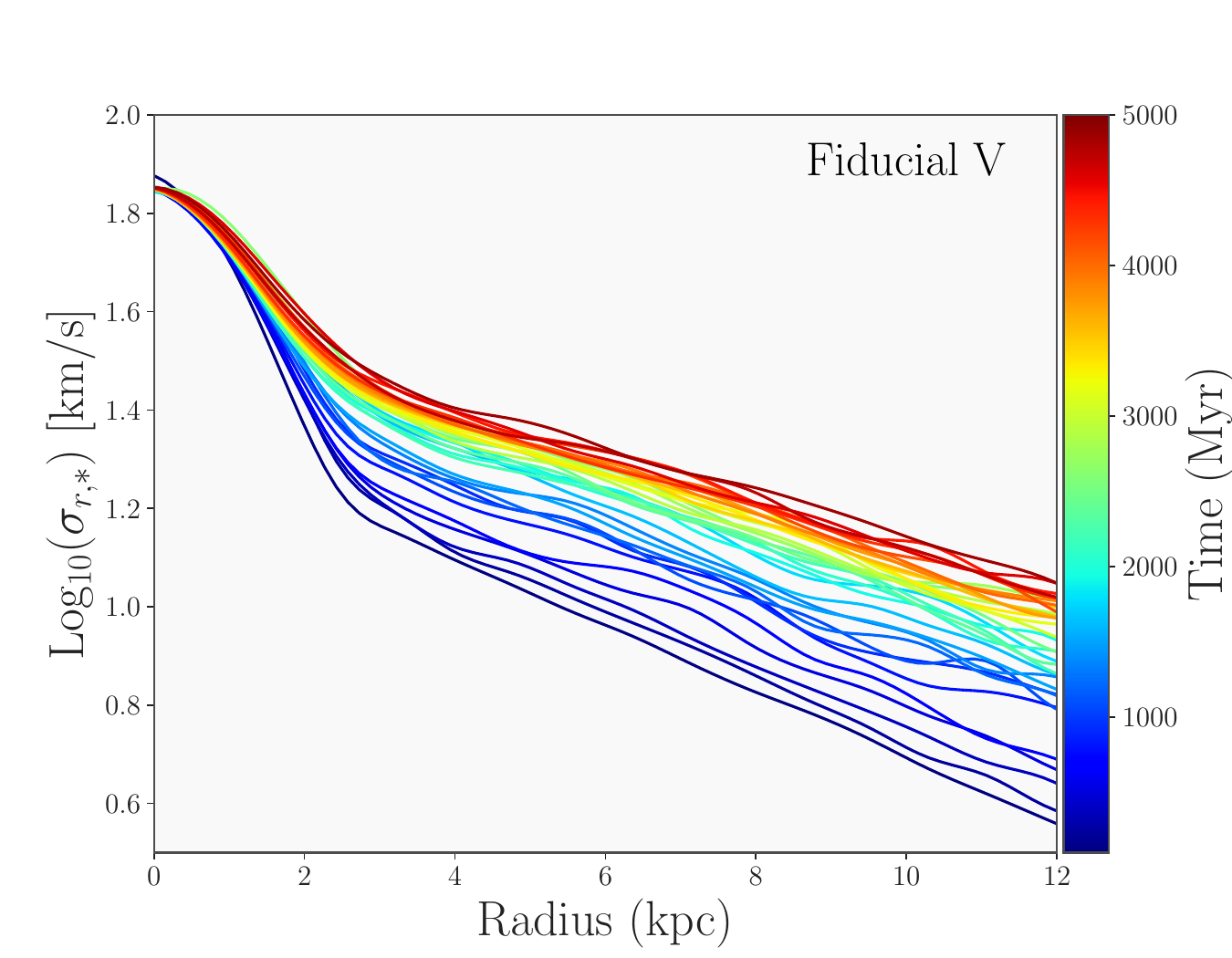}

    \caption{Evolution of the stellar surface density (left panels) and stellar velocity dispersion (right panels) of the five Fiducial Models. 
    }
    \label{fig:sigma_evolution}
\end{figure*}

Before moving on to a detailed analysis of the bar parameters, it is also interesting to investigate how the phase-space of all Fiducial Models evolves with time, which is illustrated by Fig.~\ref{fig:sigma_evolution}. Interestingly, the surface density profile and the velocity dispersion profile of the cold disc (Fiducial I) vary much more abruptly during the bar-formation phase than the profiles of the initially warmer disc (Fiducial II). The final central velocity dispersions are very similar in both models, but a much clearer ``pseudo-bulge'' structure formed in Fiducial Model I. This model has a clear Type~II profile \citep{Freeman_70}, or more precisely a Type~II.o-OLR profile \citep{Erwin2008}, with a central pseudo-bulge followed by an inner exponential and an outer exponential, the break radius between these two being located well beyond the bar radius, in the outer Lindblad resonance (OLR) region. Fiducial Model II actually presents a similar profile, although the pseudo-bulge is less marked and the scale-length slightly less flattened outside of the central region. In terms of velocity dispersions, beyond the central values being the same, the initially cold disc (Model I) ends up \emph{hotter} than the initially warm disc (Model II), due to the effect of the violent phase of bar formation. When a massive DM halo (Fiducial Model~III) or a massive bulge (Fiducial Model~V) is introduced, no bar is formed and the surface density of the cold disc remains remarkably stable, but it still heats up very efficiently thanks to the multiple spiral arms that are formed along the 5~Gyr of the simulation. For the hot disc case (Fiducial Model~IV), both the surface density and velocity dispersion profiles remain stable over 5~Gyr. 

\begin{figure*}[htbp]
    \centering
    \begin{overpic}[width=0.49\textwidth]{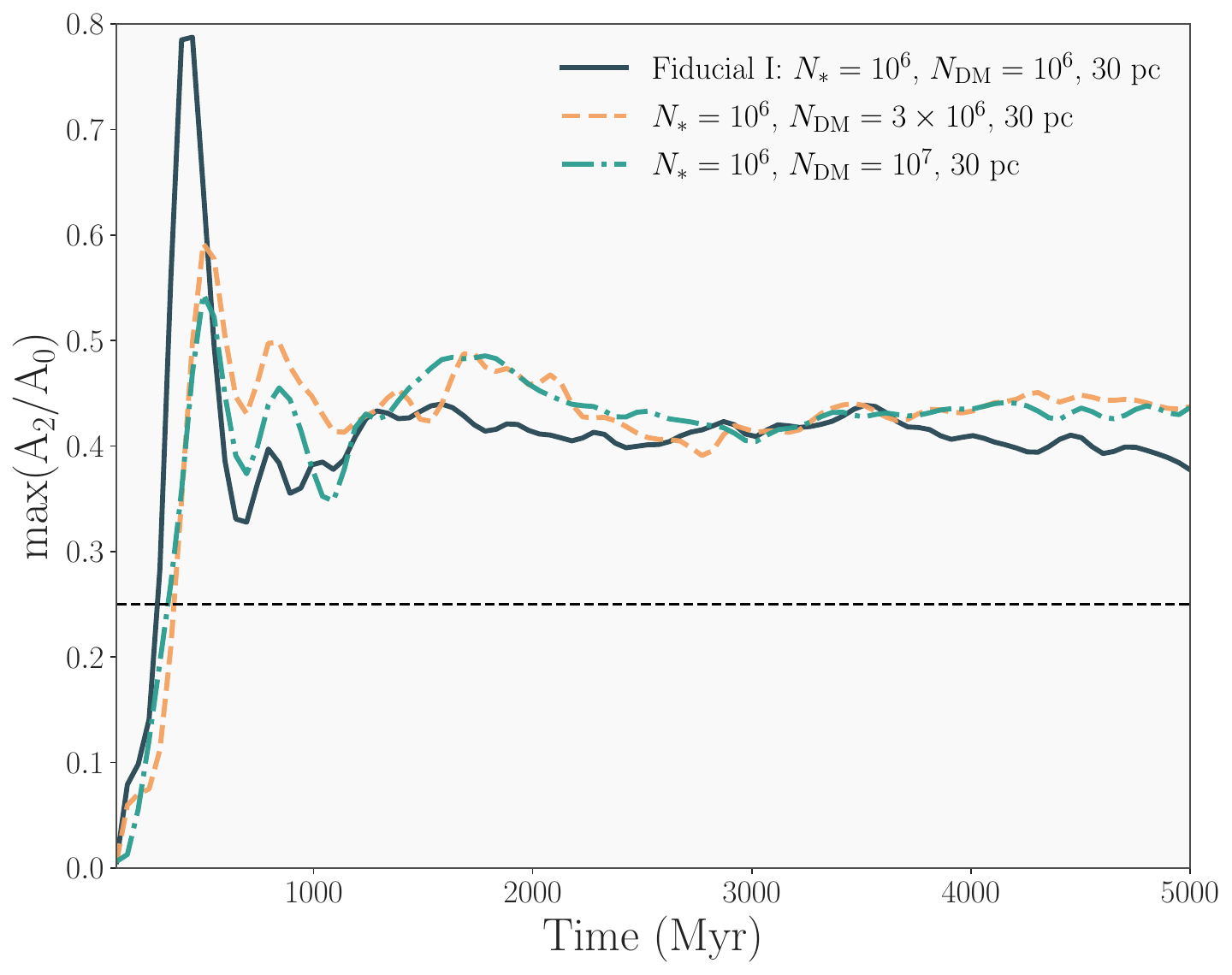}
    \end{overpic}
    \hfill
    \begin{overpic}[width=0.49\textwidth]{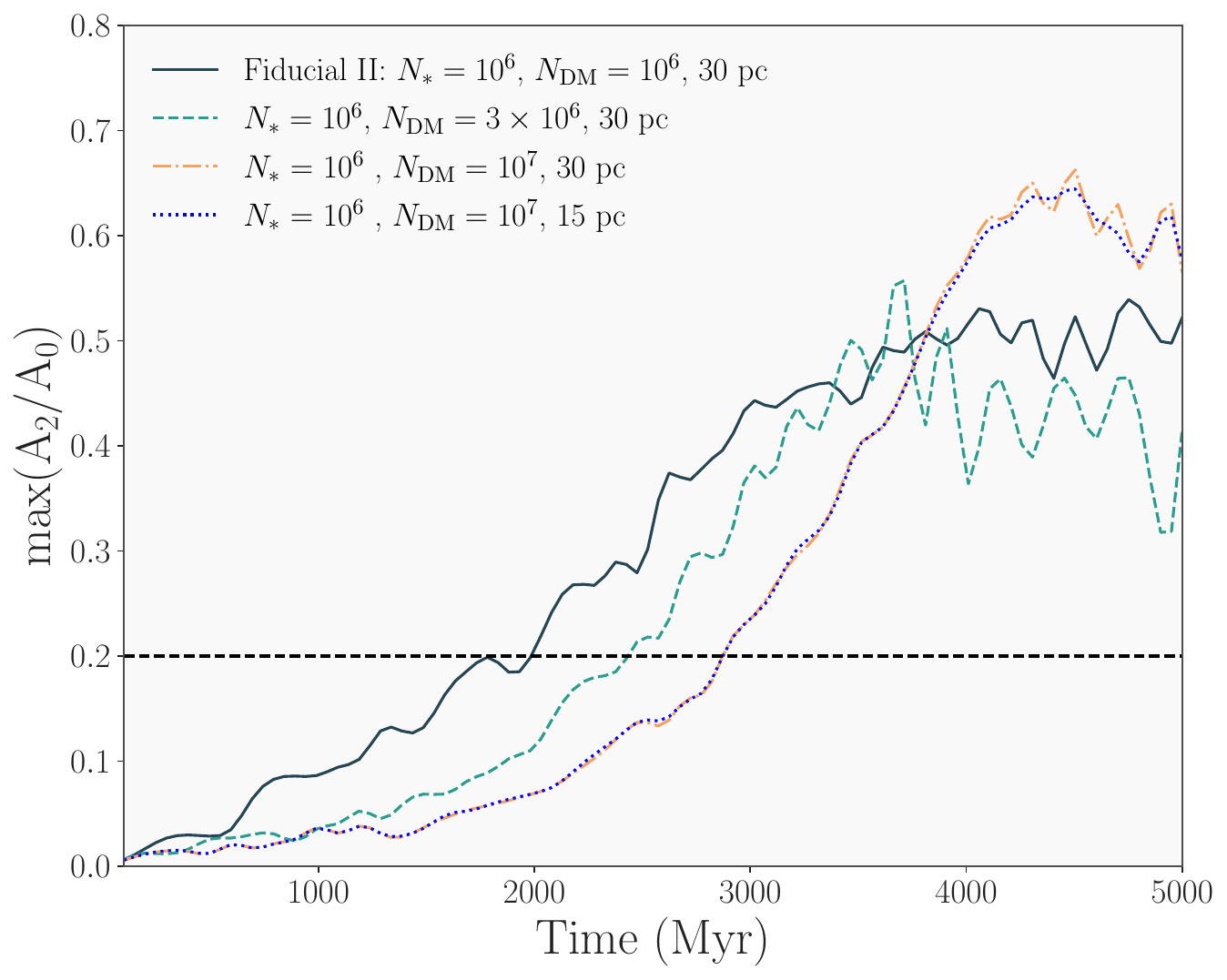}
    \end{overpic}

    \begin{overpic}[width=0.49\textwidth]{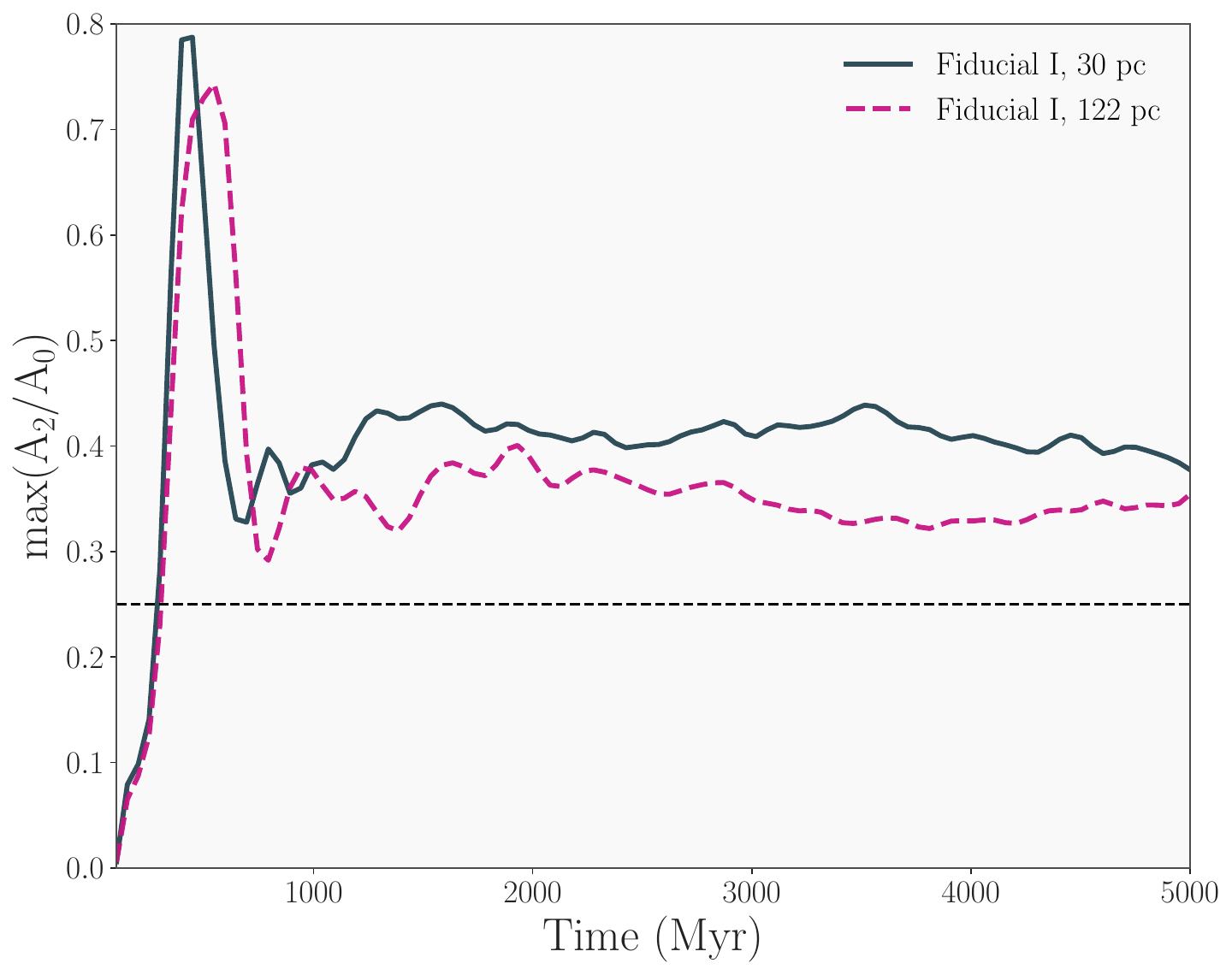}
    \end{overpic}
    \hfill
    \begin{overpic}[width=0.49\textwidth]{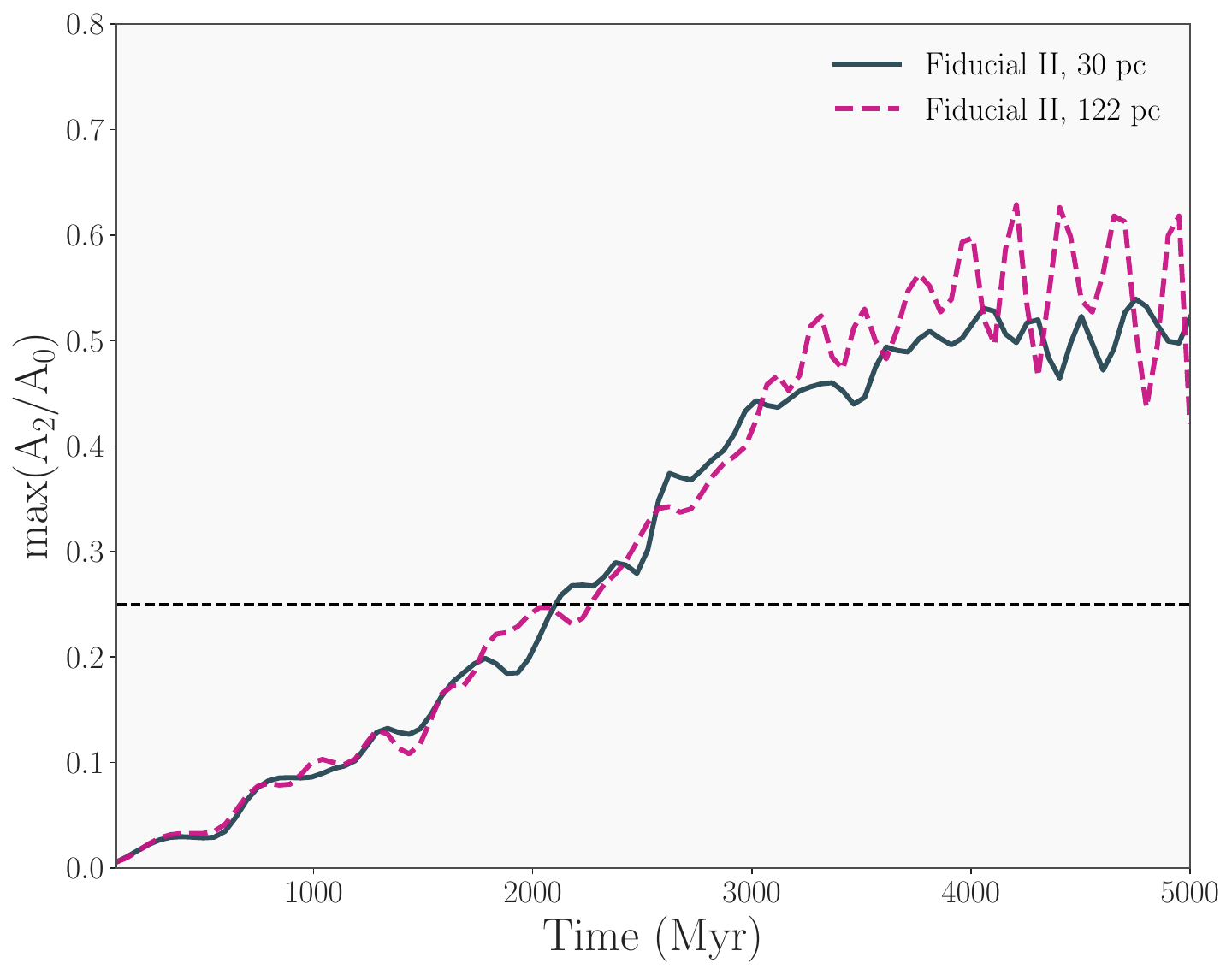}
    \end{overpic}

    \caption{Effect of the number of DM particles (upper panels) and the spatial resolution (lower panels) on the evolution of the bar amplitude for the two bar-forming Fiducial Models (Left panels: I; Right panels: II). In the upper panels, we test $N_{\rm DM} = 10^6$, $3\times 10^6$, and $10^7$ DM particles with $N_\star=10^6$ stellar particles; for the Fiducial Model II we also test increasing the maximum spatial resolution to 15~pc, with no influence on the result. In the lower panels, we compare spatial resolutions of $30\rm ~pc$ and $122\rm~ pc$ for the standard case with $N_{\rm DM} = 10^6$ and $N_\star=10^6$.}
    \label{fig:combined}
\end{figure*}

\begin{figure}
    \includegraphics[width=1\linewidth]{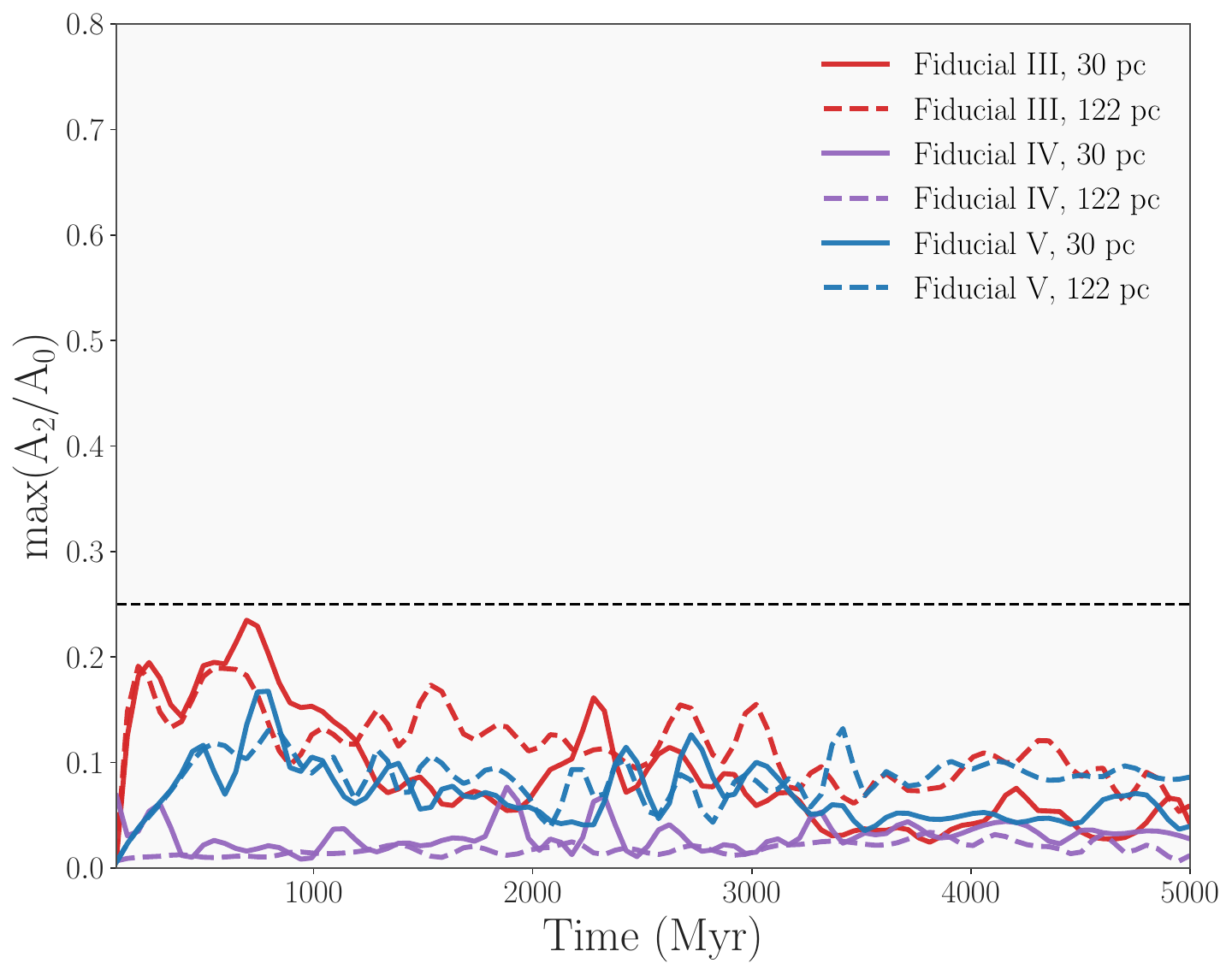}
    \caption{The $m=2$ maximum Fourier amplitude as a function of time for Fiducial Models III (massive DM halo), IV (hot stellar disc), and V (massive central bulge), for a maximum spatial resolution of 30~pc and 122~pc ($N_{\rm DM} = 10^6$ and $N_\star=10^6$). No bar forms in any of these models.}
    \label{fig:Fid_new_highres}
\end{figure}

\subsection{Measuring the bar strength and pattern speed}

To identify the bar robustly in every snapshot of every simulation, we use the method of \citet{Dehnen_2023}, which characterises the bar through both the amplitude and the phase of the $m=2$ Fourier mode. The relative amplitude of the $m=2$ to the $m=0$ component at radius $R$ at the center of an annulus of size $\Delta R$ is
\begin{eqnarray}
    A_2/A_0 ~\equiv~ \frac{\sqrt{a_2(R)^2 + b_2(R)^2}}{\Delta M(R)}\, ,
    \label{eq:A_2_def}
\end{eqnarray}
where $\Delta M(R)$ is the stellar mass in the annulus, and
\begin{eqnarray}
    a_2(R) &\equiv& \sum_{k} m_k \cos(2\varphi_k)\, , \nonumber \\
    b_2(R) &\equiv& \sum_{k} m_k \sin(2\varphi_k)\, ,
    \label{eq:A2_A0}
\end{eqnarray}
the sums running over the stellar particles $k$ in the annulus, each with masses $m_k$ and individual azimuthal angles $\varphi_k$. From this Fourier decomposition, the bar region is identified as the contiguous range over which the {\it phase} remains constant to within only $10^\circ$ and $A_2/A_0$ remains above 0.1. Using annuli of $\Delta R = 0.5$~kpc, up to a maximum radius of $10$~kpc, we additionally request for a bar to exist that $(A_2/A_0)^{\text{max}} \geq 0.25$ \citep[see the literature for the use of such a criterion with varying $A_2/A_0$ thresholds, e.g.][]{Algorry_2017, Guo_2019, Roshan2021_2, Nagesh_2023}. We also consider a stable value $0.15\leq (A_2/A_0)^{\text{max}} < 0.25$ as a weak bar (see Sect.~4.2).

We measure the bar pattern speed $\Omega_p$ with the same method of \citet{Dehnen_2023}, which is well suited to the low cadence of our outputs since it operates on individual snapshots.  Indeed, we store outputs only every 100~Myr. The method's key advantage is that it uses a window function to perform an evolving selection of particles, i.e., assuming that the density of particles in the bar region evolves. Because $\Omega_p$ is obtained from a single snapshot, the method avoids the finite-difference phase tracking that our sparse output cadence would render unreliable, and is better suited to our simulations than phase-differencing between snapshots \citep{Fragkoudi_2021} or the Tremaine-Weinberg method \citep{Tremaine_1984}. We then compute the corotation radius $R_{\rm CR}$ as the radius where the pattern speed and circular angular frequency (from the circular speed curve) are equal, while the bar length $R_{\rm bar}$ is taken to be the bar region identified by the  code \citep[for more details, the reader is referred to][]{Dehnen_2023}. 

\section{Varying number of particles and resolution}
\label{sec:num_convg}

The importance of the mass of stellar and DM particles used in simulations for the development of disc instabilities has long been known, and has for instance recently been highlighted in simulations making use of a moving mesh code \citep[{\tt AREPO},][]{Kwak2026}, concluding that a too high DM particles mass compared to the stellar particles mass could artificially weaken the bar. Our default particle mass ratio for the individual masses of DM ($m_{DM}$) and stellar ($m_*$) particles is $m_{DM}/m_* \sim 45$. We remind the reader that this ratio is similar to the one of the {\tt NewHorizon} simulation, and it is important to check that this does not artificially {\it weaken} the bar in an AMR code such as \ramses. If it did, this simple DM mass resolution problem could be the root cause of the missing bar problem at low-$z$ in {\tt NewHorizon} \citep{Reddish_2022}. 

We therefore run two simulations with the same stellar disc seed and the same refinement levels as Fiducial~I, but re-sampling the DM halo with a 3 times higher and 10 times higher number of DM particles, respectively (top-left panel of Fig.~\ref{fig:combined}). In the latter case, the particle mass ratio $m_{DM}/m_* $ drops below 5, but the final bar amplitude turns out to be essentially independent of this DM to stellar particle mass ratio. A lower mass ratio only affects the short initial phase of bar evolution by actually \emph{decreasing} the amplitude of the first peak. A slight weakening of the bar might be noticed after 4~Gyr in the Fiducial case compared to the lower mass ratio cases, but it is simply consistent with a normal fluctuation of the amplitude with time. 

We then consider the slowly growing bar fiducial model, namely Fiducial~II, and we also increase the number of DM particles by a factor 3 and 10 (top-right panel of Fig.~\ref{fig:combined}): the result of this higher number of DM particles and lower particle mass ratio is again to \emph{decrease} the growth rate of the bar within 5~Gyr. Given that increasing the number of DM particles by a factor of 10 can mean that there are sometimes more than 8 DM particles in the central cells, we also run the model with a maximum spatial resolution of 15~pc, with no influence on the result apart from a slightly less noisy evolution of the amplitude. We also run simulations with an increased number of \emph{stellar} particles in the disc, and find that this has, on the other hand, no monotonic effect on the bar growth rate in one or the other way \citep[in accordance with][]{Verwilghen}. We finally also run a simulation with as much as 68 times more DM particles and 10 times more stellar particles than the Fiducial Model~II, and get an even slower growth rate of the bar: in this case, we do not increase the force resolution, keeping a spatial resolution of 30 pc, but only make the DM and stellar `fluids' smoother.

Overall, we conclude that, within a 5~Gyr duration in a \ramses\ idealised disc galaxy simulation, lowering the ratio of the DM particles mass to stellar particles mass down to $<5$ does {\it not} speed up the formation of bars, quite the contrary. Clearly, increasing the number of DM particles has the effect of rendering the DM `fluid' smoother, and to decrease the noise, hence giving less small initial non-axisymmetric amplitude variations that can seed the bar growth. As a conclusion, a particle mass ratio as high as $m_{DM}/m_* \sim 45$ cannot be the cause of the missing bar problem in cosmological simulations. Hereafter, in order to be conservative in trying to understand which regions of parameter space dynamically prevent bars from forming in large-volume cosmological simulations, we therefore keep our fiducial particles mass in our grid of idealised simulations. If a bar does not form with such a resolution, it surely cannot form with higher resolution either.

Finally, we also tested a lower maximum spatial resolution of 122 pc for all models (bottom panels of Fig.~\ref{fig:combined}, and Fig.~\ref{fig:Fid_new_highres}). In all cases, we found that the $A_2$ amplitude converges to very similar values. Therefore, we shall now adopt a 122~pc maximum spatial resolution for our grid of simulations, in order to conveniently speed up the numerical computations.

\section{Mapping a grid of simulations to their bar status}
\label{sec:critfromgrid}
\subsection{The grid of simulations}

\begin{figure}
    \includegraphics[width=1\linewidth]{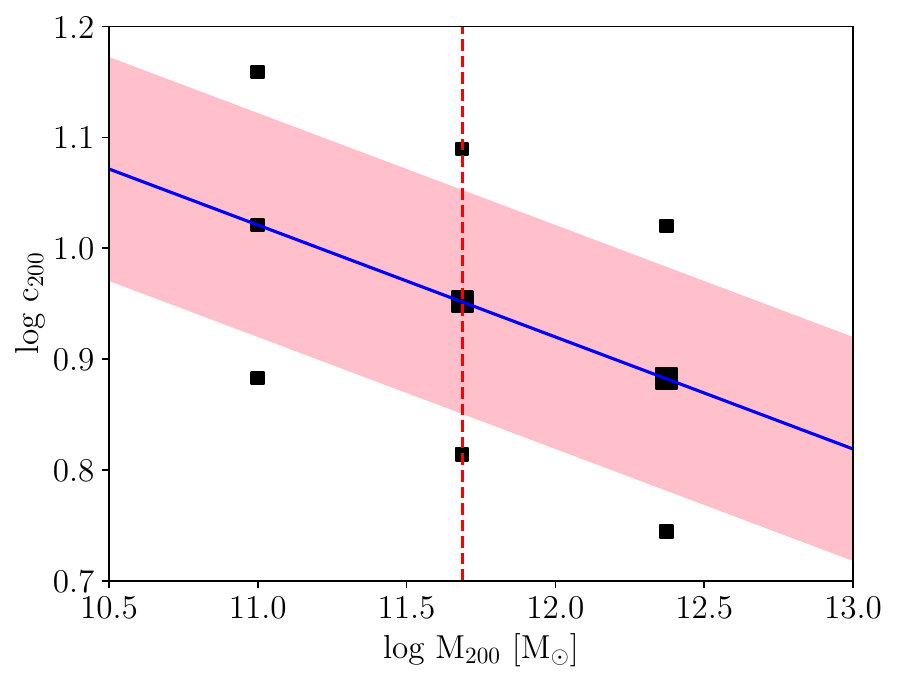}
    \caption{DM halo parameters $M_{\rm 200}$ and $c_{\rm 200}$ of the models, spanning the scatter of the stellar-to-halo mass relation \citep[$\pm 0.6 ~\rm dex$;][]{Behroozi_2013} and that of the concentration-mass relation \citep[blue line and red-shaded area; $\pm 0.13~\rm dex$; ][]{Dutton_Maccio_2014}. The dashed vertical line correspond to the halo mass sitting on the SHMR. The two larger squares denote the virial mass and concentration of the Fiducial Models I and III.}
    \label{fig:M200_C200}
\end{figure}

To investigate the role of DM haloes, gas mass, disc velocity dispersion and bulge mass, we construct a grid of 57 new models with the same stellar mass as the five Fiducial Models, but varying other parameters in a systematic way, leading to a total of 62 models: 

\begin{itemize}
\item First, in the spirit of Fiducial Model~III, we systematically vary the DM halo mass and concentration of the cold disc model in order to understand how DM can inhibit bar formation in such a highly bar-unstable cold disc: two values of $M_{200}$ deviating from the fiducial virial mass by $\pm 0.6 \rm ~dex$  are chosen to reflect the average scatter in the stellar-to-halo mass relation \citep{Behroozi_2013, Behroozi_2019, Rodriguez-Puebla2017}; for each virial mass, we then take three values of the concentration $c_{200}$ deviating from the fiducial concentration by $\pm 0.13\rm ~dex$, to reflect the scatter in the concentration mass relation \citep{Dutton_Maccio_2014}, see Fig.~\ref{fig:M200_C200}. This leads to 7 new models (on top of Fiducial Models~I and III) of cold stellar discs, all with a 20\% gas fraction (denoted as Models 6 to 12 in Table~\ref{tab:halo_params}). \\

\item For all nine models (I, III, and 6-12), we then further vary the gas mass $M_{\rm g}$, such that the gas fraction defined as $f_g= M_{\rm g}/(M_{\rm g}+M_ *)$ is 5\%, 40\%, and 60\%, in addition to the models with $f_g= $~20\%. Hence this leads to another set of 27 cold disc models (Models 13 to 39 in Table~\ref{tab:halo_params}). \\

\item To assess  the effect of gas mass on the instability of warmer stellar discs, we start from Fiducial Model~II and vary the gas mass such that $f_g= $ 5\% and 40\% but keeping other parameters equal to the fiducial ones. This leads to 2 additional warm stellar disc models (Models 40 and 41 in Table~\ref{tab:halo_params}), in addition to Fiducial Model~II with $f_g= $~20\%. \\

\item Then, in the spirit of Fiducial Model IV, to explore the role of the disc velocity dispersion profile further, we also vary in 3 additional warm disc models (Models 42 to 44 in Table~\ref{tab:halo_params}) the central value and scale-length of the radial velocity dispersion profile: we test a value of 100~km/s and 10~kpc and two models with an almost flat profile with 40~kpc scale-length, and a central dispersion of 70~km/s and 60~km/s. \\

\item We also consider two discs within a light DM halo, having a lower surface density by increasing the disc scale-length to 3~kpc. For these, we vary the velocity dispersion scale-length from the fiducial value (4~kpc) to 6~kpc (Models 45 and 46 in Table~\ref{tab:halo_params}). We also consider two discs within a halo of fiducial mass, sitting on the SHMR, similar to Fiducial Models~I and II, but also with lower surface densities by increasing the disc scale length to 4~kpc (Models 47 and 48 in Table~\ref{tab:halo_params}).\\

\item We then add bulges \citep[with shape parameters close to those in][]{Khalil_2024} with different masses. The total stellar mass is kept constant, so that stellar discs are less massive in these bulge-disc models. Here, we introduce the parameter $p$, which is defined as the usual $B/T$ ratio of bulge mass to total stellar mass of the galaxy. We consider two different values of $p$, namely $p=0.23$ and $p=0.37$. We add those bulges on top of Fiducial Model~I, as well as on its variations with gas fractions $f_g = 5$\% and $f_g=40$\%. This leads to 5 new models in addition to Fiducial Model V (Models 49 to 53 in Table~\ref{tab:halo_params}). \\

\item We then consider three cases with a massive bulge ($p=0.46$), a warm stellar velocity dispersion (input central value of 50~km/s), a large velocity dispersion scale-length of 12~kpc, and three different gas fractions, $f_g=$ 5\%, 20\% and 40\% (Models 54 to 56 in Table~\ref{tab:halo_params}). \\

\item To probe the effect of bulges further, we then consider two very cold disks, with a central velocity dispersion of only 20~km/s, but with quite massive bulges ($p=0.3$ and $p=0.42$), hence with disks that should be highly unstable in the absence of a bulge, but whose bulge might stabilize them depending on their mass (Models 57 and 58 in Table~\ref{tab:halo_params}). Conversely, we also add a very light bulge ($p=0.13$) to a slightly warmer disc with central velocity dispersion of 45~km/s (Model 59 in Table~\ref{tab:halo_params}). \\

\item Finally, to explore the effect of the disc central velocity dispersion further, we fix the $B/T$ ratio to $p=0.24$ and smoothly vary the central dispersion from 50~km/s to 60~km/s to explore when it becomes high enough to inhibit bar formation (Models 60 to 62 in Table~\ref{tab:halo_params}).
\end{itemize}

\subsection{Mapping simulations parameters to bar status}
\label{subsec:barcrit}

In the following, to ease the discussion, we introduce the following taxonomy for the bar status after 5~Gyr of evolution (bar definitions always have to rely on some arbitrary thresholds that can slightly vary in the literature): 

\begin{itemize}
\item {\it Bar with fast growth}: at the last snapshot $(A_2/A_0)^{\text{max}} \geq 0.25$ and the phase remains within $10^\circ$, while $(A_2/A_0)^{\text{max}}$ has spent more than 50\% of both the first 1.5~Gyr and of the last 1.5~Gyr above 0.25. An archetypal example is Fiducial Model I.\\
\item {\it Bar with slow growth}: at the last snapshot $(A_2/A_0)^{\text{max}} \geq 0.25$ and the phase remains within $10^\circ$, while $(A_2/A_0)^{\text{max}}$ has spent more than 50\% of the last 1.5~Gyr above 0.25, but less than 50\% of the first 1.5~Gyr. An archetypal example is Fiducial Model II. \\
\item {\it Weak bar}: at the last snapshot $(A_2/A_0)^{\text{max}} \leq 0.25$, but it has been above 0.15 for more than 50\% of the last 1.5~Gyr. The phase remains within $10^\circ$. \\
\item {\it No bar}:  at the last snapshot $(A_2/A_0)^{\text{max}} \leq 0.25$, and it has not been above 0.15 for more than 50\% of the last 1.5~Gyr.
\end{itemize}
In the following, all parameters of the \emph{initial} models that we will map onto the final bar status are taken at $t=100$~Myr. We will, in particular, systematically explore the most common diagnostics suggested in the literature, and show with concrete counter-examples, that they do not fully capture the physics of bar formation in our grid of simulations.

\subsubsection{The role of gas fraction}

\begin{figure}
    \includegraphics[width=1\linewidth]{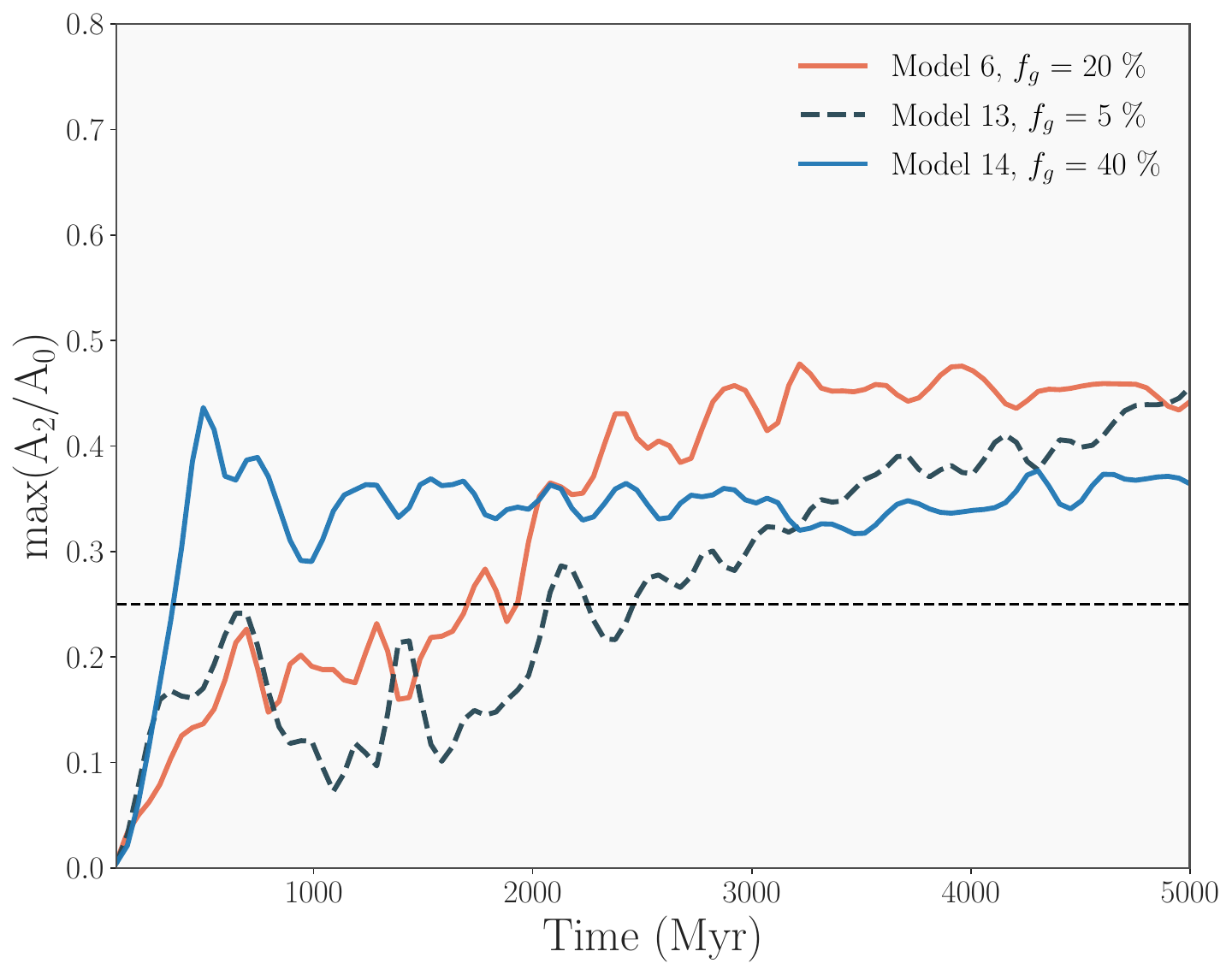}
    \caption{The $m=2$ maximum Fourier amplitude as a function of time for Model 6 ($f_g = 20\%$), 13 ($f_g = 5\%$), and 14 ($f_g = 40\%$). The bar formation timescale and the final bar strength decreases with an increase in gas fraction. Out of these, only Model 14 is labelled as 'fast growth' in our taxonomy.}
    \label{fig:Model_6_13_14}
\end{figure}

Let us first consider bar-forming models that are close to each other in every respect but their gas mass, such as Models 6, 13 and 14 (see Fig.~\ref{fig:Model_6_13_14}) or Models 49, 50 and 52. The first conclusion that one can draw from such a comparison is that a lower gas mass tends to slow down the growth rate of the bar. This is a conclusion that cannot be handled with any bar-instability parameters that would be independent of gas, as often explored in the literature \citep[e.g.,][]{Kataria_2018, Saha_2018, Jang_2025}. Let us note that \citet{Verwilghen} obtained similar results for bulgeless galaxies and for bulgy galaxies at stellar masses close to those considered here, although it seems that the trend is reversed at higher stellar masses for bulgy galaxies.

\subsubsection{The bulge-to-disc and disc-to-halo mass ratios}
As discussed in the introduction, among the main drivers of instabilities in discs are their coldness and their self-gravity. Regarding self-gravity alone, \citet{Reddish_2022} for instance studied the linear response growth rate of razor thin, cold gaseous discs embedded in Plummer halos \citep{Aoki_1979}. Inspired by these analytical insights, the two parameters that may inhibit bar formation which they explored were related to the disc self-gravity, namely the bulge-to-disc ratio and the disc-to-halo ratio. We checked the position of our models in a similar a parameter space: we did not use the $M_{200}$ values of our models but rather the enclosed mass of the DM halo within 50~kpc, motivated by the fact that the Plummer halos studied analytically in  \citet{Reddish_2022} typically yield a much lower halo mass compared to the virial mass of a corresponding NFW halo. Plotting our values on top of figure 7 from \citet{Reddish_2022}, we found various counter-examples: some models with a substantial bulge and rather massive halo within 50~kpc, such as Models 49, 50 and 52, can still form bars, meaning that considering only the enclosed mass of the halo within 50~kpc is probably not sufficient to fully grasp the angular momentum exchange mechanisms with the DM halo; conversely, Model 46, with a zero bulge-to-disc ratio and a high disc-to-halo ratio is not forming a bar, due to the fact that these two parameters do not take into account the damping of the modes by stellar random motions; even considering only cold stellar discs, Model 47 has the exact same stellar and gas disc mass, halo mass and concentration as Fiducial Model I, and only differs from it by its stellar disc scale-length, yet it does not form a bar at all, meaning that considering only mass ratios of the different components is not sufficient.

\subsubsection{The bulge force at the disc scale-length}
A relevant parameter for bar formation proposed by \citet{Kataria_2018} would imply that a bar can only form when $GM_b/(R_d V^2(R_d)) < 0.35$, where $M_b$ is the bulge mass and $V(R_d)$ is the total circular velocity at the scale-length of the disc. Such a parameter clearly could only work as a threshold above which bars are inhibited, since many hot bulgeless models, such as Fiducial Model IV, can also prevent a bar from forming. However, we also found various counter-examples where the bulge is significant and $GM_b/(R_d V^2(R_d)) > 0.35$ while the galaxy still forms a bar: for instance, Models 49, 50, 60, 61. This parameter therefore fails in our grid of simulations.

\subsubsection{The local bulge-to-disc density ratio}
In the same vein, \citet{Saha_2018} proposed that any bulge whose ratio of the mean density squared within its half mass radius by the mean density squared of the disc within that same radius is larger than $1/10$ (i.e., $\langle \rho_b \rangle^2 /\langle \rho_d \rangle^2 > 0.1$ in their Eq.~3) inhibits bar formation. Such a parameter however does not take into account the propensity of bulgy ultra-cold discs to form bars, such as Model 57 for which this ratio is much larger and which still forms a bar, thereby contradicting the capacity of the parameter to predict bar-inhibition. Conversely, it is obvious that any warm enough disc with zero bulge can also inhibit bar formation, meaning that a ratio below 0.1 is of course not a guarantee to form a bar either. 

\subsubsection{The baryonic-to-total force ratio for bulgeless models}
Another set of parameters proposed by \citet{Fujii_2018} for bulgeless galaxies combines the total circular velocity curve to that of the disc (gas+stars) at $2.2R_d$, stating that bars cannot form within a Hubble time if $f_{\rm bary} = V_d^2(2.2R_d)/V^2(2.2R_d)<0.3$. This is indeed verified by our simulations. This Fujii parameter has been generalized by \citet{Bland-Hawthorn_2023} in order to directly relate the actual growth rate of the bar to the Fujii parameter. However, let us point out that if we only consider the cold Fiducial Models I, III, and their variations (Models 6 to 39), we do not find {\it any} value of $f_{\rm bary}$ that cleanly separates the models forming bars from those that do not. For instance, $f_{\rm bary}=0.6$ for Model 39 (high mass but low concentration), which does not form a bar, while it is only 0.4 for Model 13 (high concentration) which does form a bar. This demonstrates that this rather local Fujii parameter is not enough, and that the mass of the DM halo in the outskirts can also play a significant role in damping the formation of the bar. 

\subsubsection{The role of the Toomre parameter}
The \citet{Bland-Hawthorn_2023} growth rate was further refined by \citet{Chen_2025}, who computed the bar formation timescale as being proportional to $\propto {\rm exp}(-f_{\rm bary}/0.11) \times (h_z/R_d) \times Q$, where $Q$ is some characteristic value of the stellar Toomre parameter \citep{Toomre_1964} within the disc. Indeed, as pointed out by, e.g., \citet{Athanassoula_2008}, who gave various explicit examples of idealised discs that do not form bars while being almost entirely self-gravitating, when the stellar velocity dispersion is high, it can delay or inhibit bar formation by reducing support to the bar-supporting family of orbits due to random motions, damping the reflection of the waves as they pass through the centre, and rendering the (near-) resonant materials less efficient in emitting/absorbing angular momentum \citep{Lynden-Bell_1972, Tremaine_1984, Athanassoula_Sellwood_1986}. This is typically what happens in our Fiducial Model~IV which remains remarkably axisymmetric throughout the simulation. Therefore, it is in general worth asking whether the Toomre parameter can indeed distinguish bar-forming models from models not forming bars, in combination with some other parameter taking into account the stabilization of a bulgeless disc by the DM halo. \citet{Jang_2025} and \citet{Worrakitpoonpon_2025} have proposed exactly that by combining the minimum Toomre parameter within the disc with a central mass concentration (CMC) parameter corresponding to the total mass enclosed inside some radius close to the center (typically 0.1~kpc to 0.4~kpc) relative to the disc mass. We remind that the Toomre parameter \citep{Toomre_1964} is defined locally as 
\begin{equation}
Q_* \equiv \frac{\kappa \sigma_{r,d}}{3.36 G \Sigma_d},
\label{eq:Toomre_stars}
\end{equation}
where $\kappa$ is the radial epicyclic frequency, e.g. directly given by \texttt{AGAMA}, $\sigma_{r}$ is the local radial velocity dispersion of disc stars, and $\Sigma_{d}$ is the surface density of the disc. It is originally a local stability parameter against \emph{axisymmetric} perturbations of the disc, against which the disc is stable when $Q_* > 1$. Even when $Q_* \leq 1$, the disc does not necessary fragment but can give rise to multiple non-axisymmetric modes that will quickly heat it up back to $Q_* \sim 1$. In practice, one can compute a profile $Q_*(R)$ and decide to use as a characteristic Toomre parameter, either its minimum value, or the value at some characteristic radius, such as the effective radius of the disc, the stellar half-mass radius of the galaxy (which can differ from the effective radius of the disc in the presence of a bulge), or at the peak of the baryonic rotation curve. Considering only our bulgeless models, we have looked at their position in the $Q_*-CMC$ space, for various definition of $Q_*$ as well as for various sizes of the CMC, and the bar-forming models \emph{always} overlap in this space with the bulgeless models not forming bars. Hence, these parameters are {\it not} sufficient to discriminate bar forming from non-barred models. 

As for the very general bar formation timescale proposed by \citet{Chen_2025}, generalising the Fujii parameter based on Toomre's stability parameter, we have checked that, considering the minimum $Q_*$ or any other characteristic value for it, a direct comparison of Models 13, 14, 35 and 38 {\it directly} contradicts their estimate: Models 13 and 35 should indeed have almost the same bar formation timescale, but one actually forms a bar and the other one does not within 5~Gyr; also, Model 14 should have a relatively long timescale for bar formation, not far from Model 35, but actually forms a bar very fast; Model 38 should have a shorter bar formation timescale than all three others, but does not form a bar at all. Finally, as far as the scale length is concerned, we note that the bar formation timescales of Model 46 should be shorter than Fiducial Model II, but Model 46 does not form a bar. This demonstrates that neither the Fujii parameter, nor their more recent upgrades by \citet{Bland-Hawthorn_2023} or \citet{Chen_2025}, are sufficient to identify bar-inhibiting properties of discs.
It is also clear from our grid of models that any characteristic value of the Toomre parameter alone cannot be enough to predict whether a disc forms a bar or not, since all our massive halos prevent bars from forming even in cold discs. However, we have already seen that the Fujii parameter cannot separate cold discs forming bars or not. 

\subsubsection{The role of the swing-amplification parameter}
A complementary and largely independent parameter from the original Toomre $Q$ is the swing-amplification parameter \citep{Toomre_1981}, which in principle also plays a crucial role in allowing the bar instability to grow:
\begin{eqnarray}
    X \equiv \frac{\kappa^2 R}{4\pi G \Sigma_d}\, ,
    \label{eq:X_def}
\end{eqnarray}
here written for any bi-symmetric ($m=2$) mode. Physically, this $X$ parameter measures the azimuthal wavelength of one period of an $m$-fold perturbation, $2\pi R/m$, in units of the longest wavelength that the disc self-gravity can destabilise, i.e., $\lambda_{\rm crit} = 4\pi^2 G \Sigma_d / \kappa^2$. Swing-amplification is the boost in amplitude of a non-axisymmetric density perturbation as differential rotation shears it from leading to trailing, during which its self-gravity acts in concert with the epicyclic motion of the stars to reinforce it: it is efficient only when $X$ lies in a window of order $ 1 \lesssim X \lesssim 3$. When it is too low, the perturbation is on small enough scales that random motions can act against it wherever $Q \gtrsim 1$, and when it is too high, the wavelength of the mode is so large compared to $\lambda_{\rm crit}$ that it cannot be swing-amplified. Unlike the Toomre stability parameter, which controls the damping of the amplifier through stellar random motions, $X$ is a purely structural quantity: it depends only on $\kappa$, $\Sigma_d$ and $R$, and not on the velocity dispersion. It therefore isolates the self-gravity of the disc from its temperature. It is however a purely local parameter, like $Q$, and does not distinguish the effect of a live DM halo from a static one.

For a submaximal stellar exponential disc of scale-length $R_d$ and mass $M_{\rm *,d}$, assuming the rotation curve to be locally almost flat, so that $\kappa^2 \simeq 2 V^2/R^2$, and defining
$x \equiv R/R_{\rm d}$, Eq.~(\ref{eq:X_def}) becomes
\begin{eqnarray}
    X \simeq \frac{e^{x}}{x} \left(\frac{V}{\sqrt{G M_{\rm *,d}/R_{d}}}\right)^2.
    \label{eq:X_eps}
\end{eqnarray}
This value is minimal at $R=R_d$ (x=1), and if we assume that $X_{\rm min} \sim 3-3.3$ is at the limit of swing-amplification starting to be efficient \citep{Toomre_1981}, the corresponding highest possible value at $R=2.2 R_d$ (i.e., at the maximum of the stellar rotation curve) for efficient swing-amplification in this submaximal disc with flat rotation curve becomes $X_p \sim 4.5-5$. 

Both reality and our simulations are however much more complex than the above pure exponential submaximal disc and flat rotation curve ansatz. We tried to map the value of the stellar swing-amplification parameter at its minimum, $X_{\rm min}$, and at the peak of the baryonic rotation curve, $X_p$, to the bar status of the models. The $X_{\rm min}$ value is systematically reaching below 2 in all our cold bulgeless discs, without any discriminating power. On the other hand, for the value at the peak of the baryonic rotation curve, we found out that, among our bulgeless models with $f_g \leq 40\%$, the phenomenological limit appears to be $X_p \geq 5.5$: only 1 out of 9 Models with such a high $X_p$ forms a bar (Model 14). For the reverse mapping, at low $X_p$, among cold bulgeless disc models with $f_g \leq 40\%$ and $X_p < 5.5$, only 3 out of 20 do not form a bar. The situation is obviously worse for warm disc models, and the mapping completely fails at $f_g = 60\%$ (see Fig.~\ref{fig:swing}). The latter is not surprising since the swing-amplification parameter is a purely stellar one. Concerning warm disc models, however, the situation can obviously be improved if combining the swing-amplification parameter $X_p$ with some version of the Toomre parameter $Q$, as already noted by \citet{Toomre_1981}, which we will explore below.

\subsubsection{The Romeo-Falstad stability parameter}
\begin{figure}[htbp]
    \centering

    \begin{subfigure}{0.45\textwidth}
        \centering
        \includegraphics[width=1.0\textwidth,height=0.75\textwidth]{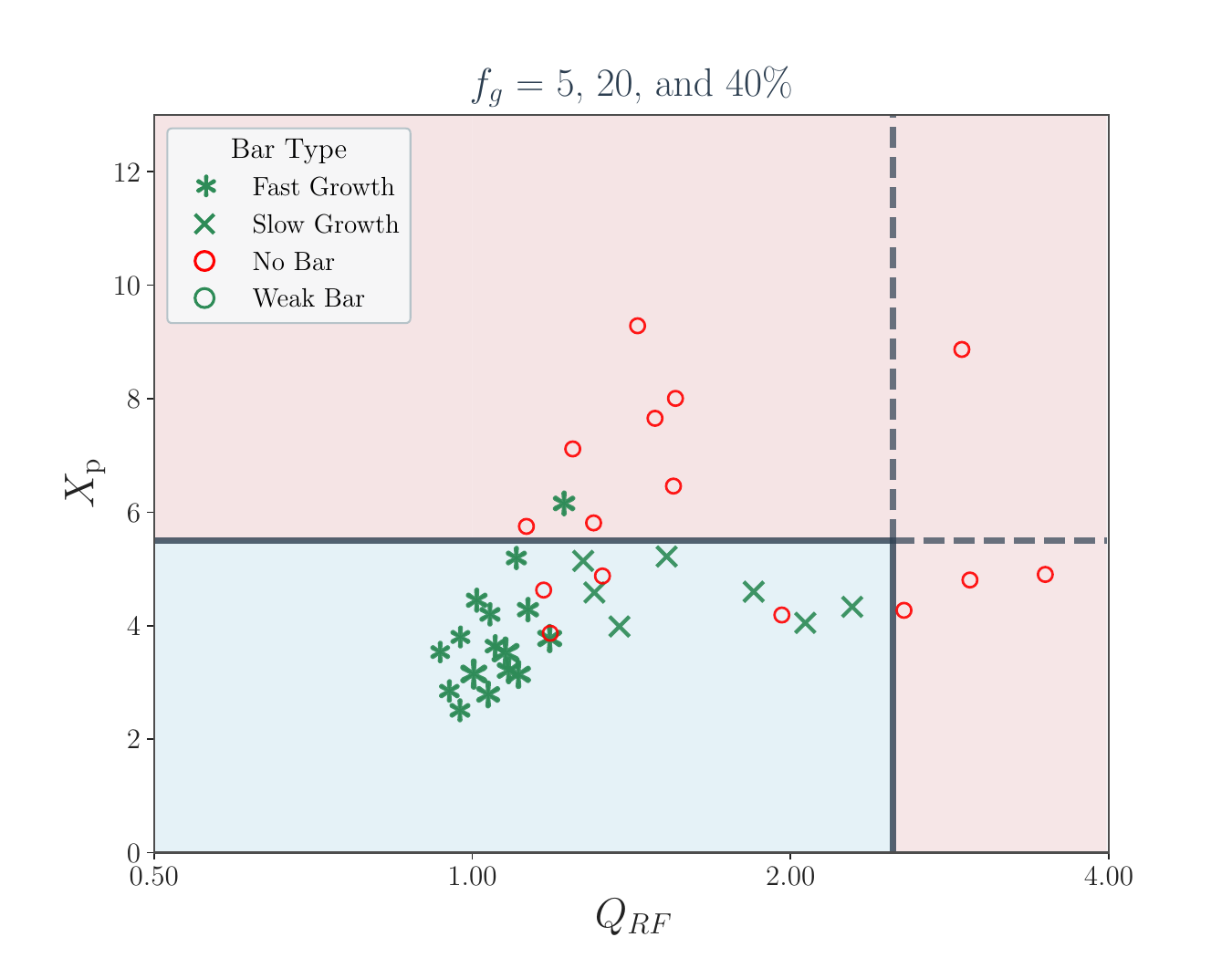}
    \end{subfigure}
    \vfill
    \begin{subfigure}{0.45\textwidth}
        \centering
        \includegraphics[width=1.0\textwidth,height=0.75\textwidth]{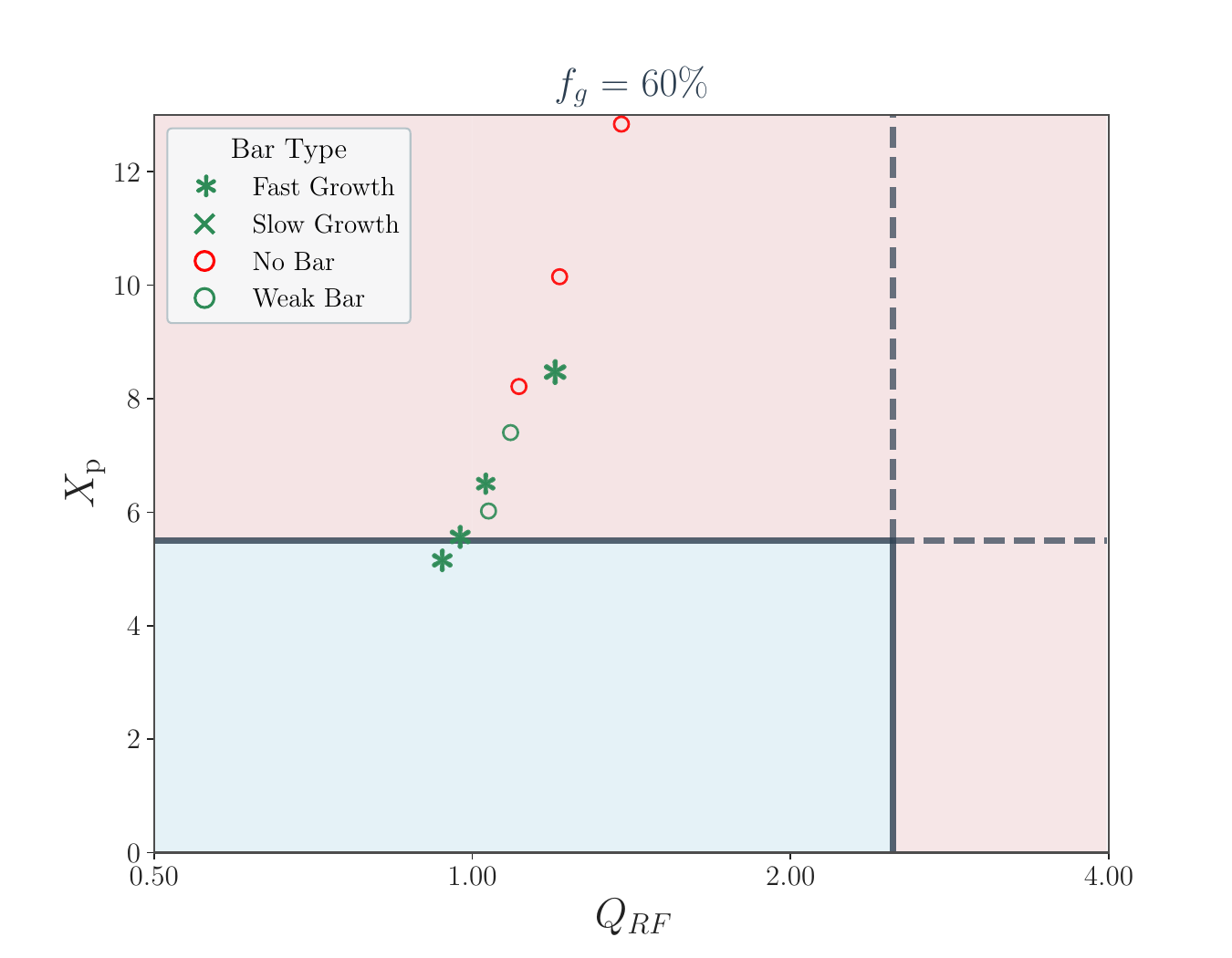}
    \end{subfigure}
    \caption{The swing-amplification parameter at the peak of the baryonic rotation curve, $X_p$, versus the Romeo-Falstad stability parameter, $Q_{\rm RF}$, for all our bulgeless models at $t=100$~Myr. Top: for $f_g \leq 40\%$. Bottom: for $f_g = 60\%$. The horizontal line is at $X_p = 5.5$ while the vertical line is at $Q_{\rm RF}=2.5$.}
    \label{fig:swing}
\end{figure}

The $Q_*$ parameter of \citet{Toomre_1964} characterises the axisymmetric stability of a razor-thin, differentially rotating purely stellar disc. Real galactic discs, however, are neither purely stellar nor infinitesimally thin: the interstellar gas contributes substantially to the local self-gravity, while the finite vertical thickness of each component weakens its self-gravity in the plane. Both effects are absent from $Q_*$, which therefore misrepresents the true stability level of a multi-phase, realistically thick disc, especially when the gas is much colder than the stellar component. Subsequent studies have therefore extended the Toomre parameter to include gaseous discs and the vertical extent of galactic discs, proposing refined stability parameters \citep[e.g.,][]{Jog1984, George_2025}. In particular, \citet{Romeo_2013} introduced a multi-component stability parameter that combines the individual components 
\begin{eqnarray}
    Q_{\rm RF} ~\equiv~ \left(\sum_i \frac{W_i}{T_i\, Q_i}\right)^{-1}\, ,
    \qquad
    \text{where} \; Q_i \equiv \frac{\kappa\,\sigma_{R,i}}{\pi\,G\,\Sigma_i}\ \ (\text{for gas}),
    \label{eq:Q_RF} \\
    \text{and} \; Q_i \equiv \frac{\kappa\,\sigma_{R,i}}{3.36\,G\,\Sigma_i}\ \ (\text{for stars})\, ,
    \nonumber
\end{eqnarray}
where the sum runs over all stellar and gaseous components, $T_i$ is a thickness correction that changes the stability parameter as a function of the ratio $\sigma_{z,i}/\sigma_{R,i}$, and the weights $W_i$, built as functions of the ratios of radial velocity dispersions, ensure that the dynamically coldest component dominates the combined stability level \citep[see][for details]{Romeo_2013}. For gas, we add in quadrature the turbulent dispersion and that coming from the actual gas temperature. Through its thickness correction, $Q_{\rm RF}$ is also naturally anchored to the scale at which the density and dispersion are actually measured. We compute this $Q_{\rm RF}$ parameter for all our bulgeless models, and identify its minimum value within the disc as its characteristic value for each galaxy. We then combine this value with the swing-amplification parameter in Fig.~\ref{fig:swing}, which demonstrates that a limiting value of about $Q_{\rm RF} \simeq 2.5$ separates well the low $X_p$ warm bulgeless models forming bars from those that do not (100\% of those with $Q_{\rm RF} > 2.5$ do not form bars). However, let us note that, probably related to the fact that we do not let the gas cool, meaning that contrary to more realistic model with cooling, star formation, and feedback, our stellar component is often cooler than the gaseous ones, simply using the original purely stellar Toomre parameter $Q_*$ also works well to delimitate the warm bar forming models from the unbarred ones, with a threshold at 2.2 instead of 2.5.

\subsubsection{The ELN parameter}

Let us now go back to Eq.~\eqref{eq:X_eps} for a pure single exponential submaximal disc with flat rotation curve. In this simplified case, the phenomenological condition $X_p \gtrsim 5.5$ from our simulations (see Fig.~\ref{fig:swing}) at the peak of the baryonic rotation curve is \emph{equivalent} to the condition $V/\sqrt{G M_{\rm *,d} / R_d} \gtrsim 1.16$ at the maximum of the stellar rotation curve in Eq.~\eqref{eq:X_eps}. This quantity, if instead taken at the total maximum circular velocity of a more general disc model , $V_{\rm max}$, is actually known as the \citet*[][ELN]{ELN_1982} parameter, $\varepsilon_{{}_{\rm ELN}}$. This phenomenological parameter, whose theoretical basis therefore lies in the swing-amplification of $m=2$ modes in pure exponential discs embedded in pure logarithmic potentials, however clearly differs from any quantification of the swing-amplification parameter in any more realistic set-up, where the rotation curve is \emph{not} flat, and where the disc is \emph{not} a single exponential, and also includes a gas component. 

The ELN parameter threshold to inhibit bar formation was originally proposed from simulations of stellar 2D exponential discs within rigid DM halos, $\varepsilon_{{}_{\rm ELN}} \equiv V_{\text{max}}/\sqrt{G M_{\rm *,d} / R_d} > 1.1$. \citet{Athanassoula_2008} notably pointed out that such a criterion based on 2D discs within rigid halos was fundamentally limited, because it neglects the angular momentum exchange with a live halo. Indeed, subsequent studies have demonstrated that the proposed threshold does {\it not} work for any rigid halo configuration. However, it apparently (and suprisingly) works again in 3D disc-halo $N$-body systems once the halo is live, as recently shown by \citet{Sellwood_2025}. This has also been qualitatively confirmed by \citet{Frosst}, showing that live halos form bars much more efficiently than rigid ones, but that the traditional ELN threshold still applies quite efficiently to live halo simulations. More precisely, they found a good correlation between the ELN parameter and the growth rate of the bar, $\tau_b$, defined such that $(A_2/A_0)^{\rm max}(t) = P \times \exp\left(t/\tau_b\right)$. Assuming initial fluctuations at the level of $P \approx 0.05$ and an orbital time at the scale-length of the disc of 75~Myr \citep[as in][]{Frosst}, 97~Myr (the minimum and typical orbital time at $R_d$ in our bulgeless models) and 146~Myr (the orbital time at $R_d$ in our most massive halo models), we obtain a \emph{maximum} ELN parameter allowing to reach $(A_2/A_0)^{\rm max}=0.25$ in 5~Gyr of 1.18, 1.14 and 1.09, respectively. For reaching $(A_2/A_0)^{\rm max}=0.25$ in 8~Gyr, the upper thresholds would be a bit larger, and range instead from 1.24 to 1.14. These values are all well in line with the typical value obtained from applying Eq.~\eqref{eq:X_eps} to the phenomenological $X_p \gtrsim 5.5$ threshold, $\varepsilon_{{}_{\rm ELN}} \gtrsim 1.16$, in the case of a pure exponential disc with an almost flat rotation curve.

Despite both parameters being derived from the same physics in the  ansatz of Eq.~\eqref{eq:X_eps}, using $X_p$ or $\varepsilon_{{}_{\rm ELN}}$ as a true diagnostic parameter in a realistic simulation is very different. Indeed, if we consider $\varepsilon_{{}_{\rm ELN}} \geq 1.14$ as a typical value above which bars would not be able to form, in line with the \citet{Frosst} estimate for our typical orbital time at $R_d$, this threshold turns out to be valid for \emph{all} our bulgeless models, including the high gas-fraction ones, contrary to $X_p$. The success of this threshold, while having no pretense to universality, should not be understated either, since the originality of our suite of simulations is that we are considering a significant sample of very cold idealised discs. Notwithstanding that this $\varepsilon_{{}_{\rm ELN}}$ threshold is obviously not a strict one, as shown by, e.g., \citet{Frosst} where the dependence on orbital time is clear, the fact that not a single of our bulgeless models akin to Fiducial Model III, with an ELN parameter above 1.14, manages to form a bar in 5~Gyr despite being extremely cold, is a significant result in itself. Moreover, simply increasing the scale-length of an extremely cold stellar disc without altering the halo mass or concentration, as in the case of Model 47 vs. Fiducial Model~I, is also sufficient to inhibit bar formation within 5~Gyr, as expected from the corresponding change in the ELN parameter. Combined with a threshold for the Romeo-Falstad parameter taken at its absolute minimum in each disc, $Q_{\rm RF}>2.5$, preventing bar formation in too warm discs even for a low $\varepsilon_{{}_{\rm ELN}}$ parameter, the propensity of our 47 bulgeless models to inhibit bar formation within 5~Gyr is then almost fully characterized, with only one exception (Model 46). Model 46 is very peculiar in the sense that it has a low minimum value of $Q_{\rm RF}$ at very large radii, while its value in the bar region is much larger than the threshold.

\subsubsection{Bulge-dependent thresholds}

Having found a relevant region of parameter space inhibiting bar formation for bulgeless models, combining their temperature (through the Romeo-Falstad parameter) and their self-gravity (through the ELN parameter), we can now study how the presence of bulges, that have long been known to be important culprits for bar inhibition mechanisms, changes the picture. 

\begin{figure*}[htbp]
    \centering
    \includegraphics[width=0.48\textwidth]{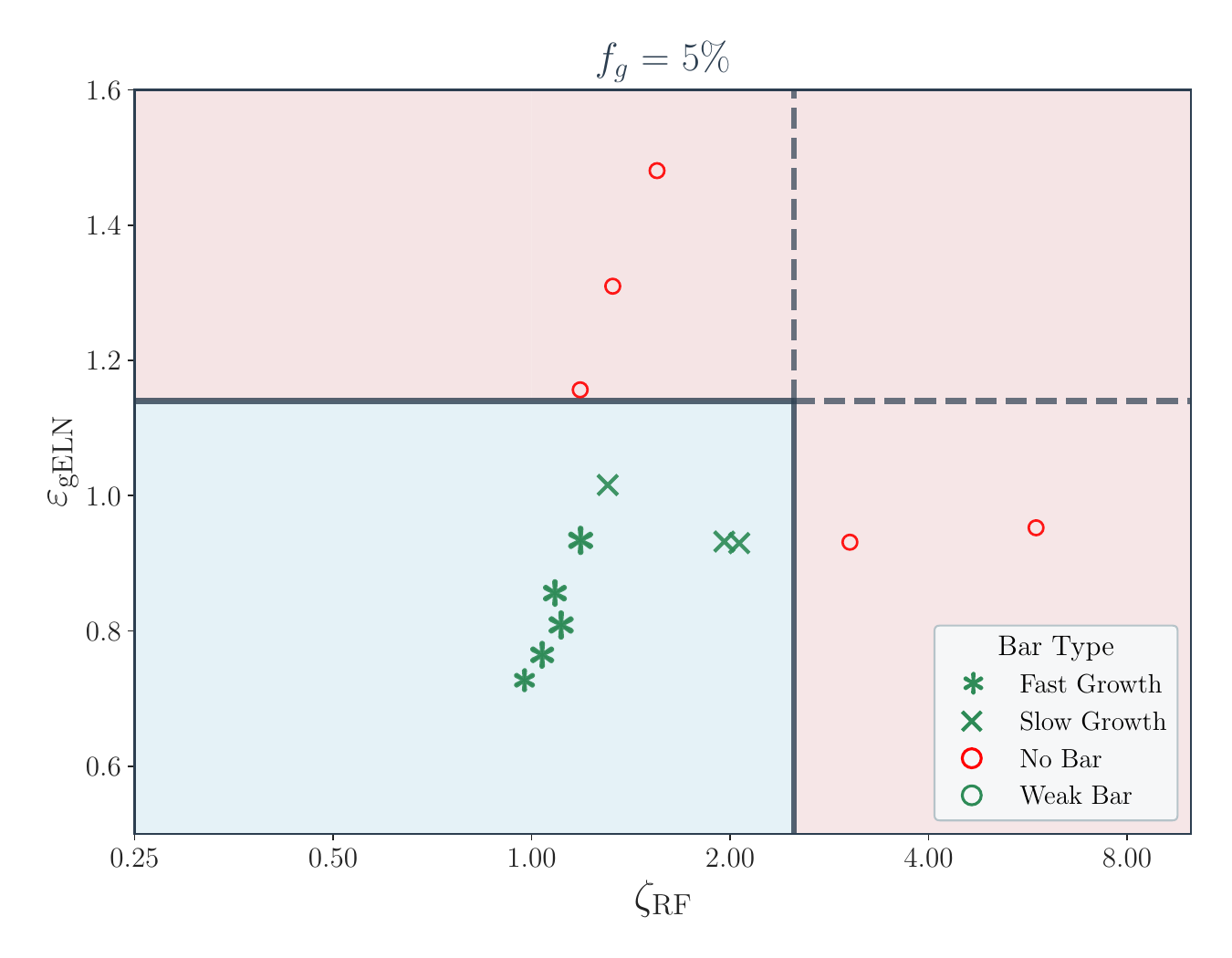}
    \hfill
    \includegraphics[width=0.48\textwidth]{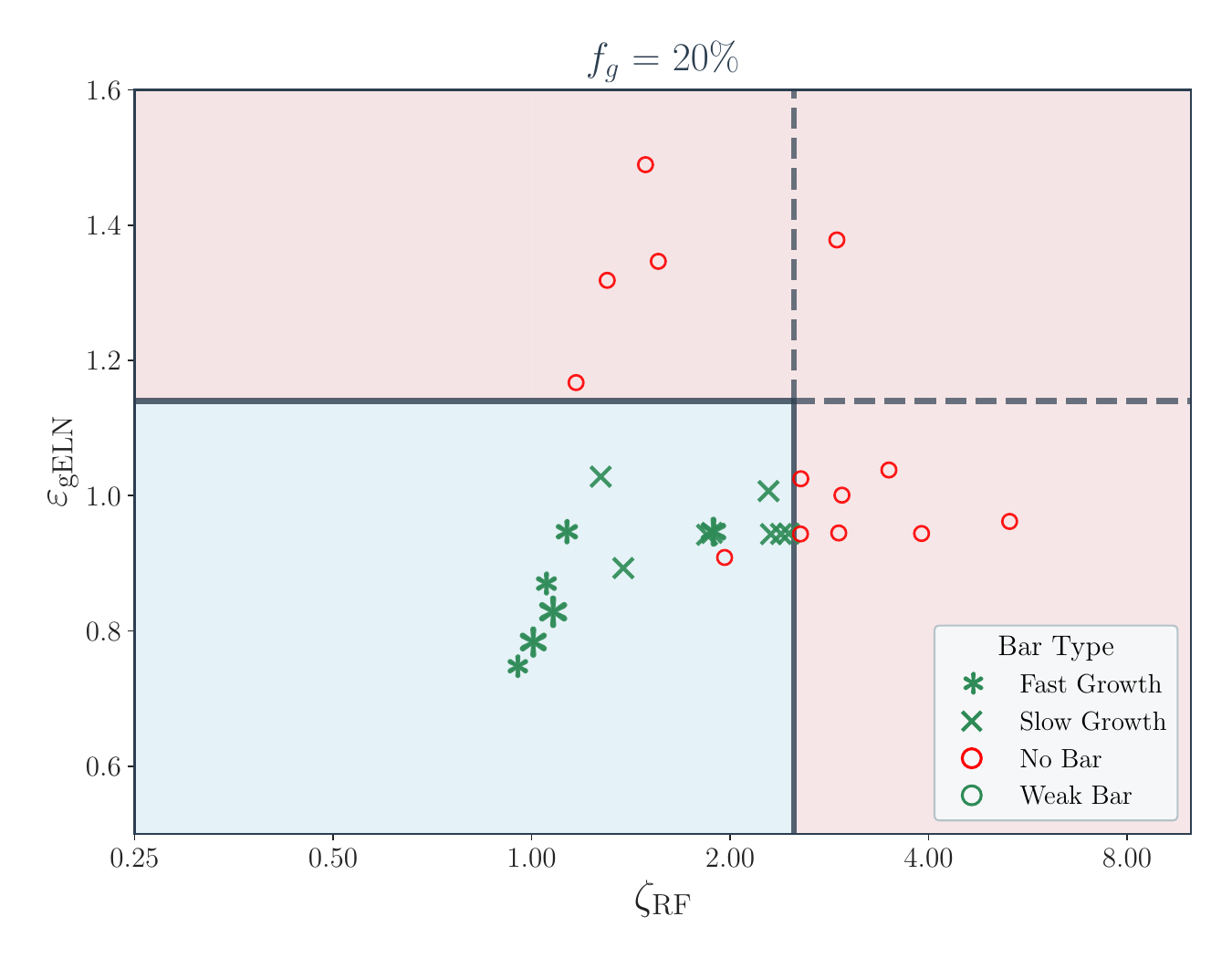}
    \\
    \includegraphics[width=0.48\textwidth]{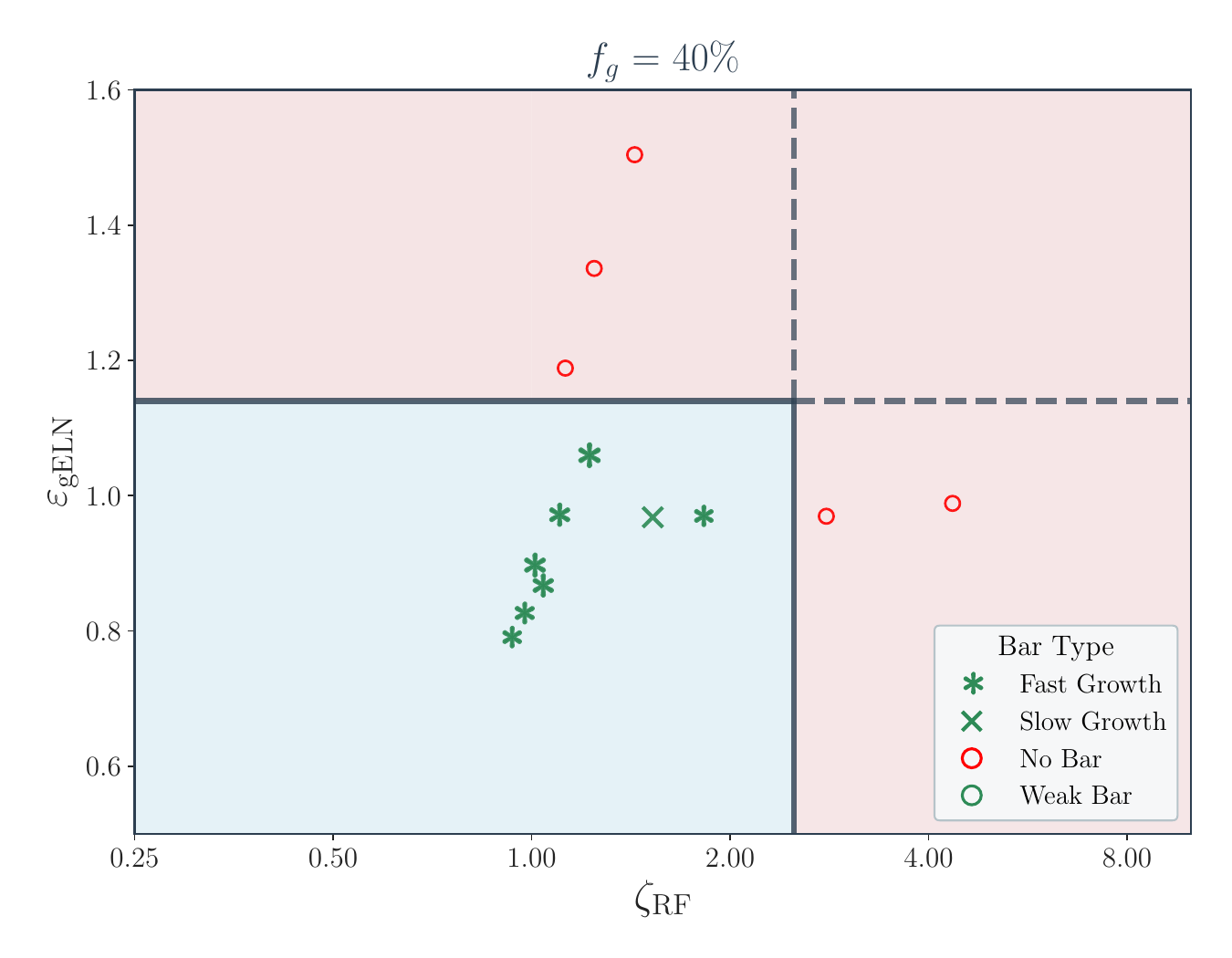}
    \hfill
    \includegraphics[width=0.48\textwidth]{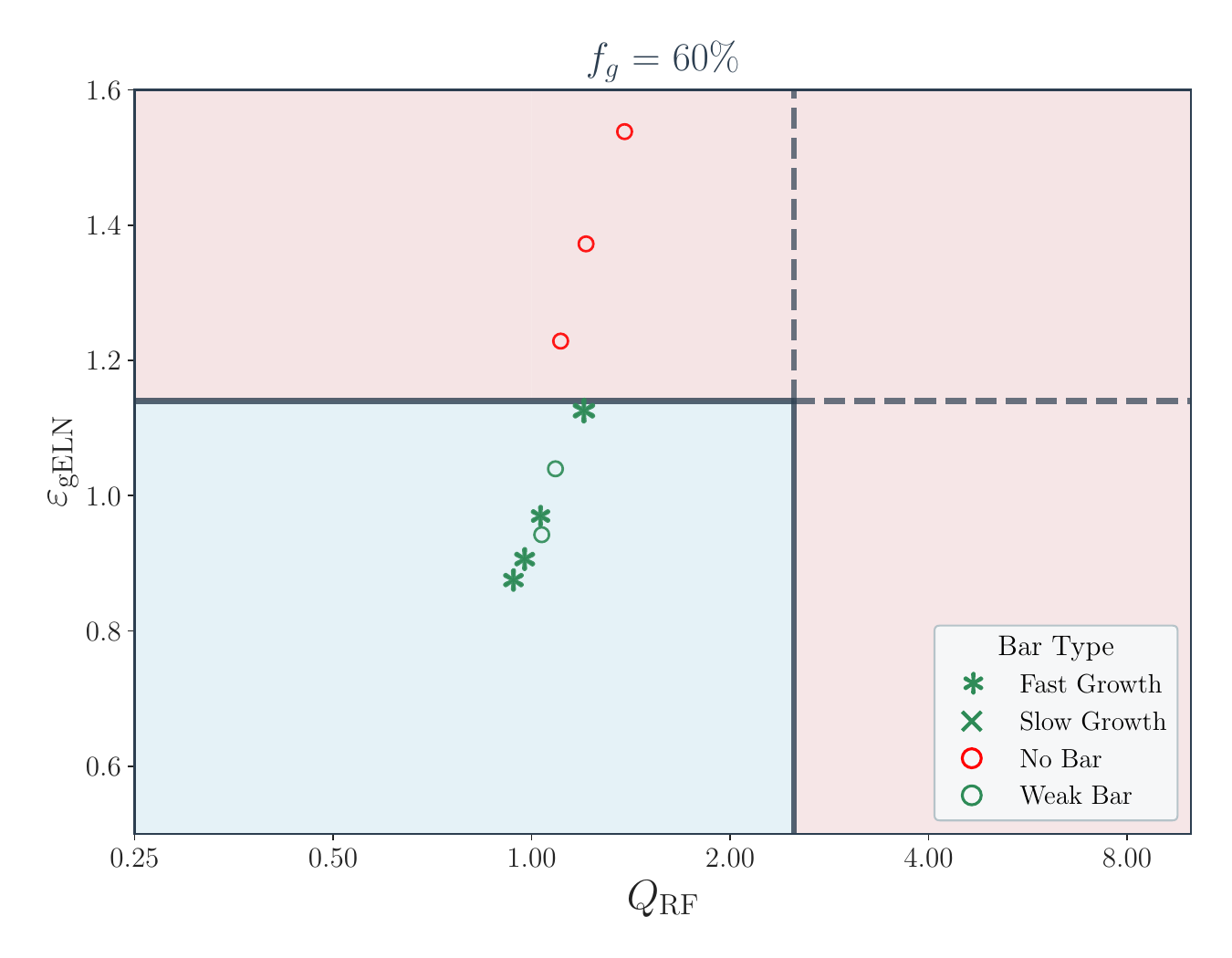}
    \caption{The generalized ELN parameter $\epsELN$ from Eq.~\eqref{eq:ELN} vs. the $\zeta_{\rm RF}$ parameter from Eq.~\eqref{eq:zeta}, combining the Romeo-Falstad $Q_{\rm RF}$ stability parameter with the square of the $p \equiv B/T$ ratio, for the different gas masses probed in our models ($f_g \equiv M_{\rm g}/(M_{\rm g}+M_*) = 5$\%, $20$\%, $40$\%, and $60$\%). Markers indicate whether a bar forms (green star in the case of a fast growth, green cross in the case of a slow growth, open green circle for a weak bar) or not (open red circle). All parameters are taken at $t=100$~Myr. Models with $\epsELN>1.14$ {\it or} $\zeta_{\rm RF}>2.5$ do \emph{not} form bars, while most others do form a bar (with only one exception -- Model 46).}
\label{fig:ELNzetafgall}
\end{figure*}

Since a bulge is just an additional spheroidal component that can in principle act to stabilize discs against bar formation, we first test whether $\varepsilon_{{}_{\rm ELN}}$ can simply be used as a relevant parameter for bulge models too, considering the bulge as another spheroidal contribution to $V_{\rm max}$. However, this does \emph{not} efficiently separate models forming bars from those that do not. For instance, Model 57 does form a bar despite having a large classic ELN ratio, about the same as that of Model 12, which does not form a bar. However, interestingly, the following condition on the \emph{generalised} ELN parameter taking into account the \emph{total} (bulge+disc) stellar mass instead of the disc stellar mass, appears to be a \emph{sufficient} condition for bar formation to be fully \emph{inhibited} over a 5~Gyr timescale in all our models (see Fig.~\ref{fig:ELNzetafgall}):
\begin{equation}
    \epsELN \equiv \frac{V_{\text{max}}}{\sqrt{G (M_{\rm *,d}+M_{\rm b}) / R_d}} \geq 1.14,
    \label{eq:ELN}
\end{equation}
where $M_{\rm *,d}+M_{\rm b}=M_*$ is the total stellar mass. As already noted above, this cannot be a strict limit, as it will depend slightly on orbital times or on the actual timescale chosen to allow for bar formation. Moreover, we have also pointed out that gas-rich models appear to be more bar-unstable than gas-poor ones, so it is likely that this limit does, in fact, also depend on the gas fraction itself. Things would also become different once gas is allowed to cool and form new stars. However, the fact that the extremely cold stellar discs of our suite of simulations do not manage to form bars for high $\epsELN$ values indicate that it likely would be hard for a bar to form in these discs under any condition, even in the presence of a colder gas component. Conversely, it means $\epsELN \lesssim 1.14$ appears to be a \emph{necessary} condition (again, with the threshold nevertheless not being a strict one) for discs to be bar-unstable. However, it is not sufficient, since discs must also be cold enough for a bar to form. 

Indeed, we have already mentioned that the minimum value of the Romeo-Falstad stability parameter of Eq.~\eqref{eq:Q_RF} within the disc allows to rather efficiently separate low-$\epsELN$ models forming bars from those that do not, with a limiting value $Q_{\rm RF} \sim 2.5$, and only one exception (Model 46 not forming a bar despite having a low minimum $Q_{\rm RF}$, which is however reached only at large radii). However, this limiting $Q_{\rm RF}$, which is a pure disc quantity, completely breaks down in the presence of bulges, which inhibit bar formation. For instance, Fiducial Model V has a minimum $Q_{\rm RF} \sim 1.6$ and still does not form a bar. We have therefore considered the locations of all bulgy models in a 2D space of the disc $Q_{\rm RF}$ vs. different parameters quantifying the bulge mass, such as the ratio of bulge mass to virial mass, the ratio of bulge mass to total baryonic mass, or the usual $B/T \equiv p$ ratio, which in our case is equivalent to the bulge mass itself since the total stellar mass is kept fixed. We found that only the latter separates clearly low-$\epsELN$ galaxies forming bars from those that do not. And an interesting finding is that it does so with a \emph{quadratic} dependence of the limiting value of the disc $Q_{\rm RF}$ on the bulge mass. Namely the sufficient condition on the  Romeo-Falstad parameter $Q_{\rm RF}$ (taken at its minimum within the disc) for bar formation to be \emph{inhibited} within our simulations in the presence of a massive and compact bulge becomes:
\begin{equation}
Q_{\rm RF} \gtrsim Q_{\rm lim} - \gamma p^2 \, ,
\label{eq:QRFlimit}
\end{equation}
with $Q_{\rm lim} = 2.5$ and $\gamma=10$. To get a non-bar-forming stable ultra-cold disk with $Q_{\rm RF} =0$, the needed intercept would be at $p \equiv B/T = 0.5$, while in the zero-bulge limit, the limit is instead at $Q_{\rm RF} = 2.5$. In combination with the $\epsELN$ threshold of Eq.~\eqref{eq:ELN}, this bulge-dependent threshold efficiently separates our bar-forming models from unbarred models, with still only one exception (Model 46). This is illustrated on Fig.~\ref{fig:ELNzetafgall}, where we plot all our models in the $\epsELN$-$\zeta_{\rm RF}$ parameter space, with
\begin{equation}
\zeta_{\rm RF} \equiv Q_{\rm RF} + 10 p^2,
\label{eq:zeta}
\end{equation}
with a separation line at $\zeta_{\rm RF} = 2.5$. We also report the parameters $\epsELN$ and $\zeta_{\rm RF}$ for all our models in Tables~\ref{tab:Fiducial_paramss} and \ref{tab:halo_params}. 

If we instead use the classic stellar Toomre parameter at its minimum value, the value of $\gamma=10$ remains unchanged but the limiting value drops to $Q_*=2.2$, due to the fact that we do not let the gas cool. In this case, we have two outliers instead of one in the $Q_{\rm RF}$ case, making the latter a better discriminant.

As a conclusion, despite most phenomenological bar-instability criteria proposed in the literature not working for our grid of simulations, we found that a (still purely phenomenological) combination of the minimum Romeo-Falstad stability parameter, of the bulge mass and of a generalised ELN parameter, does quite reliably map onto whether an initially axisymmetric idealised disc with stellar mass of about $10^{10} \, {\rm M}_\odot$ forms a bar within 5~Gyr within our own grid of simulations. The limiting values of the parameters we identified appear as sufficient conditions for the bar formation to be inhibited within a 5~Gyr timescale in our simulations, but {\it not} as sufficient conditions for bars to form in such a timescale, since we have identified one outlier (Model 46) not forming a bar despite having a low ELN and a low minimal value of the Romeo-Falstad stability parameter within the disc. We insist that the limiting values identified here should \emph{not} be seen as strict limits that would apply to any numerical configurations, codes, integration schemes, but it does give a good overall view of the trend, namely of which parameters are prone to inhibit bar formation in a pure hydrodynamic context. Even with our own numerical scheme, it is also expected that these limiting values are \emph{not} strictly sufficient to inhibit bar formation, and that a grey zone must exist around the limits, themselves depending on other parameters such as gas fraction, orbital timescales, and parameters we have not considered here, such as halo spin or variations of the bulge structural parameters (compactness in particular). Moreover, it is obvious that the limiting values must evolve if longer timescales are considered for bar formation. 

In any case, these parameters should \emph{not} be immediately translated into any kind of direct observational test for the presence of bars in observed galaxies. First of all, because real galaxies are evolving in a cosmological context, and are much more complex than these simplified hydrodynamic simulations that do not consider star formation nor feedback. Second, even when considering such simplified simulations, the diagnostic only applies to the initial state prior to bar formation but not to the final (``observable'') state of the galaxies. Indeed, let us now analyse how our simulations evolve in parameter space.

\subsection{Evolution in parameter space}
\label{subsec:evol}

\begin{figure}
    \includegraphics[width=0.48\textwidth]{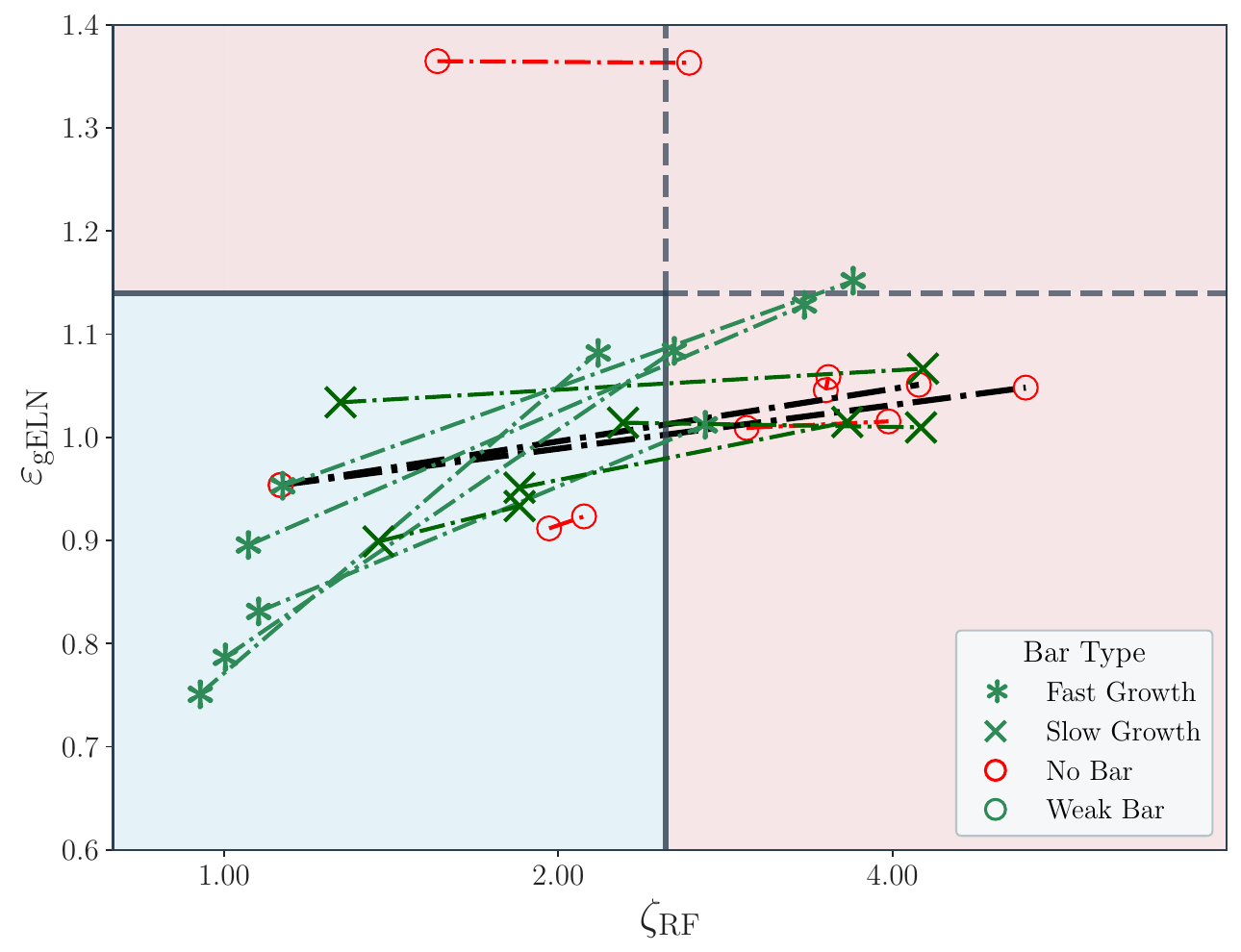}
    \caption{Evolution of models with low $\epsELN$ in the $(\epsELN, \zeta_{\rm RF})$ space, between their initial state at $t=100$~Myr and the last snapshot after 5 Gyr. Both $\epsELN$ and $\zeta_{\rm RF}$ increase, notably due to overlapping non-axisymmetric modes heating the disc and changes in the disc scale length. Once a bar has formed, the final values of $\epsELN, \zeta_{\rm RF}$ may not obey the thresholds: the proposed parameter space enables to identify bar-inhibiting axisymmetric conditions prior to bar formation, not {\it a posteriori}. The two black dot-dashed lines indicate the evolution in parameter space of the two models in which gas cooling and star formation have been turned on after the bar has formed, thereby destroying it.}
    \label{fig:ELN_Q_final}
\end{figure}

Thus far, we have inferred the zone of the $(\epsELN, \zeta_{\rm RF})$ space prone to inhibit bar formation for axisymmetric initial conditions, but in reality galaxies evolve with time, and even in such simplified simulations, the presence of a bar affects the disc properties such as its scale length and velocity dispersion which significantly impact the $\epsELN$ and $\zeta_{\rm RF}$ values. To test this, we consider the last snapshot of all the bulgeless models with low $\epsELN$, and compute the new velocity dispersion profiles, gas turbulence and scale lengths of the discs by iteratively fitting the surface density profile from outside the bar region up to the maximum allowed radius, which is $7 R_d$. 

Figure \ref{fig:ELN_Q_final} shows the $\epsELN$ and $\zeta_{\rm RF}$ values of the final snapshots of a sample of bulgeless models with 20\% gas fraction, along with their initial values, connected by a dashed line. When the initial values fall within the $(\epsELN, \zeta_{\rm RF})$  thresholds, the final values may not obey these thresholds. In all cases, the final $\zeta_{\rm RF}$ is higher than the initial $\zeta_{\rm RF}$, which is mostly due to resonant-overlapping multiple modes that naturally raise the velocity dispersion \citep[e.g.][]{Minchev_2010, Minchev_2012a}. As we do not have star formation in our simulation, the stellar mass remains strictly constant, and $V_{\rm max}$ remains roughly constant throughout, but the scale length $R_d$ changes significantly due to the presence of a bar \citep[see for example][]{Minchev_2012a}, which typically raises $\epsELN$. This shows that it \emph{is} possible to find bars in galaxies that have a high $\epsELN$ and $\zeta_{\rm RF}$. However, if the bar is destroyed while the galaxy sits there, it is likely that it cannot easily re-form a bar unless it goes back to the region of parameter space prone to bar formation. 

In summary, axisymmetric galaxies, that lie outside the $(\epsELN, \zeta_{\rm RF})$  thresholds do not form bars, but those initially within the thresholds need not lie within the thresholds once the bar has formed (Fig.~\ref{fig:ELN_Q_final}). Hence, using, for instance, the ELN threshold in observations would be highly dubious, since it should only be used to characterize the state of the disc prior to bar formation. A high ELN value should be sufficient to inhibit bar formation in an axisymmetric disc, but once a bar is formed, nothing prevents the galaxy to reach a high ELN value and keep its bar. A low ELN value, on the other hand, is not a sufficient condition for a bar to form in an axisymmetric disc. Despite these limitations, let us note that \citet[][their Eq. 10]{Romeo_2023} claimed that their {\it modified} version of the ELN parameter failed to distinguish between barred and unbarred galaxies, while \citet{Kashfi_2025} found with the original ELN parameter that only 19\% of unbarred galaxies in the Spitzer Photometry and Accurate Rotation Curves (SPARC) sample \citep{Lelli2016_SPARC} have a low ELN parameter. This fraction is actually similar to the one found in the context of hydrodynamic cosmological simulations, regardless of how badly bar statistics align with observations: \citet{Villalba_2022} indeed have shown that only 20\% of unbarred disc galaxies in TNG50 have a low ELN parameter. 

Let us however note that the heating visible in the displacement of initially cold disc models from left to right in Fig.~\ref{fig:ELN_Q_final}, is obviously not necessarily representative of what happens in real galaxy discs once gas is allowed to cool, and also to be replenished in a cosmological context. In Sect.~\ref{sec:feedback}, we will however show that $\zeta_{\rm RF}$ not only increases through heating of the disc, but can also increase through the formation of a stellar bulge in a context where gas is allowed to cool and form new stars. 

\subsection{Too short (or too slow) bars?}
\label{subsec:short}

\begin{figure*}[htbp]
    \includegraphics[width=\textwidth]{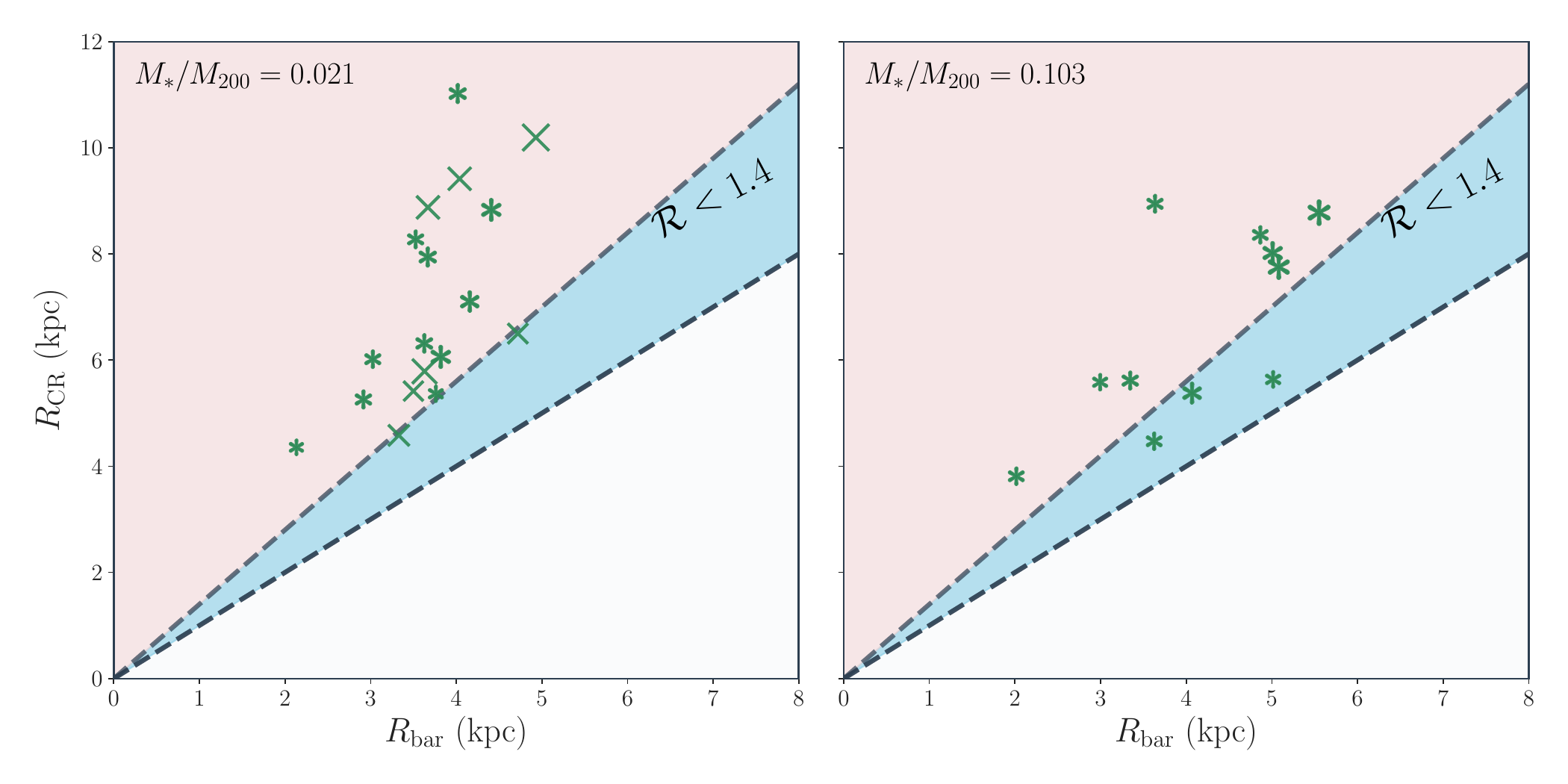}
    \caption{Corotation radius ($R_{\rm CR}$) versus bar length ($R_{\rm bar}$) for all barred galaxy models, separated according to their stellar-to-halo mass ratio ($M_\star/M_{\rm 200}=0.021$ in the left panel, $M_\star/M_{\rm 200}=0.103$ in the right panel). Models with a lower $M_\star/M_{\rm 200}=0.004$ do not form bars (cf. Table~\ref{tab:halo_params}). The fast (large) bar regime $1 < \mathcal{R} < 1.4$ is indicated by the blue shaded region, and the slow (short) bar regime $\mathcal{R} > 1.4$ is indicated by the pink shaded region. Halos hosting fast bars are all embedded in low mass halos with fiducial and low concentrations. 
    }
    \label{fig:R_parameter}
\end{figure*}

When bars form in cosmological simulations, it has long been known that they are typically too short with respect to to their corotation radius, or equivalently too slow for their size \citep[e.g.,][]{Algorry_2017, Roshan2021_2, Frankel_2022, Semczuk_2024, Ansar_2025}. This is typically parametrized by the ratio $\mathcal{R} = R_{\rm CR}/R_{\rm bar}$, where $R_{\rm CR}$ is the corotation radius, i.e. the radius where angular frequency of the bar is equal to the angular frequency of orbits of disc stars, and $R_{\rm bar}$ the bar length. Bars are called `fast' if $1 < \mathcal{R} < 1.4$, and `slow' if $\mathcal{R} > 1.4$ \citep{Debattista_2000}. If one instead considers their size at a given pattern speed, one can say they are `large' if $1 < \mathcal{R} < 1.4$, and `short' if $\mathcal{R} > 1.4$. Observations favour barred galaxies with fast (or large) bars, i.e. $\mathcal{R} < 1.4$ \citep{Corsini_2011, Guo_2019, Cuomo_2019b}. In cosmological simulations, the distribution of pattern speeds often corresponds to observations but the bars are much too short \citep{Frankel_2022}. However, \citet{Fragkoudi_2021} analysed the pattern speed of Milky Way-like galaxies in the \texttt{Auriga} cosmological zoom-in simulations \citep{Grand_2017}, and found that bars remain fast (or large with respect to their corotation radius) throughout their evolution, consistent with observations. They however noticed that their galaxies are baryon-dominated in the central regions, i.e. that they lie above the SHMR \citep[e.g.,][]{Behroozi_2013}.

In Fig.~\ref{fig:R_parameter}, we separate all our barred galaxy models in two panels, those lying on the SHMR and those that have lighter halos, and plot their corotation vs. bar length for the final snapshot of the simulations. Most of the models have $\mathcal{R} \geq 1.4$, while only 3 models have $\mathcal{R} < 1.4$. These models hosting fast bars are {\it all} embedded in low mass haloes with fiducial and low concentrations, in accordance with the results of \citet{Fragkoudi_2021}. It appears that halos of barred galaxies must systematically stand above the SHMR. While this is not a problem at high Milky Way-like stellar masses, where barred galaxies could correspond ta a baryon-dominated sub-sample, it is problematic in the mass range considered here, where observations indicate that the vast majority of galaxies are barred and large (or fast). It is not clear how to solve this problem without reducing the exchange of angular momentum between the disc and the DM halo, which could have profound implications for the nature or even the existence of DM \citep[e.g.,][]{Roshan2021_2, Roshan_2021b, Nagesh_2023}.

\section{Turning on star formation and feedback}
\label{sec:feedback}

\begin{figure*}[htbp]
    \centering
    \includegraphics[width=0.48\textwidth]{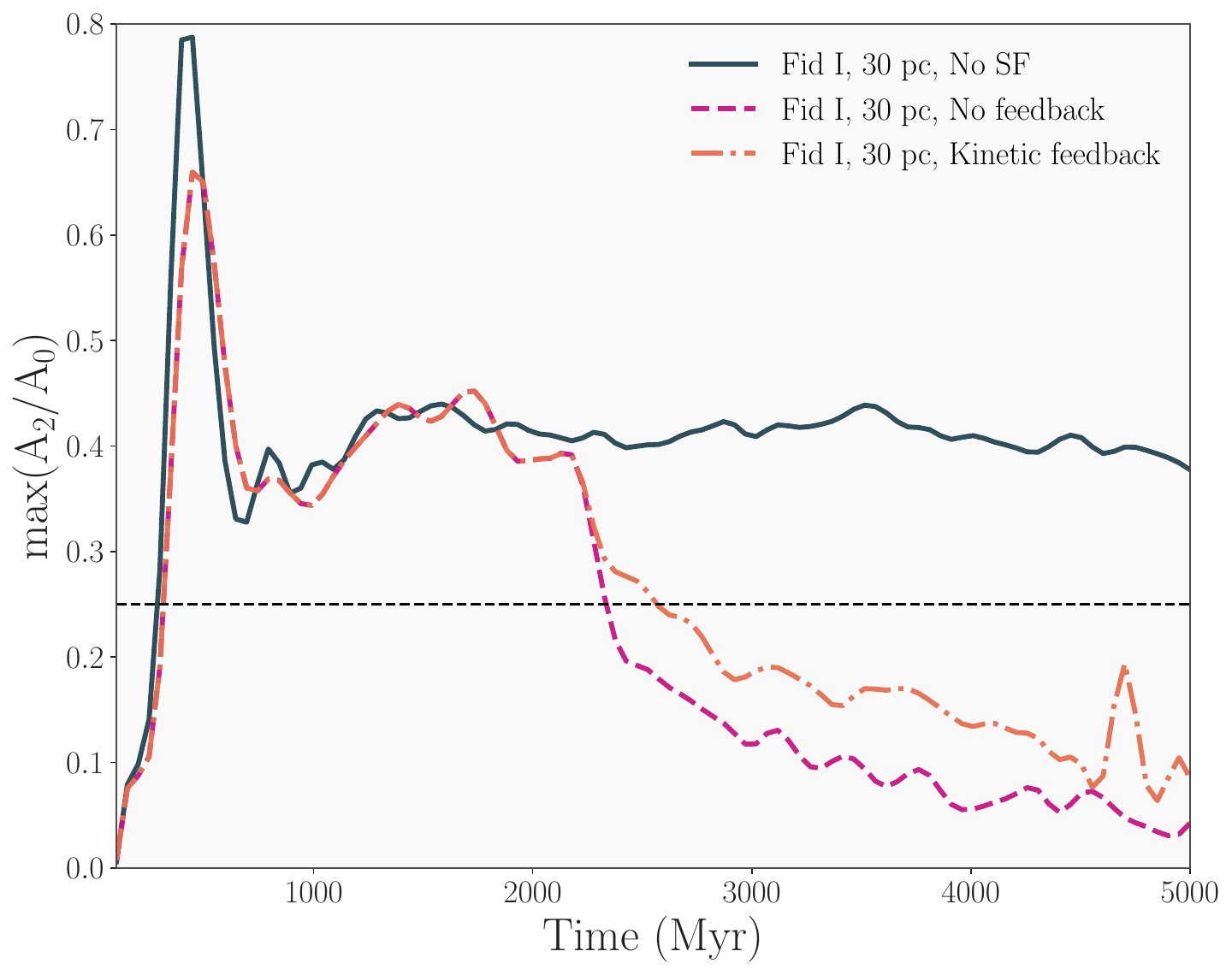}
    \hfill
    \includegraphics[width=0.48\textwidth]{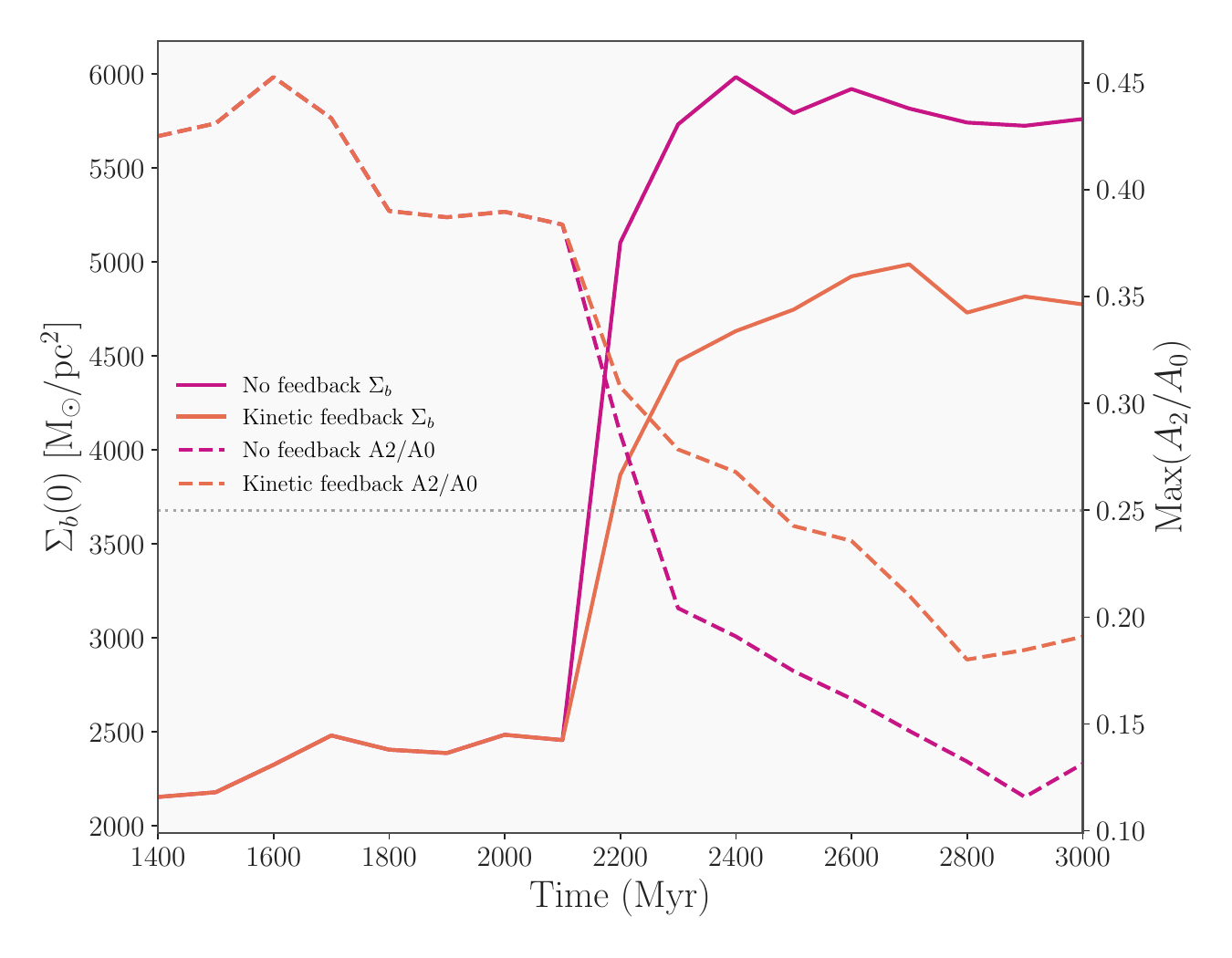}
    \caption{\textit{Left panel}: Evolution of the bar amplitude for the two simulations with full gas physics, compared to the Fiducial Model I (solid black line). Full gas physics is turned on after $\sim 2.1\rm ~Gyr$, once the bar has formed. The bar is then quickly destroyed in both the no-feedback model (dashed magenta line) and the kinetic feedback model (dash-dotted orange line). \textit{Right panel}: Comparison between the evolution of the bar amplitude between 1.4 and 3 Gyr with that of the central baryonic surface density (solid lines). The destruction of the bar coincides with a sharp increase of the central baryonic density. 
    }
\label{fig:BS_feedback}
\end{figure*}

With our improved phenomenological grasp of the conditions under which galactic bars form in a purely hydrodynamic context without star formation nor feedback, we can now speculate further on the causes of the missing bar problem in cosmological simulations. As stated above, we suspect that originally bar-unstable galaxy discs could evolve {\it after bar formation} to a region of parameter space where they cannot easily re-form if they are destroyed. This question needs to be tackled in two steps, first with idealised simulations including gas cooling, star formation and feedback, then within cosmological simulations themselves.

While this question thus deserves a full study on its own, we already ran two idealised simulations that include cooling, star formation, and feedback after a bar has formed, as an example of how a bar could be destroyed in such a more realistic setting. The parameters for the ICs of the two simulations are those of Fiducial Model I\footnote{Except that we implement an exponential vertical gaseous profile for the gaseous disc (in line with the external potential profile in \texttt{AGAMA}).}. The system is run without the implementation of subgrid recipes for the first $\sim 2.1$ Gyr to allow the bar to grow, after which, cooling, star formation, and feedback are turned on. The key difference between the two simulations resides in the feedback implemented, namely one with effectively no feedback, and one with efficient kinetic feedback. 

Star formation in \ramses\ uses a density threshold: gas in each cell is converted into stellar particles only if the density exceeds a certain threshold, thereby restricting star formation to certain cells. Here, we delay the detailed study of the impact of sub-grid recipes on bars, and adopt the default model for star formation in \ramses, following a Schmidt law. Namely, the gas is converted into stellar particles at a rate $\dot{\rho}_\star = \epsilon_\star ~\rho / t_{\rm ff}$, where $\rho$ is the gas mass density, $t_{\rm ff}= [3\pi/32G\rho]^{1/2}$ the free fall time, and $\epsilon_\star$ the star formation efficiency per free fall time (the conversion efficiency). We adopt the default density threshold of $0.1$ H/cc for both simulations. We further set $\epsilon_\star=0.02$ for the simulation with no feedback and $\epsilon_\star=  0.04$ for the simulation which adopts kinetic feedback, based on roughly matching the Kennicutt–Schmidt law. The resulting (born) stellar particles are drawn stochastically from a Poisson distribution, with the resulting mass being an integer multiple of the mass resolution \citep[see][for further details]{Rasera_2005,DT08}. We further set the cooling floor to 10K.

In the first model, thermal feedback has been formally implemented but results in effectively no feedback. The thermal energy and mass from the supernova explosions is  dumped into the cell hosting the stellar particle, followed by the removal of mass from the particle. The ejected mass is set by the supernova mass fraction parameter \texttt{eta\_sn}, which is taken as 0.1. The time after which a star particle undergoes a supernova explosion is set to be 10 Myr after its birth. With such a parametrisation, the resolution of the simulation is not high enough to resolve the Sedov-Taylor phase of the explosion and the thermal energy radiates away before the development of adiabatic shocks. This is why this thermal feedback simulation is actually equivalent to a ``no feedback'' simulation, naturally leading to the quick formation of an overcooled bulge-like central mass concentration that is expected to destroy the bar. As shown in the left panel of Fig.~\ref{fig:BS_feedback}, the bar in this simulation is indeed immediately destroyed after star formation starts.

Rather than explicitly resolving the Sedov–Taylor phase of the explosion, the kinetic feedback model of \citet{DT08} bypasses this stage by directly injecting its estimated averaged outcome. For this second simulation, we set the parameter controlling the fraction of total supernova energy released in the kinetic form $\texttt{f\_ek} = 0.4$. Since mass and momentum are injected within a bubble radius of the exploding particle, we set the parameter $\texttt{rbubble}= 150~\rm pc$. Lastly, we set the mass loading factor $\texttt{f\_{w}} = 1$. These are standard parameters for kinetic feedback in \ramses\ but, apart from slightly delaying the destruction timescale of the bar, it does not prevent it from being destroyed, as shown in the left panel of Fig.~\ref{fig:BS_feedback}. 

These two simulations suggest that the formation of a central mass concentration, fueled by standard star formation prescriptions, can be very efficient at destroying bars. Indeed, the right panel of Fig.~\ref{fig:BS_feedback} shows that the destruction of the bar coincides with a sharp increase in the central baryonic density. The correlation between these two events does not imply direct causation, but they likely have a common cause, namely the important inflow of gas related to gas cooling, followed by clump formation with fast inspiral towards the centre, with important associated gravity torques destroying the bar \citep[e.g.,][]{Bournaud_2005}. Furthermore, a standard feedback model appears insufficient to prevent the destruction of the bar in such a violent bulge-formation episode. 

It is particularly interesting that once the bar is destroyed in both simulations, the galaxy lies outside the region of parameter space favouring the re-formation of a bar (Fig.~\ref{fig:ELN_Q_final}). 
Indeed, $\zeta_{\rm RF}$ passes above its threshold due to the formation of a bulge with  $p \sim 0.35$. For the disc to become bar-unstable again, the minimum Romeo-Falstad stability parameter would need to drop back close to $\sim 1.25$; however, the resulting inside-out growth would typically increase the disc scale-length even more, thereby shifting the galaxy  outside the bar-unstable $\epsELN$ range. We suspect a similar mechanism may operate in cosmological simulations, though a detailed investigation is deferred to future work.

\section{Conclusions and discussion}
\label{sec:Conclusions}

In this study, we re-examined the conditions for bar formation and inhibition in disc galaxies, using a grid of idealised hydrodynamic simulations, all with a total stellar mass of $\sim 10^{10} M_\odot$, starting from slightly out of equilibrium conditions. The simulations were performed with \ramses, with spatial resolution and particles mass motivated by a resemblance to those of the {\tt NewHorizon} simulation \citep{Dubois_2021}. Indeed, the ultimate objective of such a study would be to gather hints on the roots of the missing bar problem in such cosmological simulations \citep{Reddish_2022}. 

We considered 62 simulations without cooling, nor star formation, nor feedback, in order to first address the mechanisms of bar formation and inhibition from a purely dynamical perspective in the presence of gas. While any detailed quantitative conclusions would obviously remain tied to the numerical schemes considered here, our actual purpose is rather to spot the general trends, giving us hints about the complex non-linear physics at play, which is still not fully under control in our present-day theoretical understanding. Since our ultimate goal is to understand the missing bar problem in cosmological simulations such as {\tt NewHorizon}, we are mostly interested into which parameter values are {\it sufficient} (but not necessary) in order to {\it inhibit} bar formation over a 5~Gyr timescale, corresponding to the time elapsed from $z \sim 1$ to $z \sim 0.2$, or equivalently which parameter values appear to be {\it necessary} (but not sufficient) for a bar to form on such a timescale. 

We started by exploring one aspect of our adopted numerical scheme, namely showing that increasing the number of DM particles in our simulations actually \emph{decreases} the amplitude and the growth rate of the bar, meaning that the individual particle mass ratio of DM and stellar particles is unlikely to be the cause of the missing bar problem in cosmological simulations. Interestingly, we also showed that, everything else being equal, decreasing the gas mass also slows down the growth of the bar. We then varied the gas disc mass, stellar disc velocity dispersion, bulge mass, DM halo mass, and DM halo concentration in a systematic way, and showed with numerous concrete counter-examples that most of the proposed diagnostics for bar formation in the literature do \emph{not} fully capture the conditions prone to bar formation or inhibition in our own grid of idealised simulations. With no pretense to find any alternative ``absolute criterion'', especially given all our simplifying assumptions (no gas cooling, non-spinning DM halos), we nevertheless identify the regions of parameter space that appear to be most efficient for dynamically inhibiting bar-formation within our own grid of simulations, i.e., under the peculiar code and integration scheme chosen in this study -- as close as possible (AMR code, similar resolution and particle mass) to the {\tt NewHorizon} cosmological simulation. 

The efficient bar-inhibiting regions of parameter space that we identified are given by Eq.~\eqref{eq:ELN} and Eq.~\eqref{eq:QRFlimit}. The first one is a generalised ELN parameter including the stellar mass of the bulge. The ELN parameter has its theoretical root into swing-amplification, as reminded in Eq.~\eqref{eq:X_eps}, but it apparently turns out to be practically \emph{more} discriminating than the swing-amplification parameter itself in our grid of simulations, probably because it is more global, and thereby implicitly better captures the effects of resonant interactions with the outskirts of the live halo. The $\varepsilon_{{}_{\rm ELN}}$ threshold of Eq.~\eqref{eq:ELN} for a pure disc however cannot be a strict one either, as already shown by, e.g., \citet{Frosst} where the dependence of the threshold on orbital time is clear, while a small dependence on gas fraction is also expected from our own results on varying the gas mass. A slight dependence on halo spin is also expected. However, since the originality of our suite of simulations compared to other published ones is that we include a significant fraction of very cold stellar discs, the fact that not a single of our bulgeless models with $\varepsilon_{{}_{\rm ELN}} > 1.14$ manages to form a bar in 5~Gyr despite being extremely cold, indicates that it is likely always difficult for a bar to form under such numerical conditions. The second relevant parameter that we identified is a limiting Romeo-Falstad stability parameter, $Q_{\rm RF}$ as defined in Eq.~\eqref{eq:Q_RF} and taken at its absolute minimum within the disc: the limiting value decreases quadratically with the bulge mass as per Eq.~\eqref{eq:QRFlimit}. In our present grid of simulations, the two identified thresholds are not tied to the 100~Myr snapshot: a very general way of phrasing the results is that every single simulated galaxy that forms a bar in our grid of simulations needs to have at least one snapshot at which it lies below the thresholds in order for the bar instability to be seeded.

The dependence of the $Q_{\rm RF}$ threshold on the bulge mass is an original finding, and deserves a tentative explanation. The formation of a bar is related to an exponentially growing loosely-wound mode, whose amplification is driven by a feedback loop in which a leading wave is swing-amplified around corotation, in the differentially rotating part of the disc, before propagating inwards as a trailing wave through the central regions, where it is reflected and emerges as a leading wave that is swing-amplified once again. Per loop, we can assume the dependence of the amplification factor on the Toomre stability parameter or on its Romeo-Falstad generalisation to be approximately exponential \citep[since the amplification factor is always a rapidly declining function of $Q$, see][]{Toomre_1981, Michikoshi, Hamilton25}, namely $\alpha_0 \approx \exp[s\,(Q_{\rm lim}-Q)] \geq  1$ for $Q \leq Q_{\rm lim}$, where $s$ is a parameter quantifying the sensitivity of the amplification to random motions, and $Q_{\rm lim} \sim 2.5$ is the value of $Q$ above which amplification ceases (i.e., $\alpha_0 = 1$).

In the presence of a bulge, the wavepacket suffers in addition a partial resonant \emph{absorption}, that we can parametrize by another factor $\alpha_1 < 1$, due to the ILR created by the presence of the central mass concentration. Since the bar sheds angular momentum while it grows, its pattern speed always decreases in the growing phase, meaning this resonant absorption does not operate from the start: as long as the pattern speed $\Omega_p$ of the growing mode exceeds the maximum of the precession frequency $\omega \equiv \Omega-\kappa/2$, no ILR exists and the mode can grow unimpeded. The $\alpha_1$ factor introduced here can thus be seen as an effective absorption factor, averaged over the growth of the mode when it still has a rather high pattern speed (meaning that the ILR, if it exists, is rather close to the centre). Then, assuming that a bar can build up only if the average net gain per loop exceeds unity, $\alpha_0 \times \alpha_1 > 1$, and taking the logarithm of this inequality, the necessary condition on the stability parameter $Q$ for a bar to form then simply becomes
\begin{eqnarray}
    Q < Q_{\rm lim} + \frac{\ln \alpha_1}{s}\, ,
    \label{eq:Qcrit}
\end{eqnarray}
where $\ln \alpha_1 < 0$ is directly related to the properties of the bulge. Placing ourselves in the situation where an ILR exists close to the centre, the strength of the resonant absorption acting as the exponent in the absorption factor $\alpha_1$ is typically governed, to lowest order, by the product of the square of the normalized potential of the mode at the ILR radius, $[\delta\Phi(R_{\rm ILR})/\delta\Phi_{\rm max}]^2$ (for a single star) and of the actual number of disc stars in resonance, $N_{\rm res}$, namely $- \ln \alpha_1 \propto N_{\rm res}\delta\Phi(R_{\rm ILR})^2$. Since the Taylor-expanded potential $\delta\Phi$ of any $m=2$ mode close to the centre scales as $\propto R^2$, the squared potential at the ILR is typically proportional to $R_{\rm ILR}^4$. In an inner disc with solid body rotation curve (i.e., with $\omega=0$), the additional presence of a compact bulge raises the precession frequency $\omega(R)$ as a monotonically increasing function of $M_{\rm b}/R^3$, so that the resonance condition implies $R_{\rm ILR}\propto M_{\rm b}^{1/3}$. Moreover, since $\omega$ depends on $R$ only through a power-law in $R^{-3}$, a resonant zone with fixed frequency width $\Delta \omega$ corresponds to a radial width $\Delta R \propto R_{\rm ILR}$, since $\Delta \omega/\Delta R \propto R_{\rm ILR}^{-1}$ at the ILR. This means that the number of disc stars in the resonant zone, $N_{\rm res}$, scales as $R_{\rm ILR}^2$, in turn meaning that $-\ln \alpha_1 \propto R_{\rm ILR}^6$, which gives
\begin{equation}
\ln \alpha_1 \propto - M_{\rm b}^{2}.  
\end{equation}
In our simulations, such a quadratic dependence on the bulge mass would translate into a quadratic dependence on $p \equiv B/T$ because the total stellar mass $M_*$ is held fixed, so that $M_{\rm b}=p\,M_*$. Nevertheless, the exact slope and power index of this dependence of the limiting disc stability parameter on the bulge mass should obviously also be related to multiple secondary factors: these include the properties of the bulge, such as its compactness, the properties of the inner disc, such as its mass and surface density, the inner rotation curve, as well as the details of the physics of the resonant absorption in the bar forming phase. Therefore, the important result here regarding the dependence of the limiting value of the stability parameter on bulge mass remains the general trend rather than the detailed numerical values in the relation, and it does motivate future theoretical explorations of the physics at play.

Regarding the size and pattern speed of bars, we also confirm with our simulations previous results indicating that only baryon-dominated discs away from the stellar-to-halo-mass relation expected from abundance matching can actually form large enough bars with respect to their corotation radius. While this is not problematic at high Milky Way-like stellar masses, where barred galaxies could correspond to a baryon-dominated sub-sample, it actually remains a problem in the mass range considered here, where observations indicate that the vast majority of galaxies are barred and large (or fast). We also remind that parameters such as those identified in Eq.~\eqref{eq:ELN} and Eq.~\eqref{eq:QRFlimit}, quantifying the state of a disc prior to bar formation in idealised simulations, cannot be immediately translated into any kind of direct observational test for the presence of bars in observed galaxies, since a galaxy can move to the bar-inhibiting region of parameter space after bar formation. We show explicitly how this can happen in our own simulations. To illustrate this further, we then carry out two additional simulations including gas cooling and star formation, in which the bar is actually destroyed in relation to important amounts of material inflow towards the centre, whilst the galaxy has already moved to the region of parameter space prone to inhibiting its re-formation. We suspect that this modern version of the angular momentum catastrophe might still be happening in large-volume cosmological simulations as a cause of the missing bar problem.

While the grid of 62+2 idealised simulations analysed here provides a dynamical baseline to study the conditions for bar formation, we acknowledge that the evolution of galaxies in a cosmological context is significantly more complex. Future studies within our research programme will focus on more realistic simulations including, in particular, DM halo spin, gas cooling, star formation and feedback, all from the onset. On the other hand, it would be important to confront the bar-inhibiting regions of parameter space identified here to cosmological simulations themselves. The closest existing attempt in the literature is that of \citet{Rosas_2025}: six Milky Way-like galaxies from TNG50 were re-simulated with the {\tt AREPO} code under seven variations of the feedback models, and their qualitative conclusion matches ours closely: the swing-amplification parameter, averaged within the stellar half-mass radius, lies in the nominally unstable window for all their models including the unbarred ones, and therefore carries no discriminating power, which mirrors what we found for $X_{\rm min}$ in our own simulations, whilst the ELN and Toomre parameters seem to separate bar forming models from unbarred models quite well. However neither of the latter two parameters on its own can fully predict the outcome. For instance, in their models without supernova winds, a bar is prevented from forming with a low classical ELN parameter, whilst in their weak-wind models the stellar Toomre parameter can be particularly high but a bar still forms. Both these failures could possibly be resolved by the genuine improvements brought by our own re-definition of the relevant parameters: the presence of a massive and compact central bulge is probably what kills the bar in their low ELN cases, while the use of the stellar Toomre parameter instead of the Romeo-Falstad parameter neglects the effect of a cold gas component. As an example, a 35\% gaseous disc component with $Q_{\rm gas} \simeq 1$, on top of a stellar disc component with $Q_* \simeq 4$, would lead to a Romeo-Falstad parameter of order $Q_{\rm RF} \simeq 1.4$, and would therefore indeed be expected to form a bar within our own numerical scheme, even in the presence of a bulge making up 30\% of the total stellar mass, leading to $\zeta_{\rm RF}=2.3$. A direct quantitative comparison with their parameters is nevertheless not straightforward since their characteristic Toomre parameter is an average computed within the stellar half-mass radius, whilst ours is the minimum of the profile within the disc. All this motivates future works where galaxy discs formed in cosmological simulations should be analysed through the lens of the bar-inhibiting parameters identified in the present study.

\begin{table*}[]
\caption{Parameters of the different models, following Table~\ref{tab:Fiducial_paramss} for the Fiducial Models, including relevant parameters dicussed in Sect.~4. The gas fraction $f_g$ is in \%, masses are is in units of $10^{10}M_\odot$, lengths $R_{\rm d}$, $R_{\sigma,d}$ in kpc, and $\sigma_{r,d,0}$ in km/s. The latter is the value after evolving the model for 100~Myr. Common parameters between all the models are discussed in section~\ref{sec:Fiducial_model}.
    }
    \label{tab:halo_params}
\begin{tabular}{lcccccccccccccl}
\hline\hline
\# & $M_{\rm 200}$ & $c_{\rm 200}$ & $M_{\rm *,d}$  & $M_{\rm g}$ & $M_{\rm b}$ & $R_{\rm d}$ & $\sigma_{r,d,0}$ & $R_{\sigma,d}$ & $f_g$ & $\epsELN$ & $Q_{\rm RF}$ & $p$ & $\zeta_{\rm RF}$ & Bar status\\ \hline
\noalign{\smallskip}  
6 & 48.6 & 12.28 & 1.03 & 0.25 & 0.00 & 2.00 & 22.82 & 4.00 & 20 & 1.03 & 1.27 & 0.00 & 1.27 & Slow growth \\ 
7 & 48.6 & 6.51 & 1.03 & 0.25 & 0.00 & 2.00 & 21.90 & 4.00 & 20 & 0.87 & 1.05 & 0.00 & 1.05 & Fast growth\\ 
 8 & 10.0 & 10.49 & 1.03 & 0.25 & 0.00 & 2.00 & 22.00 & 4.00 & 20 & 0.78 & 1.00 & 0.00 & 1.00 & Fast growth\\
 9 & 10.0 & 14.41 & 1.03 & 0.25 & 0.00 & 2.00 & 22.27 & 4.00 & 20 & 0.83 & 1.07 & 0.00 & 1.07 & Fast growth\\ 
 10 & 10.0 & 7.64 & 1.03 & 0.25 & 0.00 & 2.00 & 21.52 & 4.00 & 20 & 0.75 & 0.95 & 0.00 & 0.95 & Fast growth\\ 
 11 & 236. & 10.47 & 1.03 & 0.25 & 0.00 & 2.00 & 22.84 & 4.00 & 20 & 1.49 & 1.49 & 0.00 & 1.49 &   \\
 12 & 236. & 5.55 & 1.03 & 0.25 & 0.00 & 2.00 & 23.16 & 4.00 & 20 & 1.16 & 1.17 & 0.00 & 1.17 &   \\ 
\hline
 13 & 48.6 & 12.28 & 1.03 & 0.05 & 0.00 & 2.00 & 22.57  & 4.00 & 05 & 1.02 & 1.30 & 0.00 & 1.30 & Slow growth \\
 14 & 48.6 & 12.28 & 1.03 & 0.68 & 0.00 & 2.00 & 23.55 & 4.00 & 40 & 1.06 & 1.22 & 0.00 & 1.22 & Fast growth \\
 15 & 48.6 & 12.28 & 1.03 & 1.54 & 0.00 & 2.00 & 23.22 & 4.00 & 60 & 1.13 & 1.19 & 0.00 & 1.19 & Fast growth \\
 16 & 48.6 & 8.94 & 1.03 & 0.05 & 0.00 & 2.00 & 22.96 & 4.00 & 05 & 0.93 & 1.18 & 0.00 & 1.18 & Fast growth \\
17 & 48.6 & 8.94 & 1.03 & 0.68 & 0.00 & 2.00 & 22.61 & 4.00 & 40 & 0.97 & 1.10 & 0.00 & 1.10 & Fast growth \\
 18 & 48.6 & 8.94 & 1.03 & 1.54 & 0.00 & 2.00 & 21.96 & 4.00 & 60 & 1.04 & 1.09 & 0.00 & 1.09 & Weak bar \\
 19 & 48.6 & 6.51 & 1.03 & 0.05 & 0.00 & 2.00 & 22.17 & 4.00 & 05 & 0.85 & 1.08 & 0.00 & 1.08 & Fast growth\\
 20 & 48.6 & 6.51 & 1.03 & 0.68 & 0.00 & 2.00 & 22.03 & 4.00 & 40 & 0.90 & 1.01 & 0.00 & 1.01 & Fast growth\\
 21 & 48.6 & 6.51 & 1.03 & 1.54 & 0.00 & 2.00 & 22.52 & 4.00 & 60 & 0.97 & 1.03 & 0.00 & 1.03 & Fast growth\\
 22 & 10.0 & 14.41 & 1.03 & 0.05 & 0.00 & 2.00 & 22.10 & 4.00 & 05 & 0.81 & 1.10 & 0.00 & 1.10 & Fast growth\\
 23 & 10.0 & 14.41 & 1.03 & 0.68 & 0.00 & 2.00 & 22.27 & 4.00 & 40 & 0.87 & 1.04 & 0.00 & 1.04 & Fast growth\\
 24 & 10.0 & 14.41 & 1.03 & 1.54 & 0.00 & 2.00 & 22.60 & 4.00 & 60 & 0.94 & 1.04 & 0.00 & 1.04 & Weak bar \\
 25 & 10.0 & 10.49 & 1.03 & 0.05 & 0.00 & 2.00 & 21.57 & 4.00 & 05 & 0.76 & 1.03 & 0.00 & 1.03 & Fast growth\\
 26 & 10.0 & 10.49 & 1.03 & 0.68 & 0.00 & 2.00 & 22.29 & 4.00 & 40 & 0.83 & 0.97 & 0.00 & 0.97& Fast growth\\
 27 & 10.0 & 10.49 & 1.03 & 1.54 & 0.00 & 2.00 & 21.57 & 4.00 & 60 & 0.91 & 0.97 & 0.00 & 0.97 & Fast growth \\
 28 & 10.0 & 7.64 & 1.03 & 0.05 & 0.00 & 2.00 & 21.64 & 4.00 & 05 & 0.73 & 1.18 & 0.00 & 1.18 & Fast growth\\
 29 & 10.0 & 7.64 & 1.03 & 0.68 & 0.00 & 2.00 & 21.60 & 4.00 & 40 & 0.79 & 1.10 & 0.00 & 1.10 & Fast growth\\
 30 & 10.0 & 7.64 & 1.03 & 1.54 & 0.00 & 2.00 & 21.89& 4.00 & 60 & 0.87 & 1.09 & 0.00 & 1.09 & Fast growth \\
 31 & 236. & 10.47 & 1.03 & 0.05 & 0.00 & 2.00 & 23.10 & 4.00 & 05 & 1.48 & 1.55 & 0.00 & 1.55 &   \\
 32 & 236. & 10.47 & 1.03 & 0.68 & 0.00 & 2.00 & 24.25 & 4.00 & 40 & 1.50 & 1.43 & 0.00 & 1.43 &  \\
 33 & 236. & 10.47 & 1.03 & 1.54 & 0.00 & 2.00 & 23.09 & 4.00 & 60 & 1.54 & 1.38 & 0.00 & 1.38 &  \\
 34 & 236. & 7.62 & 1.03 & 0.05 & 0.00 & 2.00 & 22.78 & 4.00 & 05 & 1.31 & 1.33 & 0.00 & 1.33 &  \\
 35 & 236. & 7.62 & 1.03 & 0.68 & 0.00 & 2.00 & 23.41 & 4.00 & 40 & 1.33 & 1.24 & 0.00 & 1.24 &   \\
 36 & 236. & 7.62 & 1.03 & 1.54 & 0.00 & 2.00 & 21.93 & 4.00 & 60 & 1.37 & 1.21 & 0.00 & 1.21 &  \\
 37 & 236. & 5.55 & 1.03 & 0.05 & 0.00 & 2.00 & 22.18 & 4.00 & 05 & 1.15 & 1.19 & 0.00 & 1.19 &  \\
 37 & 236. & 5.55 & 1.03 & 0.68 & 0.00 & 2.00 & 22.19 & 4.00 & 40 & 1.19 & 1.13 & 0.00 & 1.13 &  \\
 39 & 236. & 5.55 & 1.03 & 1.54 & 0.00 & 2.00 & 23.08 & 4.00 & 60 & 1.23 & 1.11 & 0.00 & 1.11 &  \\
\hline
 40 & 48.6 & 8.94 & 1.03 & 0.05 & 0.00 & 2.00 & 59.74 & 4.00 & 05 & 0.93 & 2.07 & 0.00 & 2.07 & Slow growth \\
 41 & 48.6 & 8.94 & 1.03 & 0.68 & 0.00 & 2.00 & 59.03 & 4.00 & 40 & 0.97 & 1.53 & 0.00 & 1.53 & Slow growth\\
\hline 
 42 & 48.6 & 8.94 & 1.03 & 0.25 & 0.00 & 2.00 & 77.18 & 10.00 & 20 & 0.95 & 2.96 & 0.00 & 2.96 &   \\
 43 & 48.6 & 8.94 & 1.03 & 0.25 & 0.00 & 2.00 & 57.22 & 40.00 & 20 & 0.95 & 2.29 & 0.00 & 2.29 & Slow growth \\
 44 & 48.6 & 8.94 & 1.03 & 0.25 & 0.00 & 2.00 & 65.68 & 40.00 & 20 & 0.95 & 2.56 & 0.00 & 2.56 &   \\
\hline
 45 & 10.0 & 7.62 & 1.03 & 0.25 & 0.00 & 3.00 & 55.08 & 4.00 & 20 & 0.90 & 1.38 & 0.00 & 1.38& Slow growth \\
 46 & 10.0 & 7.62 & 1.03 & 0.25 & 0.00 & 3.00 & 59.71 & 6.00 & 20 & 0.90 & 1.96 & 0.00 & 
 1.96 \\
 47 & 48.6 & 8.94 & 1.03 & 0.25 & 0.00 & 4.00  & 24.14 & 8.00 & 20 & 1.34 & 1.56 & 0.00 & 1.56 & \\ 
    
 48 & 48.6 & 8.94 & 1.03 & 0.25 & 0.00 & 4.00  & 61.84 & 8.00 & 20 & 1.34 & 2.90 & 0.00 & 2.90 &  \\
\hline
 49 & 48.6 & 8.94 & 0.79 & 0.25 & 0.24 & 2.00 & 37.58 & 4.00 & 20 & 0.95 & 1.36 & 0.23 & 1.89 & Fast growth \\
 50 & 48.6 & 8.94 & 0.79 & 0.05 & 0.24 & 2.00 &  38.39 & 4.00 & 05 & 0.93 & 1.44 & 0.23 & 1.29 & Slow growth\\
 51 & 48.6 & 8.94 & 0.65 & 0.05 & 0.38 & 2.00 & 46.70 & 4.00 & 05 & 0.93 & 1.69 & 0.37 & 3.06 &  \\
 52 & 48.6 & 8.94 & 0.79 & 0.68 & 0.24 & 2.00 & 37.25 & 4.00 & 40 & 0.97 & 1.30 & 0.23 & 1.83 & Fast growth \\
 53 & 48.6 & 8.94 & 0.65 & 0.68 & 0.38 & 2.00 & 51.80 & 4.00 & 40 & 0.97 & 1.44 & 0.37 & 2.81 &  \\
\hline
 54 & 48.6 & 8.94 & 0.54 & 0.05 & 0.47 & 2.00 & 56.84 & 12.00 & 05 & 0.93 & 3.69 & 0.46 & 5.80 &  \\
 55 & 48.6 & 8.94 & 0.54 & 0.25 & 0.47 & 2.00 & 54.86 & 12.00 & 20 & 0.95 & 3.18 & 0.46 & 5.30 &   \\
 56 & 48.6 & 8.94 & 0.54 & 0.68 & 0.47 & 2.00 & 53.83 & 12.00 & 40 & 0.97 & 2.24 & 0.46 & 4.35 &   \\
\hline
 57 & 48.6 & 8.94 & 0.72 & 0.25 & 0.31 & 2.00 & 41.56 & 4.00 & 20 & 0.95 & 1.40 & 0.30 & 2.30 & Slow growth \\
58 & 48.6 & 8.94 & 0.60 & 0.25 & 0.43 & 2.00 & 49.64 & 4.00 & 20 & 0.95 & 2.11 & 0.42 & 3.87 & \\
 59 & 48.6 & 8.94 & 0.91 & 0.25 & 0.14 & 2.00 & 43.95 & 5.30 & 20 & 0.95 & 1.72 & 0.13 & 1.89 & Slow growth \\
\hline
 60 & 48.6 & 8.94 & 0.78 & 0.25 & 0.25 & 2.00 & 51.08 & 4.30 & 20 & 0.95 & 1.81 & 0.24 & 2.38 & Slow growth \\
 61 & 48.6 & 8.94 & 0.78 & 0.25 & 0.25 & 2.00 & 50.55 & 4.10 & 20 & 0.95 & 1.88 & 0.24 & 2.12 & Slow growth \\
 62 & 48.6 & 8.94 & 0.78 & 0.25 & 0.25 & 2.00 & 58.33 & 4.30 & 20 & 0.95 & 1.98 & 0.24 & 2.55 &  \\
 \hline 
    \end{tabular}
\end{table*}

\begin{acknowledgements}
     STN is very grateful to Eugene Vasiliev for his support and guidance with \agama\ and to Bharat Bhatt for his work on simulations with star formation and feedback. STN also thanks Florent Renaud, Katarina Kraljic and Chanda Jog for insightful discussions. STN initiated this work in Strasbourg, under the ERC grant No. 834148. The work of S.T.N. and Y.R. was supported by the Swiss National Science Foundation under funding reference IZURZ2 224972.
\end{acknowledgements}

\bibliographystyle{aa}
\bibliography{DM_article.bib}

%
%
\end{document}